\documentclass[aps,twocolumn,letterpaper,notitlepage]{revtex4-1}
\usepackage[T1]{fontenc}
\usepackage[latin9]{inputenc}
\usepackage{amsmath}
\usepackage{amssymb}
\usepackage{stmaryrd}
\usepackage{physics}
\usepackage[mathscr]{euscript}
\usepackage[shortlabels]{enumitem}

\addtocounter{secnumdepth}{4}

\usepackage[usenames,dvipsnames]{xcolor}
\usepackage[hyperindex,pdftex, breaklinks,colorlinks = true,linkcolor = blue,urlcolor=blue,citecolor=blue]{hyperref}

\begin{document}
	
	\title{Magnetoelectric Polarizability: A Microscopic Perspective}
	
	\author{Perry T. Mahon}
	\email{pmahon@physics.utoronto.ca}
	\affiliation{Department of Physics, University of Toronto, Toronto, Ontario M5S 1A7, Canada}
	
	\author{J. E. Sipe}
	\email{sipe@physics.utoronto.ca}
	\affiliation{Department of Physics, University of Toronto, Toronto, Ontario M5S 1A7, Canada}
	
	\date{\today}
	
\begin{abstract}
	We extend a field theoretic approach for the investigation of the electronic charge-current density response of crystalline systems to arbitrary electromagnetic fields. The approach leads to the introduction of microscopic polarization and magnetization fields, as well as free charge and current densities, the dynamics of which are described by a lattice gauge theory. The spatial averages of such quantities constitute the fields of macroscopic electrodynamics. We implement this formalism to study the modifications of the orbital electronic properties of a class of insulators due to uniform dc electric and magnetic fields, at zero temperature. To first order in the electric and magnetic fields, the free charge and current densities vanish; thus the linear effect of such fields is captured by the first-order modifications of the microscopic polarization and magnetization fields. Associated with the dipole moment of the microscopic polarization (magnetization) field is a macroscopic polarization (magnetization), for which we extract various tensors relating it to the electric and magnetic fields. We focus on the orbital magnetoelectric polarizability (OMP) tensor, and find the accepted expression as derived from the ``modern theory of polarization and magnetization.'' Since our results are based on the spatial averages of microscopic polarization and magnetization fields, we can identify the distinct contributions to the OMP tensor from the perspective of this microscopic theory, and we establish the general framework in which extensions to finite frequency can be made.
\end{abstract}

\maketitle

\section{Introduction}

Interest in describing the response of insulators to external electromagnetic fields dates back to the earliest studies of electricity and magnetism. In pioneering work near the start of the twentieth century, Lorentz \cite{Lorentz} based his definition of the macroscopic polarization and magnetization fields on a physical picture of molecules with electric and magnetic moments \cite{Jackson}, and from that perspective addressed the response of the macroscopic quantities to the electromagnetic field.

Near the end of the twentieth century a new approach, called the ``modern theory of polarization and magnetization,'' was introduced \cite{Resta1994,Resta2005,Resta2006,Niu2007}. Largely focused on the static response of crystalline materials to uniform fields, the microscopic underpinning was now electronic Bloch eigenfunctions \cite{RestaX}, or alternately the spatially localized Wannier functions that could be constructed from them. However, a macroscopic perspective was taken to define the polarization and magnetization. For example, if one imagined a slow variation in material parameters leading to a macroscopic current density $\boldsymbol{J}$, the polarization (or at least its change) could be defined through $\boldsymbol{J}=d\boldsymbol{P}/dt$ \cite{KingSmith1993}. Thus, instead of a microscopic picture of the
underlying position and motion of charges \textit{leading} to the definition of macroscopic quantities, as it did for Lorentz, links to a microscopic picture \textit{follow} from the definitions. In the ``modern theory,'' the macroscopic polarization was found to be related to the dipole moment of a Wannier function and its associated nucleus \cite{Resta1994}. The ambiguity of which nucleus to associate with a given Wannier function -- the ``closest,'' or one some number of lattice spacings away? -- leads to a ``quantum of ambiguity'' in the macroscopic polarization itself. Such ambiguities are inherent to the ``modern theory,'' and can generally be related to the behavior and description of charges and currents at the surface of a finite sample \cite{VanderbiltBook}.

We have recently argued \cite{Mahon2019} that it is useful to expand upon the approach of Lorentz by introducing \textit{microscopic} polarization and magnetization fields in bulk crystals, and defining the corresponding macroscopic fields as their spatial averages; in general, microscopic ``free'' charges and currents, and their spatial averages, are also introduced, and the resulting description takes the form of a generalized lattice gauge theory. The strategy employed is an extension of that used to introduce microscopic polarization and magnetization fields for atoms and molecules \cite{Healybook,PZW}, which itself is an extension of Lorentz' characterization of molecules by a series of multipole moments. Such microscopic polarization and magnetization fields allow for the visualization of electronic dynamics, in the sense that perturbative modifications to these microscopic fields arising due to the electromagnetic field can be found and exhibited if one has the Wannier functions in hand; existing schemes, which are primarily ab initio based \cite{WF1,WF2}, can be used to construct such Wannier functions. In the usual ``long-wavelength limit'' of optics, where the electric field is varying in time but its variation in space is neglected, we recover the standard results for crystalline solids. In other instances where comparison with the ``modern theory'' is possible, such as the modification of the polarization due to a static or uniform electric field, or similar modifications in systems expected to exhibit the quantum anomalous Hall effect, we also find agreement \cite{Mahon2019}.

The approach we implement has the advantage that it can be employed to describe the effects of spatially varying, time-dependent electromagnetic fields \footnote{While in past work \cite{Mahon2019} and in this paper we treat the electromagnetic field classically, quantum mechanical effects can, in principle, be taken into account.}. Exploring its characterization of this generalized optical response, in the linear and nonlinear regimes, is our main program. There is, however, an interesting overlap of our program with that of the ``modern theory,'' and that is in calculating the static modification of the polarization due to a uniform magnetic field, and the static modification of the magnetization due to a uniform electric field. This phenomenon is termed the \textit{magnetoelectric effect}, and for a class of insulators \footnote{This includes both ordinary and $\mathbb{Z}_2$ topological insulators. We discuss this further below.} in the ``frozen-ion'' approximation, where spin contributions are also neglected, both of these modifications are described by the orbital magnetoelectric polarizability (OMP) tensor \cite{Qi2008,Vanderbilt2009,Malashevich2010,Essin2010,Swiecicki2014,VanderbiltBook},
\begin{align}
\alpha^{il}=\left.\frac{\partial P^i}{\partial B^l}\right|_{\substack{\boldsymbol{E}=\boldsymbol{0}\\\boldsymbol{B}=\boldsymbol{0}}}=\left.\frac{\partial M^l}{\partial E^i}\right|_{\substack{\boldsymbol{E}=\boldsymbol{0}\\\boldsymbol{B}=\boldsymbol{0}}}. \label{OMP}
\end{align}
Unlike its generalization to finite frequency, the OMP tensor is nonvanishing only when both spatial-inversion and time-reversal symmetry are broken in the unperturbed system \footnote{More precisely, the OMP tensor vanishes modulo a discrete ambiguity when time-reversal or inversion symmetry are present in the unperturbed system.}, and it is composed of two distinct terms: the Chern-Simons and the cross-gap contributions. The former is isotropic and entirely a property of the subspace spanned by the cell-periodic functions associated with the originally occupied energy eigenfunctions, while the latter involves both occupied and excited eigenvectors, and the corresponding energies, of the unperturbed system. The Chern-Simons contribution has generated particular interest in the literature because of its topological features; there is a discrete ambiguity in its value, which can be used to identify $\mathbb{Z}_{2}$ topological insulators \cite{VanderbiltBook,Kane2010,Tokura2019}. As well, while the analytic structure of the cross-gap contribution is of the form one would expect to find from the usual treatment of linear response using a Kubo formalism, the Chern-Simons contribution takes a rather unexpected form.

The expression for the OMP tensor found via the ``modern theory'' is well established \cite{Essin2010,Malashevich2010}. In this paper, we present a calculation of the OMP tensor within our framework of identifying microscopic polarization and magnetization fields. It is a special case of our general approach, in which by using a set of orthogonal functions that are well-localized spatially one can associate a portion of a total quantity with the point about which each of these functions is localized; a total quantity can be decomposed into ``site'' contributions. With this, our goal is to formulate the relation of these site quantities to the electric and magnetic fields evaluated at that site. The OMP tensor is extracted upon taking these electric and magnetic fields to be uniform in space and independent of time. Our results are in complete agreement with those of the ``modern theory,'' as would be expected, but in the process we achieve some insight into the microscopic origin of the distinct contributions to (\ref{OMP}). In particular, we can compare our results for the OMP tensor with what would be expected in a ``molecular crystal limit,'' a model in which at each lattice site there is a molecule with orbitals that share no common support with the orbitals of molecules at other lattice sites. And with the development of the formalism presented here we position ourselves to extend this approach to describe material response at finite frequency.

After some preliminary discussion to begin Sec.~\ref{Sect:1}, we extend the formalism \cite{Mahon2019} where necessary in order to calculate the modification of a site quantity due to arbitrary electromagnetic fields. This is a very general development, and only in the later sections do we restrict ourselves to the limit of uniform and static electric and magnetic fields. The calculations are made in Sec.~\ref{Sect:2}; in Sec.~\ref{Sect:3}, we show that the accepted expression for the OMP tensor is reproduced. We also calculate the OMP tensor in the molecular crystal limit by two approaches. The first is a direct molecular physics calculation, and the second is by taking the appropriate limit of our general expressions; they agree, as they should. We also discuss the nature of both the Chern-Simons and cross-gap contributions from the perspective of this microscopic theory. In Sec.~\ref{Sect:Conclusion}, we conclude.

\section{Perturbative modifications of the single-particle density matrix}
\label{Sect:1}

The electronic response of a crystalline insulator is a consequence of the evolution of the fermionic electron field operator, $\widehat{\psi}(\boldsymbol{x},t)$. We assume that, in the Heisenberg picture, the dynamics of this object is governed by
\begin{align}
i\hbar\frac{\partial\widehat{\psi}(\boldsymbol{x},t)}{\partial t}=\Big[H_{0}\big(\boldsymbol{x},\boldsymbol{\mathfrak{p}}_{\text{mc}}(\boldsymbol{x},t)\big)+e\phi(\boldsymbol{x},t)\Big]\widehat{\psi}(\boldsymbol{x},t), \label{fieldEvo}
\end{align}
where $e=-|e|$ is the electron charge,
\begin{align*}
\boldsymbol{\mathfrak{p}}_{\text{mc}}(\boldsymbol{x},t)=\boldsymbol{\mathfrak{p}}(\boldsymbol{x})-\frac{e}{c}\boldsymbol{A}(\boldsymbol{x},t),
\end{align*}
and $H_0\big(\boldsymbol{x},\boldsymbol{\mathfrak{p}}(\boldsymbol{x})\big)$ is the differential operator that governs the dynamics of the electron field in the unperturbed infinite crystal. The presence of a classical electromagnetic field, described by its vector and scalar potentials, $\boldsymbol{A}(\boldsymbol{x},t)$ and $\phi(\boldsymbol{x},t)$, has been included through the usual minimal coupling prescription. In writing (\ref{fieldEvo}) the independent particle approximation is made, neglecting any interactions apart from those described by the coupling of the electron field operators to the electromagnetic field, the associated electric and magnetic fields of which are taken to be the macroscopic Maxwell fields; local field corrections are thus neglected. The Maxwell fields are assumed to be nonvanishing only at times greater than an initial time at which the system is taken to be in its unperturbed zero temperature ground state, and the expectation values of pairs of (Heisenberg) field operators $\widehat{\psi}(\boldsymbol{x},t)$ and their adjoints in the unperturbed ground state are used to construct the minimal coupling Green functions \cite{Mahon2019}.

Implementing the frozen-ion approximation, we take
\begin{align}
H_0\big(\boldsymbol{x},\boldsymbol{\mathfrak{p}}(\boldsymbol{x})\big)=\frac{\big(\boldsymbol{\mathfrak{p}}(\boldsymbol{x})\big)^{2}}{2m}+V(\boldsymbol{x}),\label{H}
\end{align} 
where $V(\boldsymbol{x})$ is the spatially periodic lattice potential that characterizes the crystal structure and satisfies $V(\boldsymbol{x})=V(\boldsymbol{x}+\boldsymbol{R})$ for all Bravais lattice vectors $\boldsymbol{R}$, and 
\begin{align}
\boldsymbol{\mathfrak{p}}(\boldsymbol{x})=\frac{\hbar}{i}\boldsymbol{\nabla}-\frac{e}{c}\boldsymbol{A}_{\text{static}}(\boldsymbol{x}).
\label{physicalP}
\end{align}
In (\ref{physicalP}) we have allowed for the presence of an ``internal,'' static, cell-periodic magnetic field described by the vector potential $\boldsymbol{A}_{\text{static}}(\boldsymbol{x})$, where $\boldsymbol{A}_{\text{static}}(\boldsymbol{x})=\boldsymbol{A}_{\text{static}}(\boldsymbol{x}+\boldsymbol{R})$. The inclusion of such an ``internal'' field respects the discrete translational symmetry of the crystal, but generically leads to a Hamiltonian (\ref{H}) with broken time-reversal symmetry, which will be important in what follows. In future publications we plan to include both the spin-orbit and Coulomb interactions that we neglect here.

In spatially periodic systems, a set of exponentially localized Wannier functions (ELWFs), $\{W_{\alpha\boldsymbol{R}}(\boldsymbol{x})\equiv\braket{\boldsymbol{x}}{\alpha\boldsymbol{R}}\}$, can generally be constructed \cite{Brouder,Marzari2012,Vanderbilt2012,Panati,Troyer2016} via
\begin{align}
\ket{\alpha\boldsymbol{R}}=\sqrt{\frac{\Omega_{uc}}{(2\pi)^3}}\int_{\text{BZ}}d\boldsymbol{k}e^{-i\boldsymbol{k}\boldsymbol{\cdot}\boldsymbol{R}}\sum_{n}U_{n\alpha}(\boldsymbol{k})\ket{\psi_{n\boldsymbol{k}}}, \label{WF}
\end{align}
where $\Omega_{uc}$ is the unit cell volume, and 
\begin{align}
\psi_{n\boldsymbol{k}}(\boldsymbol{x})\equiv\braket{\boldsymbol{x}}{\psi_{n\boldsymbol{k}}}=\frac{1}{\sqrt{(2\pi)^3}}e^{i\boldsymbol{k}\boldsymbol{\cdot}\boldsymbol{x}}u_{n\boldsymbol{k}}(\boldsymbol{x})
\end{align}
are eigenfunctions of (\ref{H}) that are normalized over the infinite crystal such that $\braket{\psi_{m\boldsymbol{k}'}}{\psi_{n\boldsymbol{k}}}=\delta_{nm}\delta(\boldsymbol{k}-\boldsymbol{k}')$. A periodic gauge choice is made such that the energy eigenvectors $\ket{\psi_{n\boldsymbol{k}}}$ and the unitary operator $U(\boldsymbol{k})$ are periodic over the first Brillouin zone \footnote{More precisely, $\ket{\psi_{n\boldsymbol{k}+\boldsymbol{G}}}=\ket{\psi_{n\boldsymbol{k}}}$ and $U(\boldsymbol{k}+\boldsymbol{G})=U(\boldsymbol{k})$ for any reciprocal lattice vector $\boldsymbol{G}$. For more details, see, e.g., Vanderbilt \cite{VanderbiltBook}.}. Associated with each Bloch eigenfunction $\psi_{n\boldsymbol{k}}(\boldsymbol{x})$ is an energy $E_{n\boldsymbol{k}}$ and a cell-periodic function $u_{n\boldsymbol{k}}(\boldsymbol{x})\equiv\braket{\boldsymbol{x}}{{n\boldsymbol{k}}}$ satisfying the orthogonality relation $\left({m\boldsymbol{k}}|{n\boldsymbol{k}}\right)=\delta_{nm}$; we adopt the notation
\begin{align}
(g|h)\equiv\frac{1}{\Omega_{uc}}\int_{\Omega_{uc}}g^{*}(\boldsymbol{x})h(\boldsymbol{x})d\boldsymbol{x}
\end{align}
for functions $g(\boldsymbol{x})\equiv\braket{\boldsymbol{x}}{g}$ and $h(\boldsymbol{x})\equiv\braket{\boldsymbol{x}}{h}$ that are periodic over a unit cell, where the integration is over any unit cell. Also, we restrict our study to three-dimensional systems \footnote{The derived expressions can later be applied to lower dimensional systems by confining the Bloch and Wannier functions to the appropriate subspace of $\mathbb{R}^3$. However, in systems with spatial dimension less than three, the Chern-Simons contribution vanishes. Thus, three-dimensional systems are of primary interest here.}. Here and below $n$ is a band index, $\alpha$ is a type index, and $\hbar\boldsymbol{k}$ denotes a crystal-momentum within the first Brillouin zone.

In this paper we initiate our considerations with the zero temperature ground state of an insulator, and consider the class of insulators for which the sets of occupied and unoccupied energy eigenfunctions each map to a set of ELWFs. This class includes ordinary insulators \footnote{By ``ordinary insulator'' we mean crystalline insulators supporting Bloch energy eigenvectors for which there exists no topological obstruction to choosing a smooth gauge that can respect some underlying symmetry of the system. For instance, there exists no obstruction to choosing a time-reversal or inversion symmetric gauge for a system with the same discrete symmetry.} and $\mathbb{Z}_2$ topological insulators \cite{Vanderbilt2012}, but excludes, for example, Chern insulators \cite{Troyer2016}. Generally a filling factor $f_n$ is associated with each $\ket{\psi_{n\boldsymbol{k}}}$ that is either $0$ or $1$, and here we introduce an analogous filling factor $f_{\alpha}$ associated with each $\ket{\alpha\boldsymbol{R}}$ that is also either $0$ or $1$; the latter can be inferred directly from the occupancy of the energy eigenvectors used in the construction of a particular ELWF. Thus, for the class of insulators we consider, $U_{n\alpha}(\boldsymbol{k})\neq0$ only if $f_n=f_{\alpha}$ \cite{WF1}.

The ELWFs (\ref{WF}) can generally be expressed as 
\begin{align*}
\braket{\boldsymbol{x}}{\alpha\boldsymbol{R}}=\sqrt{\Omega_{uc}}\int_{\text{BZ}}\frac{d\boldsymbol{k}}{(2\pi)^{3}}e^{i\boldsymbol{k}\boldsymbol{\cdot}(\boldsymbol{x}-\boldsymbol{R})}\braket{\boldsymbol{x}}{\alpha\boldsymbol{k}},
\end{align*}
where another set of cell-periodic functions, $\{\braket{\boldsymbol{x}}{\alpha\boldsymbol{k}}\}$, have been introduced and are formed via
\begin{align*}
\ket{\alpha\boldsymbol{k}}\equiv\sum_{n}U_{n\alpha}(\boldsymbol{k})\ket{n\boldsymbol{k}}.
\end{align*}
Here $\braket{\boldsymbol{x}}{\alpha\boldsymbol{k}}$ is in general \textit{not} the cell-periodic part of an energy eigenfunction. The sets of vectors $\{\ket{\alpha\boldsymbol{k}}\}$ and $\{\ket{n\boldsymbol{k}}\}$ are related by a (in general) multiband gauge transformation characterized by $U(\boldsymbol{k})$ \cite{VanderbiltBook}. Although we use Roman and Greek subscripts in the notation $U_{n\alpha}(\boldsymbol{k})$, as we will primarily be considering a transformation from the cell-periodic part of energy eigenfunctions to cell-periodic functions that are not associated with energy eigenfunctions, this need not necessarily hold. Indeed, the simplest type of gauge transformation, although of course it would not generally lead to $\ket{\alpha\boldsymbol{k}}$ associated with ELWFs, is one that involves the $\ket{n\boldsymbol{k}}$ associated with each band individually; such a transformation is achieved by taking $U_{n\alpha}(\boldsymbol{k})$ to be of the form $\delta_{n\alpha}e^{-i\lambda_{n}(\boldsymbol{k})}$, where, for a given $n$ and $\boldsymbol{k}$, $\lambda_{n}(\boldsymbol{k})\in\mathbb{R}$. Rather than a special limit of the general multiband transformation, this could be considered as simply a new choice of Bloch eigenvectors; under Bloch's theorem, energy eigenvectors are uniquely defined only within a $\boldsymbol{k}$-dependent phase \cite{Marzari2012}, even at $\boldsymbol{k}$ points where there is no degeneracy. Here, however, it is considered as one type of $U_{n\alpha}(\boldsymbol{k})$, associated with a gauge transformation of the U$(1)$ type. That is, we consider the vectors $\{\ket{n\boldsymbol{k}}\}$ fixed at the start, and use the term ``gauge dependent'' for quantities that depend generally on the $U_{n\alpha}(\boldsymbol{k})$ and their derivatives, including $U_{n\alpha}(\boldsymbol{k})$ of the U$(1)$ type.

The ELWFs are an important element of our approach, because we use them to introduce ``site'' quantities, and define the macroscopic polarization and magnetization in terms of their moments, as we discuss in detail below. Of course, not all gauge transformations of an initial set $\left\{\ket{n\boldsymbol{k}}\right\}$ will lead to ELWFs via (\ref{WF}), as indicated by the example given above. Nonetheless, the various tensors that describe the modification of the electronic quantities due to the Maxwell fields, including $\alpha^{il}$, must be such that the resulting charge and current densities in the bulk are not only invariant with respect to the choices of $U_{n\alpha}(\boldsymbol{k})$ that lead to ELWFs, but in fact to \textit{all} choices of $U_{n\alpha}(\boldsymbol{k})$; that is, the charge and current densities in the bulk must be gauge invariant. This is plausible because it would be possible -- although we would argue much less convenient and less interesting physically -- to calculate those charge and current densities directly from the minimal coupling Hamiltonian without ever introducing Wannier functions. And we shall see that this gauge invariance does indeed hold. 

A useful identity \cite{Marzari2012} is
\begin{align}
\int W^*_{\beta\boldsymbol{R}}(\boldsymbol{x})x^aW_{\alpha\boldsymbol{0}}(\boldsymbol{x})d\boldsymbol{x}=\frac{\Omega_{uc}}{(2\pi)^{3}}\int_{\text{BZ}}d\boldsymbol{k} e^{i\boldsymbol{k}\boldsymbol{\cdot}\boldsymbol{R}}\tilde{\xi}^a_{\beta\alpha}(\boldsymbol{k}),
\label{firstMoment}
\end{align}
where
\begin{align}
\tilde{\xi}^a_{\beta\alpha}(\boldsymbol{k})\equiv i\left({\beta\boldsymbol{k}}|\partial_a{\alpha\boldsymbol{k}}\right) 
\label{connectionWannier}
\end{align}
is the non-Abelian Berry connection associated with the set $\{\ket{\alpha\boldsymbol{k}}\}$; here and below, superscript indices indicate Cartesian components, repeated Cartesian components are summed over, and we adopt the shorthand $\partial_a\equiv\partial/\partial k^a$. The object (\ref{connectionWannier}) is related to the non-Abelian Berry connection associated with the set $\{\ket{n\boldsymbol{k}}\}$,
\begin{align}
\xi^a_{mn}(\boldsymbol{k})\equiv i\left({m\boldsymbol{k}}|\partial_a{n\boldsymbol{k}}\right),
\label{eq:Bloch_connection}
\end{align}
via
\begin{align}
\sum_{\alpha\beta}U_{m\beta}(\boldsymbol{k})\tilde{\xi}^a_{\beta\alpha}(\boldsymbol{k})U^{\dagger}_{\alpha n}(\boldsymbol{k})=\xi^a_{mn}(\boldsymbol{k})+\mathcal{W}^a_{mn}(\boldsymbol{k}), \label{connection}
\end{align}
Here we have defined the Hermitian matrix $\mathcal{W}^a$ \cite{VanderbiltBook}, populated by elements 
\begin{align}
\mathcal{W}^a_{mn}(\boldsymbol{k})\equiv i\sum\limits_{\alpha}\big(\partial_aU_{m\alpha}(\boldsymbol{k})\big)U^\dagger_{\alpha n}(\boldsymbol{k}), \label{W}
\end{align}
which, for the class of insulators we consider, is nonzero only if $f_m=f_n$. Under the aforementioned periodic gauge choice, all objects appearing in (\ref{connection}) are periodic over the first Brillouin zone. In what follows, the $\boldsymbol{k}$ dependence of the preceding objects is usually kept implicit.

A consequence of the Hamiltonian (\ref{H}), and the resulting dynamics (\ref{fieldEvo}) of the electron field operator and its adjoint is that the differential operators associated with the spatial components of the conserved current take the form 
\begin{align}
J^a_{\text{mc}}\big(\boldsymbol{x},\boldsymbol{\mathfrak{p}}(\boldsymbol{x});t\big)=J^a\big(\boldsymbol{x},\boldsymbol{\mathfrak{p}}_{\text{mc}}(\boldsymbol{x},t)\big)=\frac{e}{m}\mathfrak{p}^a_{\text{mc}}(\boldsymbol{x},t),
\end{align} 
in the usual fashion \footnote{See, e.g., Peskin and Schroeder \cite{Peskin}.}, where $J^a\big(\boldsymbol{x},\boldsymbol{\mathfrak{p}}(\boldsymbol{x})\big)$ are the analogous differential operators arising for the unperturbed system. As a result, another useful identity is
\begin{align}
\int\psi^*_{n'\boldsymbol{k}'}(\boldsymbol{x})\mathfrak{p}^a(\boldsymbol{x})\psi_{n\boldsymbol{k}}(\boldsymbol{x})d\boldsymbol{x}=\mathfrak{p}^a_{n'n}(\boldsymbol{k})\delta({\boldsymbol{k}-\boldsymbol{k}'}), \label{eq:full_integral}
\end{align}
where the matrix elements are
\begin{align}
&\mathfrak{p}^a_{n'n}(\boldsymbol{k})=\delta_{n'n}\frac{m}{\hbar}\partial_a E_{n\boldsymbol{k}}+\frac{im}{\hbar}\big(E_{n'\boldsymbol{k}}-E_{n\boldsymbol{k}}\big)\xi^a_{n'n}(\boldsymbol{k}). \label{pMatrixElements}
\end{align}
This can be shown by breaking the integral in (\ref{eq:full_integral}) into the sum of integrals over unit cells; the sum over Bravais lattice vectors yields the Dirac delta function in (\ref{eq:full_integral}), and the expression for (\ref{pMatrixElements}) follows from the form of the integral over the unit cell and the use of (\ref{eq:Bloch_connection}). Indeed, a more general form of (\ref{eq:full_integral},\ref{pMatrixElements}) can be derived involving the matrix elements of $\mathfrak{p}^{a}(\boldsymbol{x})$ in the basis of the cell-periodic functions $\braket{\boldsymbol{x}}{\alpha\boldsymbol{k}} $ using the same strategy \footnote{Rodrigo A. Muniz, J. L. Cheng, and J. E. Sipe, in preparation}.

In previous work \cite{Mahon2019}, we considered the calculation of the expectation values of the electronic charge and current density operators for a crystalline insulator perturbed by an electromagnetic field. Noting that the lesser, equal time single-particle Green function can be used to find such quantities, we employed a set of spatially localized, ``adjusted Wannier functions'' as a basis in which to expand the electron field operator and its adjoint in an effort to associate portions of the full electronic Green function with individual lattice sites. Upon identifying such ``site Green functions,'' and thereby identifying ``site charge and current densities,'' we defined microscopic polarization and magnetization fields associated with each lattice site using the same functions that are used in atomic and molecular physics to relate the microscopic polarization and magnetization fields of atoms and molecules to the microscopic charge and current densities \footnote{For a review and references to original work see Ref.~\cite{PZW}.}; we call these functions ``relators.'' The full microscopic polarization and magnetization fields are given by summing the respective site contributions. We then insisted that these microscopic polarization and magnetization fields, together with the electronic charge and current density expectation values, satisfy the expressions arising in classical macroscopic electrodynamics relating such quantities. This led to the identification of microscopic ``free'' electronic charge and current densities, which take predictable forms. At zero temperature the first-order modifications of both the free charge and current densities due to the Maxwell fields vanish for the class of insulators considered here, even for electromagnetic fields in the x-ray regime \cite{Mahon2019}. As a consequence, the first-order perturbative modifications to the expectation values of the electronic charge and current density operators resulting from the electromagnetic field can be found directly from the corresponding first-order modifications to the microscopic polarization and magnetization fields.

A quantity central to the calculation of both the microscopic polarization and magnetization fields is the single-particle density matrix, $\eta_{\alpha\boldsymbol{R}'';\beta\boldsymbol{R}'}(t)$. Thus, a starting point in describing the effect of an electromagnetic field to a crystalline insulator is identifying how the Maxwell fields affect this object. The single-particle density matrix evolves according to \cite{Mahon2019}
\begin{align}
 i\hbar\frac{\partial\eta_{\alpha\boldsymbol{R}'';\beta\boldsymbol{R}'}(t)}{\partial t}=\sum_{\mu\nu\boldsymbol{R}_{1}\boldsymbol{R}_{2}}\mathfrak{F}_{\alpha\boldsymbol{R}'';\beta\boldsymbol{R}'}^{\mu\boldsymbol{R}_{1};\nu\boldsymbol{R}_{2}}(t)\eta_{\mu\boldsymbol{R}_{1};\nu\boldsymbol{R}_{2}}(t),\label{EDM-eom}
\end{align}
where
\begin{align}
\mathfrak{F}_{\alpha\boldsymbol{R}'';\beta\boldsymbol{R}'}^{\mu\boldsymbol{R}_{1};\nu\boldsymbol{R}_{2}}(t)&=\delta_{\nu\beta}\delta_{\boldsymbol{R}_{2}\boldsymbol{R}'}e^{i\Delta(\boldsymbol{R}'',\boldsymbol{R}_1,\boldsymbol{R}';t)}\bar{H}_{\alpha\boldsymbol{R}'';\mu\boldsymbol{R}_{1}}(t)\nonumber\\
& -\delta_{\mu\alpha}\delta_{\boldsymbol{R}_{1}\boldsymbol{R}''}e^{i\Delta(\boldsymbol{R}'',\boldsymbol{R}_2,\boldsymbol{R}';t)}\bar{H}_{\nu\boldsymbol{R}_{2};\beta\boldsymbol{R}'}(t)\nonumber \\
& -e\Omega^0_{\boldsymbol{R}'}(\boldsymbol{R}'',t)\delta_{\nu\beta}\delta_{\mu\alpha}\delta_{\boldsymbol{R}_{2}\boldsymbol{R}'}\delta_{\boldsymbol{R}_{1}\boldsymbol{R}''}.\label{scriptF}
\end{align}
The definitions of $\Delta(\boldsymbol{R}_1,\boldsymbol{R}_2,\ldots,\boldsymbol{R}_N;t)$, $\Omega^0_{\boldsymbol{R}'}(\boldsymbol{R}'',t)$, and $\bar{H}_{\nu\boldsymbol{R}_2;\mu\boldsymbol{R}_{1}}(t)$ are as given earlier \cite{Mahon2019}, and are provided in the Appendices. The first of these quantities is related to the magnetic flux through the surface generated by connecting the points $(\boldsymbol{R}_1,\boldsymbol{R}_2,\ldots,\boldsymbol{R}_N)$ with straight lines, when the usual choice of straight-line paths for the relators is adopted. The second is related to the electric field along the path connecting points $(\boldsymbol{R}',\boldsymbol{R}'')$. The third quantity can be understood as a generalized ``hopping'' matrix element. Each of terms appearing in (\ref{scriptF}) is gauge invariant in the electromagnetic sense, and consequently so too is $\eta_{\alpha\boldsymbol{R}'';\beta\boldsymbol{R}'}(t)$. In Appendix \ref{AppendixA}, we show that
\begin{widetext}
\begin{align}
e^{i\Delta(\boldsymbol{R}'',\boldsymbol{R}_1,\boldsymbol{R}';t)}\bar{H}_{\alpha\boldsymbol{R}'';\mu\boldsymbol{R}_1}(t)=e^{i\Delta(\boldsymbol{R}'',\boldsymbol{R}_{\text{a}},\boldsymbol{R}_1,\boldsymbol{R}';t)}\bar{H}_{\alpha\boldsymbol{R}'';\mu\boldsymbol{R}_1}(\boldsymbol{R}_{\text{a}},t)-e\Omega_{\boldsymbol{R}''}^{0}(\boldsymbol{R}_{\text{a}},t)\delta_{\alpha\mu}\delta_{\boldsymbol{R}''\boldsymbol{R}_1}, \label{RaRefSite}
\end{align}
and 
\begin{align}
e^{i\Delta(\boldsymbol{R}'',\boldsymbol{R}_1,\boldsymbol{R}';t)}\bar{H}_{\mu\boldsymbol{R}_1;\beta\boldsymbol{R}'}(t)=e^{i\Delta(\boldsymbol{R}'',\boldsymbol{R}_1,\boldsymbol{R}_{\text{b}},\boldsymbol{R}';t)}\bar{H}_{\mu\boldsymbol{R}_1;\beta\boldsymbol{R}'}(\boldsymbol{R}_{\text{b}},t)-e\Omega_{\boldsymbol{R}'}^{0}(\boldsymbol{R}_{\text{b}},t)\delta_{\mu\beta}\delta_{\boldsymbol{R}'\boldsymbol{R}_1}, \label{RbRefSite}
\end{align}
for \textit{any} lattice sites $\boldsymbol{R}_{\text{a}}$ and $\boldsymbol{R}_{\text{b}}$. We have defined
\begin{align}
&\bar{H}_{\mu\boldsymbol{R}_{1};\nu\boldsymbol{R}_{2}}(\boldsymbol{R}_{\text{a}},t)\equiv\nonumber\\
&\quad\int\chi_{\mu\boldsymbol{R}_{1}}^{*}(\boldsymbol{x},t)e^{i\Delta(\boldsymbol{R}_{1},\boldsymbol{x},\boldsymbol{R}_{\text{a}};t)}\left(\mathcal{H}_{\boldsymbol{R}_{\text{a}}}(\boldsymbol{x},t)+\frac{\hbar}{2}\frac{\partial\Delta(\boldsymbol{R}_{\text{a}},\boldsymbol{x},\boldsymbol{R}_{1};t)}{\partial t}+\frac{\hbar}{2}\frac{\partial\Delta(\boldsymbol{R}_{\text{a}},\boldsymbol{x},\boldsymbol{R}_{2};t)}{\partial t}\right)e^{i\Delta(\boldsymbol{R}_{\text{a}},\boldsymbol{x},\boldsymbol{R}_{2};t)}\chi_{\nu\boldsymbol{R}_{2}}(\boldsymbol{x},t)d\boldsymbol{x}\nonumber \\
&\quad-\frac{i\hbar}{2}\int e^{i\Delta(\boldsymbol{R}_{1},\boldsymbol{x},\boldsymbol{R}_{\text{a}};t)}\left(\chi_{\mu\boldsymbol{R}_{1}}^{*}(\boldsymbol{x},t)\frac{\partial\chi_{\nu\boldsymbol{R}_{2}}(\boldsymbol{x},t)}{\partial t}-\frac{\partial\chi_{\mu\boldsymbol{R}_{1}}^{*}(\boldsymbol{x},t)}{\partial t}\chi_{\nu\boldsymbol{R}_{2}}(\boldsymbol{x},t)\right)e^{i\Delta(\boldsymbol{R}_{\text{a}},\boldsymbol{x},\boldsymbol{R}_{2};t)}d\boldsymbol{x}
\label{barHelements} 
\end{align}
\end{widetext}
and
\begin{align}
\mathcal{H}_{\boldsymbol{R}_{\text{a}}}(\boldsymbol{x},t)\equiv H_{0}\big(\boldsymbol{x},\boldsymbol{\mathfrak{p}}(\boldsymbol{x},\boldsymbol{R}_{\text{a}};t)\big)-e\Omega_{\boldsymbol{R}_{\text{a}}}^{0}(\boldsymbol{x},t), \label{Hcal}
\end{align}
where
\begin{align}
\boldsymbol{\mathfrak{p}}(\boldsymbol{x},\boldsymbol{R}_{\text{a}};t)\equiv\boldsymbol{\mathfrak{p}}(\boldsymbol{x})-\frac{e}{c}\boldsymbol{\Omega}_{\boldsymbol{R}_{\text{a}}}(\boldsymbol{x},t), \label{frakP}
\end{align}
as before \cite{Mahon2019}. Here $\boldsymbol{\Omega}_{\boldsymbol{R}_{\text{a}}}(\boldsymbol{x},t)$ is related to the Maxwell magnetic field along the path connecting points $(\boldsymbol{R}_{\text{a}},\boldsymbol{x})$, and is defined in Appendix \ref{AppendixB}. The functions in the set $\{\chi_{\alpha\boldsymbol{R}}(\boldsymbol{x},t)\}$ are generally not orthogonal, but they are related to the mutually orthogonal ``adjusted Wannier functions'' introduced earlier \cite{Mahon2019}, and they depend only on the Maxwell magnetic field and not the vector potential used to describe it. In the limit of a weak magnetic field, a perturbative expansion can be constructed for each $\chi_{\alpha\boldsymbol{R}}(\boldsymbol{x},t)$ \cite{Mahon2019}, the lowest order terms of which are 
\begin{align}
\chi_{\alpha\boldsymbol{R}}(\boldsymbol{x},t)&=W_{\alpha\boldsymbol{R}}(\boldsymbol{x})-\frac{i}{2}\sum_{\gamma\boldsymbol{R}_1}W_{\gamma\boldsymbol{R}_1}(\boldsymbol{x})\nonumber\\
&\times\Bigg[\int W_{\gamma\boldsymbol{R}_1}^{*}(\boldsymbol{y})\Delta(\boldsymbol{R}_1,\boldsymbol{y},\boldsymbol{R};t)W_{\alpha\boldsymbol{R}}(\boldsymbol{y})d\boldsymbol{y}\Bigg]+\ldots
\end{align} 
Choosing $\boldsymbol{R}_{\text{a}}=\boldsymbol{R}_{\text{b}}$, one can then re-express (\ref{scriptF}) as
\begin{align}
\mathfrak{F}_{\alpha\boldsymbol{R}'';\beta\boldsymbol{R}'}^{\mu\boldsymbol{R}_{1};\nu\boldsymbol{R}_{2}}(t)&=\delta_{\nu\beta}\delta_{\boldsymbol{R}_{2}\boldsymbol{R}'}e^{i\Delta(\boldsymbol{R}'',\boldsymbol{R}_{\text{a}},\boldsymbol{R}_{1},\boldsymbol{R}';t)}\bar{H}_{\alpha\boldsymbol{R}'';\mu\boldsymbol{R}_{1}}(\boldsymbol{R}_{\text{a}},t)\nonumber \\
&-\delta_{\mu\alpha}\delta_{\boldsymbol{R}_{1}\boldsymbol{R}''}e^{i\Delta(\boldsymbol{R}'',\boldsymbol{R}_{2},\boldsymbol{R}_{\text{a}},\boldsymbol{R}';t)}\bar{H}_{\nu\boldsymbol{R}_{2};\beta\boldsymbol{R}'}(\boldsymbol{R}_{\text{a}},t)\nonumber \\
&-\hbar\frac{\partial\Delta(\boldsymbol{R}'',\boldsymbol{R}_{\text{a}},\boldsymbol{R}';t)}{\partial t}\delta_{\nu\beta}\delta_{\mu\alpha}\delta_{\boldsymbol{R}_{2}\boldsymbol{R}'}\delta_{\boldsymbol{R}_{1}\boldsymbol{R}''}.
\end{align}

To prepare for later perturbative analysis we expand all quantities in powers of the electromagnetic field,
\begin{align*}
\eta_{\alpha\boldsymbol{R}'';\beta\boldsymbol{R}'}(t)&=\eta_{\alpha\boldsymbol{R}'';\beta\boldsymbol{R}'}^{(0)}+\eta_{\alpha\boldsymbol{R}'';\beta\boldsymbol{R}'}^{(1)}(t)+\ldots,\\
\bar{H}_{\mu\boldsymbol{R}_{1};\nu\boldsymbol{R}_{2}}(\boldsymbol{R}_{\text{a}},t)&=H_{\mu\boldsymbol{R}_{1};\nu\boldsymbol{R}_{2}}^{(0)}+\bar{H}_{\mu\boldsymbol{R}_{1};\nu\boldsymbol{R}_{2}}^{(1)}(\boldsymbol{R}_{\text{a}},t)+\ldots,
\end{align*}
etc., where the superscript $(0)$ denotes the contribution to the total quantity that is independent of the electromagnetic field, the superscript $(1)$ denotes the contribution that is first-order in the electric and magnetic fields, and so on. Using (\ref{EDM-eom}), and equating terms appearing with the same powers of the Maxwell fields, the zeroth-order term $\eta_{\alpha\boldsymbol{R}'';\beta\boldsymbol{R}'}^{(0)}$ is found to satisfy the same equation of motion as the unperturbed single-particle density matrix, and so
\begin{align}
\eta_{\alpha\boldsymbol{R}'';\beta\boldsymbol{R}'}^{(0)}=f_{\alpha}\delta_{\alpha\beta}\delta_{\boldsymbol{R}''\boldsymbol{R}'}, \label{EDMunpert}
\end{align}
while from (\ref{barHelements}) it is found that
\begin{align}
{H}^{(0)}_{\mu\boldsymbol{R}_{1};\nu\boldsymbol{R}_{2}}=\int W^*_{\mu\boldsymbol{R}_{1}}(\boldsymbol{x})H_0\big(\boldsymbol{x},\boldsymbol{\mathfrak{p}}(\boldsymbol{x})\big)W_{\nu\boldsymbol{R}_{2}}(\boldsymbol{x})d\boldsymbol{x}.
\label{unpertHelements}
\end{align} 
Next, implementing the usual Fourier series analysis via 
\begin{align}
g(t)\equiv\sum_{\omega}e^{-i\omega t}g(\omega), \label{Fouier}
\end{align}
the first-order modification to the single-particle density matrix due to the electromagnetic field can be identified, via (\ref{EDM-eom}), as
\begin{widetext}
\begin{align}
\eta_{\alpha\boldsymbol{R}'';\beta\boldsymbol{R}'}^{(1)}(\omega)&= -\sum_{\mu\nu\boldsymbol{R}_{1}\boldsymbol{R}_{2}}\sum_{mn}f_{nm}\int_{\text{BZ}}d\boldsymbol{k}d\boldsymbol{k}'\frac{\braket{\alpha\boldsymbol{R}''}{\psi_{m\boldsymbol{k}}}\braket{\psi_{m\boldsymbol{k}}}{\mu\boldsymbol{R}_{1}} H_{\mu\boldsymbol{R}_{1};\nu\boldsymbol{R}_{2}}^{(1)}(\boldsymbol{R}_{\text{a}},\omega)\braket{\nu\boldsymbol{R}_{2}}{\psi_{n\boldsymbol{k}'}}\braket{\psi_{n\boldsymbol{k}'}}{\beta\boldsymbol{R}'}}{E_{m\boldsymbol{k}}-E_{n\boldsymbol{k}'}-\hbar(\omega+i0^{+})}\nonumber\\
&+\frac{i}{2}f_{\beta\alpha}\int W_{\alpha\boldsymbol{R}''}^{*}(\boldsymbol{x})\Big(\Delta(\boldsymbol{R}'',\boldsymbol{x},\boldsymbol{R}_{\text{a}};\omega)+\Delta(\boldsymbol{R}',\boldsymbol{x},\boldsymbol{R}_{\text{a}};\omega)\Big)W_{\beta\boldsymbol{R}'}(\boldsymbol{x})d\boldsymbol{x} \label{EDM1}
\end{align}
\end{widetext}
 (Appendix \ref{AppendixD}), where $f_{nm}\equiv f_n-f_m$ and $f_{\beta\alpha}\equiv f_{\beta}-f_{\alpha}$; recall that for the class of insulators of interest here there are well defined filling factors associated with the orbital type indices, but in general this is not so. In identifying (\ref{EDM1}), we have also introduced
\begin{align}
& H_{\mu\boldsymbol{R}_{1};\nu\boldsymbol{R}_{2}}^{(1)}(\boldsymbol{R}_{\text{a}},\omega)\equiv\int W_{\mu\boldsymbol{R}_{1}}^{*}(\boldsymbol{x})\mathcal{H}_{\boldsymbol{R}_{\text{a}}}^{(1)}(\boldsymbol{x},\omega)W_{\nu\boldsymbol{R}_{2}}(\boldsymbol{x})d\boldsymbol{x}. \label{Hcorr}
\end{align}

Often in optics one is interested in the effects of macroscopic Maxwell fields that vary little over electron correlation lengths, which for the class of insulators we consider are on the order of the lattice constant. In Appendix \ref{AppendixB}, we show that if this approximation is made, and if $|\boldsymbol{y}-\boldsymbol{x}|$ and $|\boldsymbol{z}-\boldsymbol{x}|$ are on the order of lattice constants, then
\begin{align}
\Omega^a_{\boldsymbol{y}}(\boldsymbol{x},\omega)&\simeq\frac{1}{2}\epsilon^{alb}B^l(\boldsymbol{y},\omega)\big(x^b-y^b\big),\label{omegaVec}\\
\Omega_{\boldsymbol{y}}^{0}(\boldsymbol{x},\omega)&\simeq\big(x^l-y^l\big)E^l(\boldsymbol{y},\omega)\nonumber\\
&+\frac{1}{2}\big(x^{j}-y^{j}\big)\big(x^{l}-y^{l}\big)F^{jl}(\boldsymbol{y},\omega),
\label{omega0}\\
\Delta(\boldsymbol{z},\boldsymbol{x},\boldsymbol{y};\omega)&\simeq-\frac{e}{2\hbar c}\epsilon^{lab}B^l(\boldsymbol{y},\omega)\big(z^a-y^a\big)\big(x^b-y^b\big), \label{delta}
\end{align}
where $\boldsymbol{E}(\boldsymbol{y},t)$ is the Maxwell electric field, $\boldsymbol{B}(\boldsymbol{y},t)$ is the Maxwell magnetic field, 
\begin{align}
F^{jl}(\boldsymbol{y},t)\equiv\frac{1}{2}\left(\frac{\partial E^{j}(\boldsymbol{y},t)}{\partial y^{l}}+\frac{\partial E^{l}(\boldsymbol{y},t)}{\partial y^{j}}\right), \label{eq:Fdef}
\end{align}
and $\epsilon^{abd}$ is the Levi-Civita symbol. Implementing the approximate expressions (\ref{omegaVec},\ref{omega0}), as well as (\ref{H}) and (\ref{Fouier}), the first-order modification to (\ref{Hcal}) is found to be
\begin{align}
\mathcal{H}_{\boldsymbol{R}_{\text{a}}}^{(1)}(\boldsymbol{x},\omega)\simeq&-e\big(x^{l}-R_{\text{a}}^{l}\big)E^{l}(\boldsymbol{R}_{\text{a}},\omega)\nonumber\\
&-\frac{e}{2}\big(x^{j}-R_{\text{a}}^{j}\big)\big(x^{l}-R_{\text{a}}^{l}\big)F^{jl}(\boldsymbol{R}_{\text{a}},\omega)\nonumber\\
&+\frac{e}{2mc}\epsilon^{lab}B^{l}(\boldsymbol{R}_{\text{a}},\omega)\big(x^b-R_{\text{a}}^b\big)\mathfrak{p}^{a}(\boldsymbol{x}).
\label{Hcal1}
\end{align}
In the above, we have used straight-line paths in the relators \cite{Mahon2019}, and in what follows, this choice is always made. Equations (\ref{EDM1})-(\ref{Hcal1}) make clear the motivation for introducing the arbitrary lattice site $\boldsymbol{R}_{\text{a}}$; it serves as a reference site for the electromagnetic field. This will prove useful when considering the response of a quantity associated with site $\boldsymbol{R}$ to a spatially varying electromagnetic field; choosing $\boldsymbol{R}_{\text{a}}=\boldsymbol{R}$, the modification to that site quantity is related to the Maxwell field evaluated at that site.

However, in this paper we restrict ourselves to uniform electric and magnetic fields in the dc limit. Thus $F^{jl}(\boldsymbol{x},t)$ vanishes, and $\eta_{\alpha\boldsymbol{R}'';\beta\boldsymbol{R}'}^{(1)}(\omega)\neq0$ only if $\omega=0$. We retain only the nonvanishing, first-order perturbative modifications arising from the electric and magnetic fields, which we denote by $\eta_{\alpha\boldsymbol{R}'';\beta\boldsymbol{R}'}^{(E)}$ and $\eta_{\alpha\boldsymbol{R}'';\beta\boldsymbol{R}'}^{(B)}$, respectively, such that
\begin{align}
\eta_{\alpha\boldsymbol{R}'';\beta\boldsymbol{R}'}^{(1)}(\omega=0)=\eta_{\alpha\boldsymbol{R}'';\beta\boldsymbol{R}'}^{(E)}+\eta_{\alpha\boldsymbol{R}'';\beta\boldsymbol{R}'}^{(B)}. \label{EDMlinear}
\end{align}
\begin{widetext}
Implementing (\ref{EDM1}), the first-order perturbative modification to the single-particle density matrix due to a uniform dc electric field is found to be
\begin{align}
\eta_{\alpha\boldsymbol{R}'';\beta\boldsymbol{R}'}^{(E)}=e\Omega_{uc}E^l\sum_{mn}f_{nm}\int_{\text{BZ}}\frac{d\boldsymbol{k}}{(2\pi)^3}\frac{e^{i\boldsymbol{k}\boldsymbol{\cdot}(\boldsymbol{R}''-\boldsymbol{R}')}U^\dagger_{\alpha m}\xi^l_{mn}U_{n\beta}}{E_{m\boldsymbol{k}}-E_{n\boldsymbol{k}}} \label{EDMe}
\end{align}
(Appendix \ref{AppendixD}), where $\boldsymbol{E}\equiv\boldsymbol{E}(\boldsymbol{R}_{\text{a}},\omega=0)$, for any $\boldsymbol{R}_{\text{a}}$, is now the uniform dc electric field. The result (\ref{EDMe}) is consistent with previous work \cite{Mahon2019}, where this expression was derived via a different method. Notably (\ref{EDMe}) is written as a single Brillouin zone integral, unlike (\ref{EDM1}). This feature is expected upon comparison to the usual perturbative treatment, in which, when spatial variation of the electric field is neglected, a $\boldsymbol{k}$-conserving interaction term arises that results in perturbative modifications being given by single Brillouin zone integrals \cite{SipeBook}. While here we assume the macroscopic Maxwell electric field is uniform, more generally this reduction to a single $\boldsymbol{k}$ integral holds as a consequence of the approximation that the electric field is to vary little over electron correlation lengths, allowing a Taylor series expansion about each lattice site. In the limit of uniform fields of interest here, the $\boldsymbol{R}_{\text{a}}$ dependence of (\ref{EDMe}) vanishes.

There are two distinct contributions to the first-order perturbative modification due to a uniform dc magnetic field, 
\begin{align}
\eta_{\alpha\boldsymbol{R}'';\beta\boldsymbol{R}'}^{(B)}&=\frac{e\Omega_{uc}}{4\hbar c}\epsilon^{lab}B^l\sum_{mn}f_{nm}\int_{\text{BZ}}\frac{d\boldsymbol{k}}{(2\pi)^{3}}e^{i\boldsymbol{k}\boldsymbol{\cdot}(\boldsymbol{R}''-\boldsymbol{R}')}U^\dagger_{\alpha m}{\mathscr{B}}^{ab}_{mn}(\boldsymbol{k})U_{n\beta}\nonumber\\
&+\frac{e\Omega_{uc}}{4\hbar c}\epsilon^{lab}B^l\sum_{mn}f_{nm}\int_{\text{BZ}}\frac{d\boldsymbol{k}}{(2\pi)^{3}}e^{i\boldsymbol{k}\boldsymbol{\cdot}(\boldsymbol{R}''-\boldsymbol{R}')}\Big\{\big(\partial_a U^\dagger_{\alpha m}\big)U_{n \beta}-U^\dagger_{\alpha m}\big(\partial_a U_{n\beta}\big)\Big\}\xi^b_{mn}, \label{EDMb2}
\end{align}
where we have defined
\begin{align}
{\mathscr{B}}^{ab}_{mn}(\boldsymbol{k})&\equiv i\sum_{s}\left\{\frac{E_{s\boldsymbol{k}}-E_{n\boldsymbol{k}}}{E_{m\boldsymbol{k}}-E_{n\boldsymbol{k}}}\xi^a_{ms}\xi^b_{sn}+\frac{E_{s\boldsymbol{k}}-E_{m\boldsymbol{k}}}{E_{m\boldsymbol{k}}-E_{n\boldsymbol{k}}}\xi^a_{ms}\xi^b_{sn}
\right\}-2\frac{\partial_a(E_{m\boldsymbol{k}}+E_{n\boldsymbol{k}})}{E_{m\boldsymbol{k}}-E_{n\boldsymbol{k}}}\xi^b_{mn},
\end{align}
and $\boldsymbol{B}\equiv\boldsymbol{B}(\boldsymbol{R}_{\text{a}},\omega=0)$, for any $\boldsymbol{R}_{\text{a}}$, is the uniform dc magnetic field. We mention that the two terms appearing in (\ref{EDMb2}) are not
simply the individual contributions of the terms of (\ref{EDM1}). Moreover, in the first-order perturbative modifications of the quantities considered below, the first term of (\ref{EDMb2}) gives rise to gauge invariant contributions, while the final term will give rise to gauge dependent contributions \footnote{As previously discussed, in this paper, we move the gauge freedom of the energy eigenvectors, and thus the gauge dependence of the connections $\xi^a_{mn}$, into the $U_{n\alpha}$ matrices.}.
\end{widetext}

\section{Modification of $\boldsymbol{P}$ and $\boldsymbol{M}$ due to uniform dc $\boldsymbol{E}$ and $\boldsymbol{B}$ fields}
\label{Sect:2}

By implementing the perturbative modifications of the single-particle density matrix (\ref{EDMlinear}), we now calculate the first-order modifications of the electric and magnetic dipole moments -- found from the microscopic polarization and magnetization fields, respectively -- due to uniform dc electric and magnetic fields. In this section we restrict earlier \cite{Mahon2019}, more general expressions to this limit. Thus a function previously written in terms of frequency components $g(\omega)$ (\ref{Fouier}), will simply be given by the single nonvanishing component $g\equiv g(\omega=0)$. Furthermore, as previously mentioned, to first-order in the electromagnetic field, the free charge and current densities vanish at zero temperature for the class of insulators we consider \cite{Mahon2019}, as would be expected physically, and so those quantities do not appear here.

\subsection{Summary of formalism}

Introducing a set of ``adjusted Wannier functions'' that is an orthonormal basis of the electronic Hilbert space allows for the lesser, equal time single-particle Green function to be exactly decomposed as a sum of site contributions \cite{Mahon2019}. Consequently, so too can be the microscopic polarization and magnetization fields, such that
\begin{align}
\boldsymbol{p}(\boldsymbol{x})&=\sum_{\boldsymbol{R}}\boldsymbol{p}_{\boldsymbol{R}}(\boldsymbol{x}), \nonumber\\
\boldsymbol{m}(\boldsymbol{x})&=\sum_{\boldsymbol{R}}\boldsymbol{m}_{\boldsymbol{R}}(\boldsymbol{x}), \label{pm}
\end{align}
where the sums range over all Bravais lattice vectors $\boldsymbol{R}$. In the limit of uniform dc electric and magnetic fields treated perturbatively, the discrete translational symmetry that was present in the unperturbed Hamiltonian (\ref{H}) is now lost (cf.~(\ref{fieldEvo})), but the polarization and magnetization fields associated with each lattice site remain physically equivalent; that is
\begin{align*}
\boldsymbol{p}_{\boldsymbol{R}+\boldsymbol{R}'}(\boldsymbol{x})&=\boldsymbol{p}_{\boldsymbol{R}}(\boldsymbol{x}-\boldsymbol{R}'), \\
\boldsymbol{m}_{\boldsymbol{R}+\boldsymbol{R}'}(\boldsymbol{x})&=\boldsymbol{m}_{\boldsymbol{R}}(\boldsymbol{x}-\boldsymbol{R}'),
\end{align*}
for any $\boldsymbol{R}$ and $\boldsymbol{R}'$. This is a result of the fact that it is the electric and magnetic fields, \textit{not} the vector and scalar potentials, that enter in the expressions that follow. Thus, in its perturbative modifications, the system retains its periodic nature in the limit of uniform dc Maxwell fields. 

Each site polarization field, $\boldsymbol{p}_{\boldsymbol{R}}(\boldsymbol{x})$, is related to the electronic charge density associated with that site, $\rho_{\boldsymbol{R}}(\boldsymbol{x})$, via the ``relator'' ${s}^i(\boldsymbol{x};\boldsymbol{y},\boldsymbol{R})$ (see Appendix \ref{AppendixB}), as shown previously \cite{Mahon2019}. It is given by
\begin{align}
{p}^i_{\boldsymbol{R}}(\boldsymbol{x})& \equiv\sum_{\alpha\beta\boldsymbol{R}'\boldsymbol{R}''}\left[\int{s}^i(\boldsymbol{x};\boldsymbol{y},\boldsymbol{R})\rho_{\beta\boldsymbol{R}';\alpha\boldsymbol{R}''}(\boldsymbol{y},\boldsymbol{R})d\boldsymbol{y}\right]\nonumber \\
&\quad\qquad\qquad\times\eta_{\alpha\boldsymbol{R}'';\beta\boldsymbol{R}'},\label{siteP}
\end{align}
where above and below the sums range over all lattice vectors and orbital types, and 
\begin{align}
\rho_{\beta\boldsymbol{R}';\alpha\boldsymbol{R}''}(\boldsymbol{x},\boldsymbol{R})&=\frac{e}{2}\big(\delta_{\boldsymbol{R}\boldsymbol{R}'}+\delta_{\boldsymbol{R}\boldsymbol{R}''}\big)e^{i\Delta(\boldsymbol{R}',\boldsymbol{x},\boldsymbol{R}'')}\nonumber \\
&\qquad\times\chi_{\beta\boldsymbol{R}'}^{*}(\boldsymbol{x})\chi_{\alpha\boldsymbol{R}''}(\boldsymbol{x}). \label{rhoElements}
\end{align}
Meanwhile, there are two contributions to each site magnetization field,
\begin{align}
\boldsymbol{m}_{\boldsymbol{R}}(\boldsymbol{x})=\bar{\boldsymbol{m}}_{\boldsymbol{R}}(\boldsymbol{x})+\tilde{\boldsymbol{m}}_{\boldsymbol{R}}(\boldsymbol{x}).\label{siteM}
\end{align}
The first of these, $\bar{\boldsymbol{m}}_{\boldsymbol{R}}(\boldsymbol{x})$, corresponds to the ``local'' or ``atomic-like'' contribution to each site magnetization field, and is related to the electronic current density associated with that site, $\boldsymbol{j}_{\boldsymbol{R}}(\boldsymbol{x})$, via the relator $\alpha^{ib}(\boldsymbol{x};\boldsymbol{y},\boldsymbol{R})$ (see Appendix \ref{AppendixB}) \footnote{The relator $\alpha^{ib}(\boldsymbol{x};\boldsymbol{y},\boldsymbol{R})$ is not to be confused with the OMP tensor $\alpha^{il}$ introduced in Eq.~(\ref{OMP}).}. It is given by
\begin{align}
\bar{m}_{\boldsymbol{R}}^{i}(\boldsymbol{x})&\equiv\frac{1}{c}\sum_{\alpha\beta\boldsymbol{R}'\boldsymbol{R}''}\left[\int\alpha^{ib}(\boldsymbol{x};\boldsymbol{y},\boldsymbol{R})j_{\beta\boldsymbol{R}';\alpha\boldsymbol{R}''}^{b}(\boldsymbol{y},\boldsymbol{R})d\boldsymbol{y}\right]\nonumber\\
&\quad\qquad\qquad\times\eta_{\alpha\boldsymbol{R}'';\beta\boldsymbol{R}'},\label{siteMbar}
\end{align}
where 
\begin{align}
	&\boldsymbol{j}_{\beta\boldsymbol{R}';\alpha\boldsymbol{R}''}(\boldsymbol{x},\boldsymbol{R})=\nonumber\\ &\frac{1}{4}\delta_{\boldsymbol{R}\boldsymbol{R}''}e^{i\Delta(\boldsymbol{R}',\boldsymbol{x},\boldsymbol{R}'')}\chi_{\beta\boldsymbol{R}'}^{*}(\boldsymbol{x})\Big[\boldsymbol{J}\big(\boldsymbol{x},\boldsymbol{\mathfrak{p}}(\boldsymbol{x},\boldsymbol{R})\big)\chi_{\alpha\boldsymbol{R}''}(\boldsymbol{x})\Big]\nonumber\\
	& +\frac{1}{4}\delta_{\boldsymbol{R}\boldsymbol{R}'}\Big[\boldsymbol{J}^{*}\big(\boldsymbol{x},\boldsymbol{\mathfrak{p}}(\boldsymbol{x},\boldsymbol{R})\big)\chi_{\beta\boldsymbol{R}'}^{*}(\boldsymbol{x})\Big]e^{i\Delta(\boldsymbol{R}',\boldsymbol{x},\boldsymbol{R}'')}\chi_{\alpha\boldsymbol{R}''}(\boldsymbol{x})\nonumber \\
	& +\frac{1}{4}\delta_{\boldsymbol{R}\boldsymbol{R}''}\left[\boldsymbol{J}^{*}\big(\boldsymbol{x},\boldsymbol{\mathfrak{p}}(\boldsymbol{x},\boldsymbol{R})\big)\Big(e^{i\Delta(\boldsymbol{R}',\boldsymbol{x},\boldsymbol{R}'')}\chi_{\beta\boldsymbol{R}'}^{*}(\boldsymbol{x})\Big)\right]\chi_{\alpha\boldsymbol{R}''}(\boldsymbol{x})\nonumber \\
	& +\frac{1}{4}\delta_{\boldsymbol{R}\boldsymbol{R}'}\chi_{\beta\boldsymbol{R}'}^{*}(\boldsymbol{x})\left[\boldsymbol{J}\big(\boldsymbol{x},\boldsymbol{\mathfrak{p}}(\boldsymbol{x},\boldsymbol{R})\big)\Big(e^{i\Delta(\boldsymbol{R}',\boldsymbol{x},\boldsymbol{R}'')}\chi_{\alpha\boldsymbol{R}''}(\boldsymbol{x})\Big)\right].\label{jElements} 
\end{align}
It is not obvious that there is another contribution to each site magnetization field, as each atomic-like contribution is found from a site electronic current density, and these collectively compose the total current. However, in extended systems where the charge-current density associated with each site need \textit{not} be conserved, there is in general an additional term, $\tilde{\boldsymbol{m}}_{\boldsymbol{R}}(\boldsymbol{x})$ \cite{Mahon2019}. Adopting the terminology of the ``modern theory,'' it corresponds to the ``itinerant'' contribution to each site magnetization field, and is given by
\begin{align}
	\tilde{m}_{\boldsymbol{R}}^{i}(\boldsymbol{x})&\equiv\frac{1}{c}\sum_{\alpha\beta\boldsymbol{R}'\boldsymbol{R}''}\left[\int\alpha^{ib}(\boldsymbol{x};\boldsymbol{y},\boldsymbol{R})\tilde{j}_{\beta\boldsymbol{R}';\alpha\boldsymbol{R}''}^{b}(\boldsymbol{y},\boldsymbol{R})d\boldsymbol{y}\right]\nonumber\\
	&\quad\qquad\qquad\times\eta_{\alpha\boldsymbol{R}'';\beta\boldsymbol{R}'}. \label{siteMtilde}
\end{align}
Here
\begin{align}
& \boldsymbol{\tilde{j}}_{\beta\boldsymbol{R}';\alpha\boldsymbol{R}''}(\boldsymbol{x},\boldsymbol{R})=\frac{1}{2}\big(\delta_{\boldsymbol{R}\boldsymbol{R}'}+\delta_{\boldsymbol{R}\boldsymbol{R}''}\big)\boldsymbol{\tilde{j}}_{\beta\boldsymbol{R}';\alpha\boldsymbol{R}''}(\boldsymbol{x}), \label{jTildeSite}
\end{align}
with
\begin{align}
{\tilde{j}}^b_{\beta\boldsymbol{R}';\alpha\boldsymbol{R}''}(\boldsymbol{x}) & =-\sum_{\boldsymbol{R}_3}\int{s}^b(\boldsymbol{x};\boldsymbol{z},\boldsymbol{R}_3)\Gamma_{\boldsymbol{R}_3}^{\alpha\boldsymbol{R}'';\beta\boldsymbol{R}'}(\boldsymbol{z})d\boldsymbol{z}\nonumber \\
& -\frac{1}{2}\sum_{\boldsymbol{R}_1\boldsymbol{R}_2}{s}^b(\boldsymbol{x};\boldsymbol{R}_2,\boldsymbol{R}_1)\varsigma_{\boldsymbol{R}_2\boldsymbol{R}_1}^{\alpha\boldsymbol{R}'';\beta\boldsymbol{R}'},\label{jTilde}
\end{align}
where
\begin{align}
\varsigma_{\boldsymbol{R}_2\boldsymbol{R}_1}^{\alpha\boldsymbol{R}'';\beta\boldsymbol{R}'} &=\frac{e}{i\hbar}\Big(\delta_{\boldsymbol{R}'\boldsymbol{R}_2}\delta_{\boldsymbol{R}''\boldsymbol{R}_1}\bar{H}_{\beta\boldsymbol{R}_2;\alpha\boldsymbol{R}_1} \nonumber\\
& \quad\qquad-\delta_{\boldsymbol{R}''\boldsymbol{R}_2}\delta_{\boldsymbol{R}'\boldsymbol{R}_1}\bar{H}_{\beta\boldsymbol{R}_1;\alpha\boldsymbol{R}_2}\Big),
\end{align}
and
\begin{align}
\Gamma_{\boldsymbol{R}_3}^{\alpha\boldsymbol{R}'';\beta\boldsymbol{R}'}(\boldsymbol{x})&=\boldsymbol{\nabla}\boldsymbol{\cdot}\boldsymbol{j}_{\beta\boldsymbol{R}';\alpha\boldsymbol{R}''}(\boldsymbol{x},\boldsymbol{R}_3)\nonumber\\
&+\frac{1}{i\hbar}\sum_{\mu\nu\boldsymbol{R}_{1}\boldsymbol{R}_{2}}\rho_{\nu\boldsymbol{R}_{2};\mu\boldsymbol{R}_{1}}(\boldsymbol{x},\boldsymbol{R}_3)\mathfrak{F}_{\mu\boldsymbol{R}_{1};\nu\boldsymbol{R}_{2}}^{\alpha\boldsymbol{R}'';\beta\boldsymbol{R}'}. \label{Gamma}
\end{align}

Typically it is the modifications of the electric and magnetic dipole moments due to electromagnetic perturbations that one is interested in studying; these correspond to the spatial integrals of the microscopic polarization and magnetization fields \cite{Mahon2019}, respectively, 
\begin{align}
\mu^i_{\boldsymbol{R}}\equiv\int p^i_{\boldsymbol{R}}(\boldsymbol{x})d\boldsymbol{x} \text{\space ,\space} \nu^i_{\boldsymbol{R}}\equiv\int m^i_{\boldsymbol{R}}(\boldsymbol{x})d\boldsymbol{x}.
\end{align}
From (\ref{siteP}), we find
\begin{align}
\mu^i_{\boldsymbol{R}}&=\sum_{\alpha\beta\boldsymbol{R}'\boldsymbol{R}''}\left[\int\big(y^i-R^i\big)\rho_{\beta\boldsymbol{R}';\alpha\boldsymbol{R}''}(\boldsymbol{y},\boldsymbol{R})d\boldsymbol{y}\right]\eta_{\alpha\boldsymbol{R}'';\beta\boldsymbol{R}'}, \label{mu}
\end{align}
and $\boldsymbol{\nu}_{\boldsymbol{R}}=\bar{\boldsymbol{\nu}}_{\boldsymbol{R}}+\tilde{\boldsymbol{\nu}}_{\boldsymbol{R}}$, where, from (\ref{siteMbar}),
\begin{align}
\bar{\nu}^i_{\boldsymbol{R}}&=\sum_{\alpha\beta\boldsymbol{R}'\boldsymbol{R}''}\left[\frac{\epsilon^{iab}}{2c}\int\big(y^a-R^a\big)j_{\beta\boldsymbol{R}';\alpha\boldsymbol{R}''}^{b}(\boldsymbol{y},\boldsymbol{R})d\boldsymbol{y}\right]\nonumber\\
&\quad\qquad\qquad\times\eta_{\alpha\boldsymbol{R}'';\beta\boldsymbol{R}'} \label{nuBar}
\end{align}
is the atomic-like contribution to the site magnetic dipole moment and, from (\ref{siteMtilde}),
\begin{align}
\tilde{\nu}^i_{\boldsymbol{R}}&=\sum_{\alpha\beta\boldsymbol{R}'\boldsymbol{R}''}\left[\frac{\epsilon^{iab}}{2c}\int\big(y^a-R^a\big)\tilde{j}_{\beta\boldsymbol{R}';\alpha\boldsymbol{R}''}^{b}(\boldsymbol{y},\boldsymbol{R})d\boldsymbol{y}\right]\nonumber\\
&\quad\qquad\qquad\times\eta_{\alpha\boldsymbol{R}'';\beta\boldsymbol{R}'} \label{nuTilde}
\end{align}
is the itinerant contribution to the site magnetic dipole moment. The macroscopic polarization and magnetization can be found from their respective site dipole moments introduced above, and are taken to be
\begin{align}
\boldsymbol{P}\equiv\frac{\boldsymbol{\mu_R}}{\Omega_{uc}} \text{\space,\space} \boldsymbol{M}\equiv\frac{\boldsymbol{\nu_R}}{\Omega_{uc}}. \label{PM}
\end{align}
In the limit of uniform dc Maxwell fields, both the site electric and magnetic dipole moments are independent of $\boldsymbol{R}$, and as a consequence, the macroscopic polarization and magnetization fields are uniform.

\subsection{Unperturbed expressions}

We begin by confirming that our microscopic treatment yields the standard expressions for the unperturbed ground state macroscopic polarization and magnetization, more usually constructed from macroscopic arguments \cite{Resta2010}. While (\ref{mu},\ref{nuBar},\ref{nuTilde}) have been defined to include only valence and conduction electron contributions, the contributions from ion cores can be identified as well (see Mahon \textit{et al.}~\cite{Mahon2019}). However, we focus only on the former contributions here. 

Expanding (\ref{mu}) in powers of the electromagnetic field, we find the zeroth-order term to be
\begin{align}
\mu^{i(0)}_{\boldsymbol{R}}&=\sum\limits_{\alpha\beta\boldsymbol{R}'\boldsymbol{R}''}\left[\int \big(y^i-R^i\big)\rho^{(0)}_{\beta\boldsymbol{R}';\alpha\boldsymbol{R}''}(\boldsymbol{y},\boldsymbol{R})d\boldsymbol{y} \right]\eta^{(0)}_{\alpha\boldsymbol{R}'';\beta\boldsymbol{R}'} \nonumber\\
&=e\sum_{\alpha}f_{\alpha}\int W^*_{\alpha\boldsymbol{0}}(\boldsymbol{x})x^iW_{\alpha\boldsymbol{0}}(\boldsymbol{x})d\boldsymbol{x} \nonumber\\
&=e\Omega_{uc}\sum_{n}f_{n}\int_{\text{BZ}}\frac{d\boldsymbol{k}}{(2\pi)^3}\left(\xi^i_{nn}+\mathcal{W}^i_{nn}\right), \label{unpertP}
\end{align}
where we have used (\ref{firstMoment})-(\ref{connection}). This corresponds to the unperturbed ground state site electric dipole moment, and upon implementing (\ref{PM}), the usual expression \cite{Resta1994} for $\boldsymbol{P}^{(0)}$ is reproduced. It is well known that the unperturbed macroscopic polarization is unique modulo a ``quantum of ambiguity.'' This ambiguity originates from the gauge dependence of (\ref{unpertP}), and it has been shown that the gauge dependent term of (\ref{unpertP}) contributes only to this ``quantum'' \cite{Resta1994}. Importantly, it is only the diagonal elements of the $\mathcal{W}^i$ matrix that appear in (\ref{unpertP}), and as a result, even a U$(1)$ gauge transformation can give rise to this ``quantum of ambiguity.'' This is discussed further in Sec.~\ref{Sect:3}.

Turning to the site magnetic dipole moment, we expand (\ref{nuBar}) and (\ref{nuTilde}) in powers of the electromagnetic field. The zeroth-order terms are found to be
\begin{align}
\bar{\nu}^{i(0)}_{\boldsymbol{R}}&=\sum_{\alpha\boldsymbol{R}'}f_{\alpha}\left[\frac{\epsilon^{iab}}{2c}\int\big(y^a-R^a\big)j^{b(0)}_{\alpha\boldsymbol{R}';\alpha\boldsymbol{R}'}(\boldsymbol{y},\boldsymbol{R})d\boldsymbol{y}\right] \nonumber \\
&=\frac{e}{2mc}\epsilon^{iab}\sum_{\alpha}f_{\alpha}\int W^*_{\alpha\boldsymbol{0}}(\boldsymbol{x})x^a\mathfrak{p}^b(\boldsymbol{x})W_{\alpha\boldsymbol{0}}(\boldsymbol{x})d\boldsymbol{x}, \label{nuBarUnpert}
\end{align}
and
\begin{align}
\tilde{\nu}^{i(0)}_{\boldsymbol{R}}&=\sum_{\alpha\boldsymbol{R}'}f_{\alpha}\left[\frac{\epsilon^{iab}}{2c}\int\big(y^a-R^a\big)\tilde{j}^{b(0)}_{\alpha\boldsymbol{R}';\alpha\boldsymbol{R}'}(\boldsymbol{y},\boldsymbol{R})d\boldsymbol{y}\right] \nonumber \\
&=\frac{e}{2\hbar c}\epsilon^{iab}\sum_{\alpha\gamma\boldsymbol{R}_1}f_{\alpha}R_1^a\nonumber\\
&\quad\times\text{Im}\Big[H^{(0)}_{\alpha\boldsymbol{0};\gamma\boldsymbol{R}_1}\int W^*_{\gamma\boldsymbol{R}_1}(\boldsymbol{x})x^bW_{\alpha\boldsymbol{0}}(\boldsymbol{x})d\boldsymbol{x}\Big], \label{nuTildeUnpert}
\end{align}
which, together, form the unperturbed ground state site magnetic dipole moment, $\boldsymbol{\nu}^{(0)}_{\boldsymbol{R}}=\bar{\boldsymbol{\nu}}^{(0)}_{\boldsymbol{R}}+\tilde{\boldsymbol{\nu}}^{(0)}_{\boldsymbol{R}}$. Separately (\ref{nuBarUnpert}) and (\ref{nuTildeUnpert}) are ``multiband gauge dependent;'' we adopt this phrase to describe quantities that are gauge dependent only if there are degenerate Bloch energy eigenvectors. Here (\ref{nuBarUnpert}) and (\ref{nuTildeUnpert}) are in fact only gauge dependent if there are degenerate occupied energy eigenvectors, and in the limit of isolated valence bands both (\ref{nuBarUnpert}) and (\ref{nuTildeUnpert}) become gauge invariant individually. This is in agreement with past results \cite{Resta2006}. Nonetheless, even if the occupied bands are not isolated the sum of (\ref{nuBarUnpert},\ref{nuTildeUnpert}) is generally a gauge invariant quantity and thus there is no ambiguity in the value of the unperturbed ground state macroscopic magnetization. Implementing (\ref{PM}), the usual expression \cite{Resta2005,Resta2006} for $\boldsymbol{M}^{(0)}$ is reproduced.

\subsection{First-order perturbative modifications}
\label{Sect:3c}

We now turn to the first-order modifications of the Cartesian components of $\boldsymbol{\mu}_{\boldsymbol{R}}$ and $\boldsymbol{\nu}_{\boldsymbol{R}}$ and thus, through (\ref{PM}), to the components of $\boldsymbol{P}$ and $\boldsymbol{M}$, due to an electromagnetic field. Generally, the site quantities we consider are of the form
\begin{align}
\Lambda_{\boldsymbol{R}}\equiv\sum_{\alpha\beta\boldsymbol{R}'\boldsymbol{R}''}\Lambda_{\beta\boldsymbol{R}';\alpha\boldsymbol{R}''}(\boldsymbol{R})\eta_{\alpha\boldsymbol{R}'';\beta\boldsymbol{R}'}, \label{siteQuantity}
\end{align}
where $\Lambda_{\boldsymbol{R}}$ indicates one of the components of $\boldsymbol{\mu}_{\boldsymbol{R}}$ or $\boldsymbol{\nu}_{\boldsymbol{R}}$, $\eta_{\alpha\boldsymbol{R}'';\beta\boldsymbol{R}'}$ is the single-particle density matrix, and $\Lambda_{\beta\boldsymbol{R}';\alpha\boldsymbol{R}''}(\boldsymbol{R})$, which we call a ``site quantity matrix element,'' is of the form
\begin{align}
\Lambda_{\beta\boldsymbol{R}';\alpha\boldsymbol{R}''}(\boldsymbol{R})=\frac{1}{2}\big(\delta_{\boldsymbol{RR}'}+\delta_{\boldsymbol{RR}''}\big)\Lambda_{\beta\boldsymbol{R}';\alpha\boldsymbol{R}''}; \label{siteElements}
\end{align}
see (\ref{rhoElements},\ref{mu}) for $\boldsymbol{\mu}_{\boldsymbol{R}}$, and (\ref{jElements},\ref{jTildeSite},\ref{nuBar},\ref{nuTilde}) for $\boldsymbol{\nu}_{\boldsymbol{R}}$.

\subsubsection{Dynamical and compositional modifications}

The first-order modification to (\ref{siteQuantity}) due to an electromagnetic field thus has two types of contributions,
\begin{align}
\Lambda_{\boldsymbol{R}}^{(1)}=\Lambda_{\boldsymbol{R}}^{(1;\text{I})}+\Lambda_{\boldsymbol{R}}^{(1;\text{II})},\label{siteModification}
\end{align}
the first arising from the combination of the unperturbed expression for $\Lambda_{\beta\boldsymbol{R}';\alpha\boldsymbol{R}''}(\boldsymbol{R})$ and the first-order modification of $\eta_{\alpha\boldsymbol{R}'';\beta\boldsymbol{R}'}$ due to the Maxwell fields,
\begin{align}
\Lambda_{\boldsymbol{R}}^{(1;\text{I})}\equiv\frac{1}{2}\sum_{\alpha\beta\boldsymbol{R}'}\left(\Lambda_{\alpha\boldsymbol{R};\beta\boldsymbol{R}'}^{(0)}\eta_{\beta\boldsymbol{R}';\alpha\boldsymbol{R}}^{(1)}+\Lambda_{\beta\boldsymbol{R}';\alpha\boldsymbol{R}}^{(0)}\eta_{\alpha\boldsymbol{R};\beta\boldsymbol{R}'}^{(1)}\right),\label{lambdaDynamical}
\end{align}
and the second from the combination of the first-order modification of $\Lambda_{\beta\boldsymbol{R}';\alpha\boldsymbol{R}''}(\boldsymbol{R})$ due to the Maxwell fields
and the unperturbed expression for $\eta_{\alpha\boldsymbol{R}'';\beta\boldsymbol{R}'}$, (\ref{EDMunpert}),
\begin{align}
\Lambda_{\boldsymbol{R}}^{(1;\text{II})}\equiv\sum_{\alpha\boldsymbol{R}'}f_{\alpha}\Lambda_{\alpha\boldsymbol{R}';\alpha\boldsymbol{R}'}^{(1)}(\boldsymbol{R})=\sum_{\alpha}f_{\alpha}\Lambda_{\alpha\boldsymbol{R};\alpha\boldsymbol{R}}^{(1)}.\label{lambdaSite}
\end{align}

We refer to the first contribution (\ref{lambdaDynamical}) as ``dynamical'' because it arises from modifications of the single-particle density matrix, which captures the electronic transition amplitude between various lattice sites and orbital types, due to the electromagnetic field. Notably in this type of modification of a site quantity associated with $\boldsymbol{R}$, that lattice vector always appears as at least one of the lattice vector indices identifying the relevant single-particle density matrix elements. This is expected physically; a site quantity associated with $\boldsymbol{R}$ is affected by electrons moving between different orbital types at that lattice site, and by electrons moving from the region ``nearest'' $\boldsymbol{R}$ to regions ``nearest'' other $\boldsymbol{R}'$. Contributions to (\ref{lambdaDynamical}) arising from $\eta_{\beta\boldsymbol{R}';\alpha\boldsymbol{R}}$ and $\eta_{\alpha\boldsymbol{R};\beta\boldsymbol{R}'}$, with $\boldsymbol{R}'\neq\boldsymbol{R}$, are a consequence of the extended nature of the system, in which electrons are not confined to regions of space; such contributions vanish in the limit that the crystalline solid is considered simply as a periodic array of ``isolated molecules,'' which we call the ``molecular crystal limit'' (see Sec.~\ref{Sect:limits}). Conversely, contributions to (\ref{lambdaDynamical}) arising from $\eta_{\alpha\boldsymbol{R};\beta\boldsymbol{R}}$ take the form of single-site modifications.

The second contribution (\ref{lambdaSite}) to the first-order modification of a site quantity associated with
$\boldsymbol{R}$ depends on the first-order modification of the site quantity matrix element associated
only with lattice site $\boldsymbol{R}$, $\Lambda_{\alpha\boldsymbol{R};\alpha\boldsymbol{R}}^{(1)}$, and with orbital types $\alpha$ that are originally occupied. It is not associated with any change in the single-particle density matrix, but rather with the dependence of the associated site quantity matrix elements on the electromagnetic field itself; thus we call it a ``compositional'' modification. It leads to a dependence of the moments $\boldsymbol{\mu}_{\boldsymbol{R}}$ and $\boldsymbol{\nu}_{\boldsymbol{R}}$ on those fields, even though in this contribution (\ref{lambdaSite}) the electron populations remain as they were before the system was perturbed. There is a familiar analog to this in the response of an atom to the Maxwell magnetic field. Considering a single electron, the initial operator for the magnetic dipole moment $\boldsymbol{\nu}_{\text{atom}}=(e/2mc)\boldsymbol{\mathfrak{X}}\cross\boldsymbol{\mathfrak{P}},$ where here $\boldsymbol{\mathfrak{X}}$ and $\boldsymbol{\mathfrak{P}}$ are the position and momentum operators of the electron, becomes $\boldsymbol{\nu}_{\text{atom}}\rightarrow(e/2mc)\boldsymbol{\mathfrak{X}}\cross(\boldsymbol{\mathfrak{P}}-e\boldsymbol{A}(\boldsymbol{\mathfrak{X}})/c)$ when the magnetic field is nonvanishing. For a uniform magnetic field, we can take $\boldsymbol{A}(\boldsymbol{\mathfrak{X}})=(\boldsymbol{B}\cross\boldsymbol{\mathfrak{X}})/2$, giving 
\begin{align}
\boldsymbol{\nu}_{\text{atom}}=\frac{e}{2mc}\boldsymbol{\mathfrak{X}}\cross\boldsymbol{\mathfrak{P}}-\frac{e^{2}}{4mc^{2}}\Big(\mathfrak{X}^{2}\boldsymbol{B}-(\boldsymbol{\mathfrak{X}}\boldsymbol{\cdot}\boldsymbol{B})\boldsymbol{\mathfrak{X}}\Big). \label{eq:diamagnetic}
\end{align}
The second term gives a contribution when the expectation value is taken, even in the ground state (say a $1s$ orbital), and gives the diamagnetic response of the atom. The contributions from the compositional modification to (\ref{siteModification}) are of this form. Nonetheless, in the extended systems we consider it is important to note that during the perturbative analysis many lattice sites and all orbital types may be involved; for instance, observe that (\ref{jTildeSite})-(\ref{Gamma}) would be used in constructing (\ref{nuTilde}).

We also distinguish between the first-order modifications arising from the uniform dc Maxwell electric and magnetic fields individually, such that 
\begin{align}
\Lambda_{\boldsymbol{R}}^{(1)}=\Lambda_{\boldsymbol{R}}^{(E)}+\Lambda_{\boldsymbol{R}}^{(B)},
\end{align}
where each of these modifications is composed of a dynamical and a compositional term, 
\begin{align*}
\Lambda_{\boldsymbol{R}}^{(E)}&=\Lambda_{\boldsymbol{R}}^{(E;\text{I})}+\Lambda_{\boldsymbol{R}}^{(E;\text{II})}, \\ \Lambda_{\boldsymbol{R}}^{(B)}&=\Lambda_{\boldsymbol{R}}^{(B;\text{I})}+\Lambda_{\boldsymbol{R}}^{(B;\text{II})}.
\end{align*}

\subsubsection{Induced polarization}

\paragraph{Modification due to the electric field.}
We begin by considering modifications due to the electric field. From (\ref{rhoElements}), and the fact that $\chi_{\alpha\boldsymbol{R}}(\boldsymbol{x})$ only depends on the magnetic field and $\Delta(\boldsymbol{R},\boldsymbol{x},\boldsymbol{R})=0$, it is clear that 
\begin{align}
\rho^{(E)}_{\alpha\boldsymbol{R}';\alpha\boldsymbol{R}'}(\boldsymbol{y},\boldsymbol{R})=0, \label{rhoE}
\end{align}
and so there is no compositional modification to $\boldsymbol{\mu}^{(E)}_{\boldsymbol{R}}$. The first-order modification is entirely dynamical, as described above, and given by
\begin{align}
\mu_{\boldsymbol{R}}^{i(E)}&=\sum\limits_{\alpha\beta\boldsymbol{R}'\boldsymbol{R}''}\left[\int \big(y^i-R^i\big)\rho^{(0)}_{\beta\boldsymbol{R}';\alpha\boldsymbol{R}''}(\boldsymbol{y},\boldsymbol{R})d\boldsymbol{y} \right]\eta^{(E)}_{\alpha\boldsymbol{R}'';\beta\boldsymbol{R}'}\nonumber\\
&=e^2\Omega_{uc}E^l\sum_{mn}f_{nm}\int_{\text{BZ}}\frac{d\boldsymbol{k}}{(2\pi)^3}\frac{\xi^l_{mn}\xi^i_{nm}}{E_{m\boldsymbol{k}}-E_{n\boldsymbol{k}}}.\label{muE}
\end{align}
Implementing (\ref{PM}), we find the usual result from perturbation theory \cite{Aversa}. This modification is gauge invariant, in that the final line of (\ref{muE}) is independent of the unitary transformation $U_{n\alpha}(\boldsymbol{k})$ \footnote{As previously discussed, in this paper, we move the gauge freedom of the energy eigenvectors, and thus the gauge dependence of the connections $\xi^a_{mn}$, into the $U_{n\alpha}$ matrices.}. Also, the modification of $\mu_{\boldsymbol{R}}^{i(E)}$ due to $E^l$ is the same as that of $\mu_{\boldsymbol{R}}^{l(E)}$ due to $E^i$.

\paragraph{Modification due the magnetic field.}
The first-order modification due to the magnetic field has nonvanishing dynamical and compositional modifications. Considering first the compositional modification, we find
\begin{align}
\mu_{\boldsymbol{R}}^{i(B;\text{II})}
&=\sum\limits_{\alpha\boldsymbol{R}'}f_\alpha\left[\int \big(y^i-R^i\big)\rho^{(B)}_{\alpha\boldsymbol{R}';\alpha\boldsymbol{R}'}(\boldsymbol{y},\boldsymbol{R})d\boldsymbol{y}\right]\nonumber\\
&=\frac{e^2\Omega_{uc}}{2\hbar c}\epsilon^{lab}B^l\sum_{\alpha\gamma}f_\alpha\int_{\text{BZ}}\frac{d\boldsymbol{k}}{(2\pi)^3} \text{Re}\big[\tilde{\xi}^i_{\alpha\gamma}\partial_b \tilde{\xi}^a_{\gamma\alpha}\big].
\label{muBb}
\end{align}
We later simplify this term; notably it is gauge dependent. Recalling the modification of the single-particle density matrix due to the magnetic field (\ref{EDMb2}), we find
\begin{align}
&\mu_{\boldsymbol{R}}^{i(B;\text{I})}\nonumber\\
&=\sum\limits_{\alpha\beta\boldsymbol{R}'\boldsymbol{R}''}\left[\int \big(y^i-R^i\big)\rho^{(0)}_{\beta\boldsymbol{R}';\alpha\boldsymbol{R}''}(\boldsymbol{y},\boldsymbol{R})d\boldsymbol{y} \right]\eta^{(B)}_{\alpha\boldsymbol{R}'';\beta\boldsymbol{R}'} \nonumber\\
&=\frac{e^2\Omega_{uc}}{2\hbar c}\epsilon^{lab}B^l\sum_{mn}f_{nm}\int_{\text{BZ}}\frac{d\boldsymbol{k}}{(2\pi)^3}\Bigg\{\frac{\partial_b(E_{m\boldsymbol{k}}+E_{n\boldsymbol{k}})}{E_{m\boldsymbol{k}}-E_{n\boldsymbol{k}}}\xi^a_{mn}\xi^i_{nm}\nonumber\\
&\qquad\qquad+\sum_{s}\frac{E_{s\boldsymbol{k}}-E_{n\boldsymbol{k}}}{E_{m\boldsymbol{k}}-E_{n\boldsymbol{k}}}\text{Re}\big[i\xi^a_{ms}\xi^b_{sn}\xi^i_{nm}\big]\Bigg\} \nonumber\\
&+\frac{e^2\Omega_{uc}}{2\hbar c}\epsilon^{lab}B^l\sum_{mns}f_{nm}\int_{\text{BZ}}\frac{d\boldsymbol{k}}{(2\pi)^3}\text{Re}\big[i\xi^i_{ns}\mathcal{W}^a_{sm}\xi^b_{mn}\big],\label{muBa}
\end{align}
which is also gauge dependent. The two separate terms appearing in the final equality of (\ref{muBa}) originate individually from the two terms of (\ref{EDMb2}). In going from the first to the final equality we have implemented (\ref{firstMoment},\ref{connection}), and used $\mathcal{W}^i_{mn}\neq 0$ only if $f_m= f_n$, which holds for the class of insulators considered here. Similar arguments are used in the following subsection when finding (\ref{nuBarE},\ref{nuTildeEa}). In Sec.~\ref{Sect:3}, we explicitly combine (\ref{muBb}) and (\ref{muBa}), and show that the usual OMP tensor \cite{Essin2010} is reproduced.

\subsubsection{Induced magnetization}

In this work, we only consider modifications of the site magnetic dipole moment arising due to the electric field. We defer to a later study the modification of the magnetic dipole moment due to the magnetic field, as considered in this framework.

\paragraph{Modification of the atomic-like contribution due to the electric field.}
This involves the first-order modification of (\ref{nuBar}) due to the electric field. The compositional modification vanishes, as from (\ref{jElements}) it is clear that
\begin{align}
j^{b(E)}_{\alpha\boldsymbol{R}';\alpha\boldsymbol{R}'}(\boldsymbol{y},\boldsymbol{R})=0,
\end{align}
following the argument leading to (\ref{rhoE}), and so this modification is entirely dynamical. Using (\ref{EDMe}) we find
\begin{align}
&\bar{\nu}_{\boldsymbol{R}}^{i(E)}= \nonumber \\
&\frac{e^2\Omega_{uc}}{4\hbar c}\epsilon^{iab}E^l\sum_{mn}f_{nm}\int_{\text{BZ}}\frac{d\boldsymbol{k}}{(2\pi)^3}\Bigg\{\frac{\partial_b(E_{m\boldsymbol{k}}+E_{n\boldsymbol{k}})}{E_{m\boldsymbol{k}}-E_{n\boldsymbol{k}}}\xi^a_{nm}\xi^l_{mn}\nonumber\\
&\qquad\qquad+2\sum_{s}\frac{E_{s\boldsymbol{k}}-E_{m\boldsymbol{k}}}{E_{m\boldsymbol{k}}-E_{n\boldsymbol{k}}}\text{Re}\big[i\xi^a_{ns}\xi^b_{sm}\xi^l_{mn}\big]\Bigg\} \nonumber\\
&+\frac{e^2\Omega_{uc}}{2\hbar c}\epsilon^{iab}E^l\sum_{mns}f_{nm}\int_{\text{BZ}}\frac{d\boldsymbol{k}}{(2\pi)^3}\Bigg\{\text{Re}\big[i\xi^l_{mn}\mathcal{W}^b_{ns}\xi^a_{sm}\big]\nonumber\\
&\qquad\qquad+\frac{E_{s\boldsymbol{k}}-E_{n\boldsymbol{k}}}{E_{m\boldsymbol{k}}-E_{n\boldsymbol{k}}}\text{Re}\big[i\xi^l_{mn}\mathcal{W}^a_{ns}\xi^b_{sm}\big]\Bigg\}. \label{nuBarE}
\end{align}

\paragraph{Modification of the itinerant contribution due to the electric field.}
This involves the first-order modification of (\ref{nuTilde}) due to the electric field, and has nonvanishing modifications of both compositional and dynamical origin. The compositional modification is
\begin{align}
&\tilde{\nu}_{\boldsymbol{R}}^{i(E;\text{II})}=\frac{e^2\Omega_{uc}}{2\hbar c}\epsilon^{iab}E^l\sum_{\alpha\gamma}f_\alpha\int_{\text{BZ}}\frac{d\boldsymbol{k}}{(2\pi)^3} \text{Re}\big[\tilde{\xi}^l_{\alpha\gamma}\partial_b \tilde{\xi}^a_{\gamma\alpha}\big],\label{nuTildeEb}
\end{align}
and the dynamical modification is
\begin{align}
&\tilde{\nu}_{\boldsymbol{R}}^{i(E;\text{I})}=\nonumber\\
&\frac{e^2\Omega_{uc}}{4\hbar c}\epsilon^{iab}E^l\sum_{mn}f_{nm}\int_{\text{BZ}}\frac{d\boldsymbol{k}}{(2\pi)^3}\Bigg\{\frac{\partial_b(E_{m\boldsymbol{k}}+E_{n\boldsymbol{k}})}{E_{m\boldsymbol{k}}-E_{n\boldsymbol{k}}}\xi^a_{nm}\xi^l_{mn}\nonumber\\
&\qquad\qquad-2\sum_{s}\frac{E_{s\boldsymbol{k}}-E_{n\boldsymbol{k}}}{E_{m\boldsymbol{k}}-E_{n\boldsymbol{k}}}\text{Re}\big[i\xi^l_{mn}\mathcal{W}^a_{ns}\xi^b_{sm}\big]\Bigg\}. \label{nuTildeEa}
\end{align}

Both (\ref{nuBarE}) and (\ref{nuTildeEb}) are gauge dependent in general, while (\ref{nuTildeEa}) is multiband gauge dependent. Very generally, there is a simplification that occurs when (\ref{nuBarE},\ref{nuTildeEb},\ref{nuTildeEa}) are summed to form the total site magnetic dipole moment: the term appearing in the final line of (\ref{nuTildeEa}) cancels with the term appearing in the final line of (\ref{nuBarE}), and as a result the gauge dependent terms appearing in the total $\boldsymbol{\nu}_{\boldsymbol{R}}^{(E)}$ do not explicitly depend on the energies $E_{n\boldsymbol{k}}$.

\section{Microscopic origin of contributions to the OMP tensor}
\label{Sect:3}

\subsection{Constructing the OMP tensor}

The OMP tensor, which describes the first-order modification of the macroscopic polarization due to a uniform dc magnetic field, is defined through
\begin{align*}
P^{i(B)}=\alpha^{il}B^l,
\end{align*}
and, from (\ref{PM}), (\ref{muBb}), and (\ref{muBa}), is found to be
\begin{align}
\alpha^{il}&=\frac{e^2}{2\hbar c}\epsilon^{lab}\sum_{mn}f_{nm}\int_{\text{BZ}}\frac{d\boldsymbol{k}}{(2\pi)^3}\Bigg\{\frac{\partial_b(E_{m\boldsymbol{k}}+E_{n\boldsymbol{k}})}{E_{m\boldsymbol{k}}-E_{n\boldsymbol{k}}}\xi^a_{nm}\xi^i_{mn}\nonumber\\
&\qquad\qquad+\sum_{s}\frac{E_{s\boldsymbol{k}}-E_{m\boldsymbol{k}}}{E_{m\boldsymbol{k}}-E_{n\boldsymbol{k}}}\text{Re}\big[i\xi^a_{ns}\xi^b_{sm}\xi^i_{mn}\big]\Bigg\} \nonumber\\
&+\frac{e^2}{2\hbar c}\epsilon^{lab}\sum_{mns}f_{nm}\int_{\text{BZ}}\frac{d\boldsymbol{k}}{(2\pi)^3}\text{Re}\big[i\xi^i_{mn}\mathcal{W}^b_{ns}\xi^a_{sm}\big]\nonumber\\
&+\frac{e^2}{2\hbar c}\epsilon^{lab}\sum_{\alpha\gamma}f_\alpha \int_{\text{BZ}}\frac{d\boldsymbol{k}}{(2\pi)^3}\text{Re}\big[\tilde{\xi}^i_{\alpha\gamma}\partial_b \tilde{\xi}^a_{\gamma\alpha}\big],\label{alpha}
\end{align}
after some manipulation of the band indices. If we now define an analogous tensor describing the first-order modification of the macroscopic magnetization due to a uniform dc electric field, we find that, upon combining (\ref{nuBarE}), (\ref{nuTildeEb}), and (\ref{nuTildeEa}), this modification is described by the same $\alpha^{il}$ tensor introduced above, but with the order of the indices switched, such that 
\begin{align*}
M^{i(E)}=\alpha^{li}E^l.
\end{align*}
This is the usual result in the $\hbar\omega\ll E_{\text{gap}}$ limit, which is effectively the condition we have initially assumed.

In the previous section we made a concerted effort to identify dynamical and compositional contributions to the various first-order modifications; distinguishing between these proves useful here. Because the $\alpha^{il}$ tensor describes the first-order modification of both $P^i$ due to $B^l$, and $M^l$ due to $E^i$, we can focus on the terminology associated with only the magnetization. In what follows, we illustrate how the three types of nonvanishing terms -- the dynamical atomic-like, the compositional itinerant, and the dynamical itinerant modifications -- combine to give an OMP tensor having the usual form
\begin{align}
\alpha^{il}={\alpha}_{\text{G}}^{il}+\alpha_{\text{CS}}^{il}, \label{alphaDecomp}
\end{align}
where ${\alpha}_{\text{G}}^{il}$ contains only cross-gap contributions and $\alpha_{\text{CS}}^{il}$, the Chern-Simons contribution, is a property of the subspace spanned by the originally occupied $\ket{n\boldsymbol{k}}$. We find that the cross-gap contribution, ${\alpha}_{\text{G}}^{il}$, originates from a combination of the dynamical atomic-like (\ref{nuBarE}) and dynamical itinerant (\ref{nuTildeEa}) terms, whereas $\alpha_{\text{CS}}^{il}$ has contributions from the dynamical atomic-like term as well, but also from the compositional itinerant term (\ref{nuTildeEb}).

In order to compare with past results \cite{Malashevich2010,Essin2010}, we will re-express (\ref{alpha}) in terms the cell-periodic functions $\braket{\boldsymbol{x}}{n\boldsymbol{k}}$, and re-write the sums to be over occupied states (the set $\{\ket{v\boldsymbol{k}}\}$) and unoccupied states (the set $\{\ket{c\boldsymbol{k}}\}$). We first consider the term in (\ref{alpha}) involving the ratio of energy differences, which can be traced back to (\ref{nuBarE}), and adopt the shorthand $\ket{n}\equiv\ket{{n\boldsymbol{k}}}$; we find
\begin{widetext}
\begin{align}
&\epsilon^{lab}\sum_{mns}f_{nm}\frac{E_{s\boldsymbol{k}}-E_{m\boldsymbol{k}}}{E_{m\boldsymbol{k}}-E_{n\boldsymbol{k}}}\text{Re}\big[i\xi^a_{ns}\xi^b_{sm}\xi^i_{mn}\big]= \nonumber\\
&\quad 2\epsilon^{lab}\Bigg\llbracket-\sum_{cvv'}\frac{E_{v\boldsymbol{k}}-E_{v'\boldsymbol{k}}}{E_{v\boldsymbol{k}}-E_{c\boldsymbol{k}}}\text{Re}\big[\left(\partial_bv|v'\right)\left(\partial_av'|c\right)\left(c|\partial_iv\right)\big]+\sum_{cc'v}\frac{E_{c\boldsymbol{k}}-E_{c'\boldsymbol{k}}}{E_{v\boldsymbol{k}}-E_{c\boldsymbol{k}}}\text{Re}\big[\left(\partial_bv|c'\right)\left(c'|\partial_ac\right)\left(c|\partial_iv\right)\big]\Bigg\rrbracket\nonumber\\
&\quad+\epsilon^{lab}\Bigg\llbracket2\sum_{cvv'}\text{Re}\big[\left(\partial_iv|c\right)\left(c|\partial_av'\right)\left(v'|\partial_bv\right)\big]+\sum_{cv}\text{Re}\big[\left(\partial_iv|c\right)\left(\partial_ac|\partial_bv\right)\big]\Bigg\rrbracket.
\label{eq:work1}
\end{align}
The first set of $\llbracket...\rrbracket$'s are identified as cross-gap contributions and will be included in ${\alpha}_{\text{G}}^{il}$. The second set of $\llbracket...\rrbracket$'s, together with the penultimate and final terms of (\ref{alpha}), form $\alpha_{\text{CS}}^{il}$; the penultimate term of (\ref{alpha}) originates from the gauge dependent term arising in the dynamical atomic-like modification (\ref{nuBarE}) that does not explicitly depend on energy, and the final term of (\ref{alpha}) originates from the compositional itinerant modification (\ref{nuTildeEb}). We find 
\begin{align}
\alpha^{il}_{\text{CS}}&=\frac{e^2}{2\hbar c}\epsilon^{lab}\int_{\text{BZ}}\frac{d\boldsymbol{k}}{(2\pi)^3}\Bigg\llbracket2\sum_{cvv'}\text{Re}\big[\left(\partial_iv|c\right)\left(c|\partial_av'\right)\left(v'|\partial_bv\right)\big]+\sum_{cv}\text{Re}\big[\left(\partial_iv|c\right)\left(\partial_ac|\partial_bv\right)\big]\Bigg\rrbracket\nonumber\\
&+\frac{e^2}{2\hbar c}\epsilon^{lab}\sum_{mns}f_{nm}\int_{\text{BZ}}\frac{d\boldsymbol{k}}{(2\pi)^3}\text{Re}\big[i\xi^i_{mn}\mathcal{W}^b_{ns}\xi^a_{sm}\big]+\frac{e^2}{2\hbar c}\epsilon^{lab}\sum_{\alpha\gamma}f_\alpha \int_{\text{BZ}}\frac{d\boldsymbol{k}}{(2\pi)^3}\text{Re}\big[\tilde{\xi}^i_{\alpha\gamma}\partial_b \tilde{\xi}^a_{\gamma\alpha}\big] \nonumber\\
&=-\delta^{il}\frac{e^2}{2\hbar c}\epsilon^{abc}\int_{\text{BZ}}\frac{d\boldsymbol{k}}{(2\pi)^3}\Bigg\{\left(\sum_{vv'}\xi^a_{vv'}\partial_b\xi^c_{v'v}-\frac{2i}{3}\sum_{vv'v_1}\xi^a_{vv'}\xi^b_{v'v_1}\xi^c_{v_1v}\right)+\sum_{vv'}(\partial_b\mathcal{W}^a_{vv'})\mathcal{W}^c_{v'v}-\frac{2i}{3}\sum_{vv'v_1}\mathcal{W}^a_{vv'}\mathcal{W}^b_{v'v_1}\mathcal{W}^c_{v_1v}\Bigg\}
\label{alphaCS}
\end{align}
(Appendix \ref{AppendixCS}), which is the usual Chern-Simons contribution to the OMP tensor \cite{Qi2008,Vanderbilt2009,Essin2010,Malashevich2010}, and is multiband gauge dependent in the sense introduced after Eq.~(\ref{nuTildeUnpert}). Due to this gauge dependence, this contribution to the OMP tensor is multivalued, but, like $\boldsymbol{P}^{(0)}$, has been shown to be unique modulo a quantum of ambiguity \cite{VanderbiltBook}. Furthermore, this contribution is isotropic, and the corresponding quantity vanishes in systems of spatial dimension less than three.

The remaining terms compose ${\alpha}^{il}_{\text{G}}$, in accordance with (\ref{alphaDecomp}). These terms originate from the dynamical atomic-like modification (\ref{nuBarE}), and the dynamical itinerant modification (\ref{nuTildeEa}). We find
\begin{align}
{\alpha}^{il}_{\text{G}}&=\frac{e^2}{\hbar c}\epsilon^{lab}\int_{\text{BZ}}\frac{d\boldsymbol{k}}{(2\pi)^3}\Bigg\{-\sum_{cv}\frac{\partial_b( E_{c\boldsymbol{k}}+E_{v\boldsymbol{k}})}{E_{v\boldsymbol{k}}-E_{c\boldsymbol{k}}}\text{Re}\big[\left(\partial_av|c\right)\left(c|\partial_iv\right)\big]-\sum_{cvv'}\frac{E_{v\boldsymbol{k}}-E_{v'\boldsymbol{k}}}{E_{v\boldsymbol{k}}-E_{c\boldsymbol{k}}}\text{Re}\big[\left(\partial_bv|v'\right)\left(\partial_av'|c\right)\left(c|\partial_iv\right)\big]\nonumber\\
&\qquad\qquad\qquad\qquad\qquad+\sum_{cc'v}\frac{E_{c\boldsymbol{k}}-E_{c'\boldsymbol{k}}}{E_{v\boldsymbol{k}}-E_{c\boldsymbol{k}}}\text{Re}\big[\left(\partial_bv|c'\right)\left(c'|\partial_ac\right)\left(c|\partial_iv\right)\big]\Bigg\},
\label{alphaG}
\end{align}
\end{widetext}
which is in agreement with the usual expression for the cross-gap contribution \cite{Essin2010,Malashevich2010}. Importantly this contribution is found to be gauge invariant; this is a result of the fact that the multiband gauge dependent terms appearing in (\ref{nuBarE}) and (\ref{nuTildeEa}) canceled one another.

It has been pointed out that the net gauge dependence of the OMP tensor, through the Chern-Simons contribution, has no effect on the induced charge-current density in the bulk \cite{Vanderbilt2009,Essin2010}. We present a slightly different formulation of that argument here, but starting with the same assumption used in earlier arguments: While the derivation we have presented here, as well as that in the approach of the ``modern theory,'' holds strictly only for uniform dc fields, we can expect that if the macroscopic Maxwell electric and magnetic fields that we consider are varying very slowly in both space and time, the same tensors $\alpha_{\text{CS}}^{il}$ and $\alpha_{\text{G}}^{il}$ can be used to good approximation \footnote{See, e.g., Zhong \textit{et al.}~\cite{Souza2016}.}. For such slowly varying fields, the first-order response to the electromagnetic field in the bulk of a medium can be expressed as 
\begin{align}
\boldsymbol{P}^{(1)}(\boldsymbol{x},t)&=\boldsymbol{P}'^{(1)}(\boldsymbol{x},t)+\alpha_{\text{CS}}\boldsymbol{B}(\boldsymbol{x},t),\nonumber\\
\boldsymbol{M}^{(1)}(\boldsymbol{x},t)&=\boldsymbol{M}'^{(1)}(\boldsymbol{x},t)+\alpha_{\text{CS}}\boldsymbol{E}(\boldsymbol{x},t), \label{eq:CSrelations} 
\end{align}
where we have taken $\alpha_{\text{CS}}^{il}=\delta^{il}\alpha_{\text{CS}}$, and at the level of linear response we have 
\begin{align*}
P'^{i(1)}(\boldsymbol{x},t)&=\chi_{E}^{il}E^{l}(\boldsymbol{x},t)+\alpha_{\text{G}}^{il}B^{l}(\boldsymbol{x},t),\\
M'^{i(1)}(\boldsymbol{x},t)&=\chi_{B}^{il}B^{l}(\boldsymbol{x},t)+\alpha_{\text{G}}^{li}E^{l}(\boldsymbol{x},t),
\end{align*}
where $\chi_{E}^{il}$ and $\chi_{B}^{il}$ are the usual linear electric
and magnetic susceptibilities. Inside the bulk material we then immediately
find that the induced macroscopic charge and current densities that result,
\begin{align}
\rho^{(1)}(\boldsymbol{x},t)&=-\boldsymbol{\nabla}\boldsymbol{\cdot}\boldsymbol{P}^{(1)}(\boldsymbol{x},t),\nonumber\\
\boldsymbol{J}^{(1)}(\boldsymbol{x},t)&=\frac{\partial\boldsymbol{P}^{(1)}(\boldsymbol{x},t)}{\partial t}+c\boldsymbol{\nabla}\cross\boldsymbol{M}^{(1)}(\boldsymbol{x},t), \label{eq:Jrho_PM}
\end{align}
can also be written as 
\begin{align}
\rho^{(1)}(\boldsymbol{x},t)&=-\boldsymbol{\nabla}\boldsymbol{\cdot}\boldsymbol{P}'^{(1)}(\boldsymbol{x},t),\nonumber\\
\boldsymbol{J}^{(1)}(\boldsymbol{x},t)&=\frac{\partial\boldsymbol{P}'^{(1)}(\boldsymbol{x},t)}{\partial t}+c\boldsymbol{\nabla}\cross\boldsymbol{M}'^{(1)}(\boldsymbol{x},t). \label{eq:Jrho_PMtilde}
\end{align}
That is, the contributions from the Chern-Simons coefficient,
$\alpha_{\text{CS}}$, completely cancel each other. In deriving (\ref{eq:Jrho_PMtilde}) from
(\ref{eq:Jrho_PM}) we have used Faraday's law and Gauss' law for magnetism, 
\begin{gather}
c\boldsymbol{\nabla}\cross\boldsymbol{E}(\boldsymbol{x},t)+\frac{\partial\boldsymbol{B}(\boldsymbol{x},t)}{\partial t}=0, \nonumber\\
\boldsymbol{\nabla}\boldsymbol{\cdot}\boldsymbol{B}(\boldsymbol{x},t)=0, \label{eq:FG} 
\end{gather}
which of course must be assumed to hold for $\boldsymbol{E}(\boldsymbol{x},t)$
and $\boldsymbol{B}(\boldsymbol{x},t)$, no matter how slowly they
are varying in space and time. Interestingly, the relation between
$\{ \boldsymbol{P}^{(1)}(\boldsymbol{x},t),\boldsymbol{M}^{(1)}(\boldsymbol{x},t)\} $
and $\{ \boldsymbol{P}'^{(1)}(\boldsymbol{x},t),\boldsymbol{M}'^{(1)}(\boldsymbol{x},t)\} $
can be understood as another kind of ``gauge dependence.'' Very
generally, such sets of fields lead to the same induced charge-current densities when
\begin{align}
\boldsymbol{P}'^{(1)}(\boldsymbol{x},t)&=\boldsymbol{P}^{(1)}(\boldsymbol{x},t)+\boldsymbol{\nabla}\cross\boldsymbol{a}(\boldsymbol{x},t),\nonumber\\
\boldsymbol{M}'^{(1)}(\boldsymbol{x},t)&=\boldsymbol{M}^{(1)}(\boldsymbol{x},t)-\frac{1}{c}\frac{\partial\boldsymbol{a}(\boldsymbol{x},t)}{\partial t}+\boldsymbol{\nabla}b(\boldsymbol{x},t),
\end{align}
for a general vector field $\boldsymbol{a}(\boldsymbol{x},t)$ and
a general scalar field $b(\boldsymbol{x},t);$ here the sets of fields $\{\boldsymbol{P}^{(1)}(\boldsymbol{x},t),\boldsymbol{M}^{(1)}(\boldsymbol{x},t)\}$
and $\{\boldsymbol{P}'^{(1)}(\boldsymbol{x},t),\boldsymbol{M}'^{(1)}(\boldsymbol{x},t)\}$
of (\ref{eq:CSrelations}) are related by 
\begin{align}
b(\boldsymbol{x},t)&=0,\nonumber\\ \boldsymbol{a}(\boldsymbol{x},t)&=c\alpha_{\text{CS}}\int_{-\infty}^{t}\boldsymbol{E}(\boldsymbol{x},t')dt', 
\end{align}
which can be easily confirmed using (\ref{eq:FG}), under the condition that at $t=-\infty$ the system is unperturbed.

In the past, the origin of the ambiguity arising in the OMP tensor has, like that in the unperturbed macroscopic polarization, been understood in the context of finite sized systems \cite{VanderbiltBook,Lee2012,Souza2017}. At a surface where $\alpha^{il}$ cannot be treated as uniform the argument in the above paragraph breaks down, and the surface current that will arise shows an ambiguity that reflects the well-known gauge dependence of $\alpha_{\text{CS}}$ \cite{Qi2008,Vanderbilt2009}. However, it seems there should be an equivalent bulk interpretation. That certainly holds for the unperturbed macroscopic polarization; in a calculation where the energy eigenvectors are chosen and fixed at the start, a ``quantum of ambiguity'' arises from the gauge dependent term of (\ref{unpertP}). This term has been shown \cite{Resta1994} to depend only on the phase of the determinant of $U(\boldsymbol{k})$, and as such, is not qualitatively different whether the bands are isolated or not. Indeed, this ambiguity can be understood at the level of a U$(1)$ gauge transformation, which in the simplest of cases takes $U_{n\alpha}(\boldsymbol{k})$ of form $\delta_{n\alpha}e^{-i\boldsymbol{k}\boldsymbol{\cdot}\boldsymbol{R}}$; this corresponds to changing the site with which each Wannier function is associated and, in turn, this changes $\boldsymbol{P}^{(0)}$ by a discrete amount, proportional to $\boldsymbol{R}$. However, such an interpretation cannot be used to understand the ambiguity associated with the OMP tensor; the gauge dependent term of (\ref{alphaCS}) vanishes in the limit of isolated bands, and thus must interpreted on the more general grounds of a multiband gauge transformation. Yet, at some level these ambiguities appear to be related, since the terms giving rise to them are constructed from the same object, the $\mathcal{W}^a$ matrix; (\ref{unpertP}) contains only diagonal matrix elements, while (\ref{alphaCS}) contains off-diagonal matrix elements as well. Perhaps it is from this perspective that a bulk interpretation of the discrete ambiguity associated (\ref{alphaCS}) can be formulated.

\subsection{Limiting cases}
\label{Sect:limits}

In this section we explore the magnetoelectric effect in the limit of isolated molecules. First we construct the relevant tensor for a single molecule, and then use that to construct the OMP tensor of a crystal in the ``molecular crystal limit,'' which we take to be a model where there is a molecule at each lattice site with orbitals that have no common support with orbitals of molecules associated with other lattice sites. Finally, we show that the molecular crystal limit so obtained is in agreement with the appropriate limit of our general expressions (\ref{alphaCS}) and (\ref{alphaG}).

\subsubsection{A single molecule}

As pointed out earlier (Appendix D of Mahon \textit{el al.}~\cite{Mahon2019}),
the response of a molecule to an electromagnetic field can
be treated via the same approach we have used here for a crystal, by
constructing microscopic polarization and magnetization fields from
the electronic Green function, with the expectation values of the
electric and magnetic dipole moments following from the single-particle
density matrix (\ref{mu},\ref{nuBar},\ref{nuTilde}). However, for
a molecule (or an atom), it is also possible to follow a more common strategy
in molecular physics \cite{Healybook,PZW}, where microscopic polarization
and magnetization \textit{operators} are introduced, leading to operators associated
with the electric and magnetic dipole moments. We present that approach
here (Appendix C of Mahon \textit{et al.}~\cite{Mahon2019}) to better make
the connection between this calculation and molecular physics.

We take the initially unperturbed system to be described by (\ref{H},\ref{physicalP}),
where we include an $\boldsymbol{A}_{\text{static}}(\boldsymbol{x})$ --
which of course need not be periodic, since we are considering a localized
system -- to guarantee the breaking of time-reversal symmetry, and
consider a $V(\boldsymbol{x})$ that vanishes as $|\boldsymbol{x}|\rightarrow\infty$ and that does not satisfy inversion symmetry
about any point. The latter condition could arise in a molecule because
of a noncentrosymmetric configuration of the nuclei, or even in an
atomic system because of an imposed dc electric field that is considered
part of the unperturbed Hamiltonian. We consider a ``special point'' $\boldsymbol{R}=\boldsymbol{0}$,
which for a molecule could be taken to be, say, the center of mass
of the ion cores and for an atom could be taken as the position of
the ion core. In the frozen-ion approximation the contribution of
the ions to the multipole moments of a molecule will not affect the
perturbative response calculation we make, so we neglect them. We
introduce a ``special point'' electron field operator (see Appendix C of Mahon \textit{et al.}~\cite{Mahon2019}) $\widehat{\psi}_{\text{sp}}(\boldsymbol{x},t)$,
\begin{align*}
& \widehat{\psi}_{\text{sp}}(\boldsymbol{x},t)=e^{-i\Phi(\boldsymbol{x},\boldsymbol{0};t)}\widehat{\psi}(\boldsymbol{x},t),
\end{align*}
where 
\begin{align*}
& \Phi(\boldsymbol{x},\boldsymbol{0};t)\equiv\frac{e}{\hbar c}\int s^{a}(\boldsymbol{w};\boldsymbol{x},\boldsymbol{0})A^{a}(\boldsymbol{w},t)d\boldsymbol{w},
\end{align*}
and $\boldsymbol{A}(\boldsymbol{x},t)$ is again the vector potential describing the electromagnetic field. Then the Hamiltonian operator governing the evolution of $\widehat{\psi}_{\text{sp}}(\boldsymbol{x},t)$ is 
\begin{align}
&\mathsf{\widehat{H}}_{\text{sp}}(t)=\int\widehat{\psi}_{\text{sp}}^{\dagger}(\boldsymbol{x},t)\Big(H_{0}\big(\boldsymbol{x},\boldsymbol{\mathfrak{p}}(\boldsymbol{x},\boldsymbol{0};t)\big)\nonumber\\
&\qquad\qquad\qquad\qquad\qquad-e\Omega_{\boldsymbol{0}}^{0}(\boldsymbol{x},t)\Big)\widehat{\psi}_{\text{sp}}(\boldsymbol{x},t)d\boldsymbol{x},
\label{eq:Hsp}
\end{align}
where $\boldsymbol{\mathfrak{p}}(\boldsymbol{x},\boldsymbol{0};t)$ is given by (\ref{frakP}) and $\Omega_{\boldsymbol{0}}^{0}(\boldsymbol{x},t)$ by (\ref{eq:Omega_0}); we have
\begin{align*}
 i\hbar\frac{\partial\widehat{\psi}_{\text{sp}}(\boldsymbol{x},t)}{\partial t}=\left[\widehat{\psi}_{\text{sp}}(\boldsymbol{x},t),\mathsf{\widehat{H}}_{\text{sp}}(t)\right].
\end{align*}
We begin the evolution at a time $t_{0}$ before the electromagnetic field is nonzero, taking as the (Heisenberg) ket the ground state $\ket{G}$; at this time we have $\widehat{\psi}_{\text{sp}}(\boldsymbol{x},t_{0})=\widehat{\psi}(\boldsymbol{x},t_{0})\equiv\widehat{\psi}(\boldsymbol{x})$. The dynamics can equivalently be described in a Schr\"{o}dinger picture where the field operator is fixed at $\widehat{\psi}(\boldsymbol{x}$) and the ket evolves from $\ket{G}$ at $t_{0}$ according to a Hamiltonian operator $\mathsf{\widehat{H}}(t)$ having the form of (\ref{eq:Hsp}), but with $\widehat{\psi}_{\text{sp}}(\boldsymbol{x},t)$ replaced by $\widehat{\psi}(\boldsymbol{x})$. Using the approximate expressions (\ref{omegaVec},\ref{omega0}) for $\boldsymbol{\Omega_{0}}(\boldsymbol{x},t)$ and $\Omega_{\boldsymbol{0}}^{0}(\boldsymbol{x},t)$, and neglecting the variation of the electric and magnetic fields over the atom or molecule, we can write $\widehat{\mathsf{H}}(t)$ as
\begin{align} \widehat{\mathsf{H}}(t)=\widehat{\mathsf{H}}^{0}-\widehat{\boldsymbol{\mu}}\boldsymbol{\cdot}\boldsymbol{E}(t)-\widehat{\boldsymbol{\nu}}_{\text{P}}\boldsymbol{\cdot}\boldsymbol{B}(t)-\frac{1}{2}\widehat{\boldsymbol{\nu}}_{\text{D}}(t)\boldsymbol{\cdot}\boldsymbol{B}(t),\label{eq:Hatom_use}
\end{align}
where $\boldsymbol{E}(t)\equiv\boldsymbol{E}(\boldsymbol{0},t)$ and
$\boldsymbol{B}(t)\equiv\boldsymbol{B}(\boldsymbol{0},t)$,
\begin{align*} \widehat{\mathsf{H}}^{0}=\int\widehat{\psi}^{\dagger}(\boldsymbol{x})H_{0}\big(\boldsymbol{x},\boldsymbol{\mathfrak{p}}(\boldsymbol{x})\big)\widehat{\psi}(\boldsymbol{x})d\boldsymbol{x},
\end{align*}
and the operator for the electric dipole moment $\widehat{\boldsymbol{\mu}}$, and operators for the paramagnetic ($\widehat{\boldsymbol{\nu}}_{\text{P}}$) and diamagnetic ($\widehat{\boldsymbol{\nu}}_{\text{D}}(t))$ contributions to the magnetic dipole moment operator $\widehat{\boldsymbol{\nu}}(t)=\widehat{\boldsymbol{\nu}}_{\text{P}}+\widehat{\boldsymbol{\nu}} _{\text{D}}(t)$ are given by
\begin{align*}
\widehat{\boldsymbol{\mu}}&=e\int\widehat{\psi}^{\dagger}(\boldsymbol{x})\boldsymbol{x}\widehat{\psi}(\boldsymbol{x})d\boldsymbol{x},\\
\widehat{\boldsymbol{\nu}}_{\text{P}}&=\frac{e}{2mc}\int\widehat{\psi}^{\dagger}(\boldsymbol{x})\Big(\boldsymbol{x}\cross\boldsymbol{\mathfrak{p}}(\boldsymbol{x})\Big)\widehat{\psi}(\boldsymbol{x})d\boldsymbol{x},\\
\widehat{\boldsymbol{\nu}}_{\text{D}}(t)&=-\frac{e^{2}}{4mc^{2}}\int\widehat{\psi}^{\dagger}(\boldsymbol{x})\Big((\boldsymbol{x}\boldsymbol{\cdot}\boldsymbol{x})\boldsymbol{B}(t)\\
&\qquad\qquad\qquad\qquad\qquad-\big(\boldsymbol{x}\boldsymbol{\cdot}\boldsymbol{B}(t)\big)\boldsymbol{x}\Big)\widehat{\psi}(\boldsymbol{x})d\boldsymbol{x}.
\end{align*}
Note that the last expression is the field theoretic analog of (\ref{eq:diamagnetic}). Taking $\widehat{\mathsf{H}}^{0}\ket{G}=E_{G}\ket{G}$, we then calculate the first-order perturbative modifications due to static fields $\boldsymbol{E}$ and $\boldsymbol{B}$. The diamagnetic term in (\ref{eq:Hatom_use}) will make no contribution to first order, and if we denote the excited states by $\ket{H}$ (with energies $E_{H}$), then by standard perturbation theory the modification of the expectation value of the electric dipole moment operator due to the magnetic field, $\expval{\widehat{\boldsymbol{\mu}}}_{\text{atom}}^{(B)}$, and the modification of the expectation value of the magnetic dipole moment operator due to the electric field, $\expval{\widehat{\boldsymbol{\nu}}} _{\text{atom}}^{(E)}$, are given by 
\begin{align}
& \expval{\widehat{\mu}^{i}} _{\text{atom}}^{(B)}=\check{\alpha}^{il}B^{l},\nonumber\\
& \expval{\widehat{\nu}^{i}} _{\text{atom}}^{(E)}=\check{\alpha}^{li}E^{l},
\label{eq:atom_moments} 
\end{align}
where 
\begin{align}
& \check{\alpha}^{il}=2\text{Re}\sum_{H}\frac{\bra{G}\widehat{\mu}^{i}\ket{H}\bra{H}\widehat{\nu}_{\text{P}}^{l}\ket{G}}{E_{H}-E_{G}}.\label{eq:alpha_first}
\end{align}
We expand the field operator $\widehat{\psi}(\boldsymbol{x})$ in terms of orbitals $W_{v}(\boldsymbol{x})$ that are occupied in the ground state and orbitals $W_{c}(\boldsymbol{x})$ that are not, 
\begin{align}
\widehat{\psi}(\boldsymbol{x})=\sum_{v}\widehat{c}_{v}W_{v}(\boldsymbol{x})+\sum_{c}\widehat{c}_{c}W_{c}(\boldsymbol{x}),\label{eq:orbital_expand}
\end{align}
where $\widehat{c}_{v}$ and $\widehat{c}_{c}$ are fermionic annihilation operators generating single-particle eigenfunctions of $H_{0}(\boldsymbol{x},\mathfrak{p}(\boldsymbol{x}))$ with energies $E_{v}$ and $E_{c}$ respectively, with 
\begin{align*}
\widehat{c}_{c}\ket{G}=0 \text{\space,\space} \widehat{c}_{v}^{\dagger}\ket{G}=0.
\end{align*}
Then the expression (\ref{eq:alpha_first}) becomes 
\begin{align}
\check{\alpha}^{il}=2\text{Re}\sum_{c,v}\frac{\left(\mu^{i}\right)_{vc}\left(\nu_{\text{P}}^{l}\right)_{cv}}{E_{c}-E_{v}},\label{eq:alpha_atom2}
\end{align}
where 
\begin{align*}
\left(\mu^{i}\right)_{cv}=ex_{cv}^{i},
\end{align*}
with 
\begin{align}
x_{cv}^{i}\equiv\int W_{c}^{*}(\boldsymbol{x})x^iW_{v}(\boldsymbol{x})d\boldsymbol{x},\label{eq:xcv}
\end{align}
and 
\begin{align*}
\left(\nu_{\text{P}}^{l}\right)_{cv}&=\frac{e}{2mc}\epsilon^{lab}\int W_{c}^{*}(\boldsymbol{x})x^{a}\mathfrak{p}^{b}(\boldsymbol{x})W_{v}(\boldsymbol{x})d\boldsymbol{x}\\
& =\frac{e}{2mc}\epsilon^{lab}\sum_{n}x_{cn}^{a}\int W_{n}^{*}(\boldsymbol{x})\mathfrak{p}^{b}(\boldsymbol{x})W_{v}(\boldsymbol{x})d\boldsymbol{x} \\
&=\frac{ie}{2\hbar c}\epsilon^{lab}\sum_{n}(E_{n}-E_{v})x_{cn}^{a}x_{nv}^{b}.
\end{align*}
Here we have inserted a complete set of states $\{\ket{n}\}$, with single-particle energies $\{E_{n}\}$, into the first expression, where the label $n$ ranges over all $v$ and $c$, and in going to the third line we have used the commutation relation of $\boldsymbol{x}$ and $H_{0}(\boldsymbol{x},\mathfrak{p}(\boldsymbol{x}))$ to write the matrix element of $\mathfrak{p}^{b}(\boldsymbol{x})$ in terms of $x_{nv}^{b}$ in the usual way. From (\ref{eq:alpha_atom2}) we then have
\begin{align}
\check{\alpha}^{il}=\frac{e^{2}}{\hbar c}\epsilon^{lab}\sum_{vcn}\frac{E_{n}-E_{v}}{E_{c}-E_{v}}\text{Re}\left[ix_{vc}^{i}x_{cn}^{a}x_{nv}^{b}\right].\label{eq:alphabar_work}
\end{align}
As in a solid, this vanishes unless both time-reversal symmetry and inversion symmetry are broken: For if there is time-reversal symmetry the orbitals $W_{n}(\boldsymbol{x})$ can be chosen to be real and the quantity inside the brackets is purely imaginary, while if there is inversion symmetry the matrix elements $x_{vc}^{i}$ themselves vanish.

Splitting the sum over $n$ in (\ref{eq:alphabar_work}) into a sum over filled states $v'$ and a sum over empty states $c'$ we can write
\begin{align}
\check{\alpha}^{il}=\check{\alpha}_{\text{G}}^{il}+\check{\alpha}_{\text{CS}}^{il},\label{eq:alphabar_decomp}
\end{align}
where 
\begin{align}
\check{\alpha}_{\text{G}}^{il}&=\frac{e^{2}}{\hbar c}\epsilon^{lab}\sum_{vcc'}\frac{E_{c'}-E_{c}}{E_{c}-E_{v}}\text{Re}\left[ix_{vc}^{i}x_{cc'}^{a}x_{c'v}^{b}\right]\nonumber\\
& +\frac{e^{2}}{\hbar c}\epsilon^{lab}\sum_{vv'c}\frac{E_{v'}-E_{v}}{E_{c}-E_{v}}\text{Re}\left[ix_{vc}^{i}x_{cv'}^{a}x_{v'v}^{b}\right]\label{eq:alphabar_CG} 
\end{align}
and 
\begin{align*}
\check{\alpha}_{\text{CS}}^{il}=\frac{e^{2}}{\hbar c}\epsilon^{lab}\text{Re}\left[i\sum_{vcc'}x_{vc}^{i}x_{cc'}^{a}x_{c'v}^{b}\right].
\end{align*}
In the second of these expressions we sequentially put 
\begin{align*}
\sum_{c'}=\sum_{n}-\sum_{v'} \text{\space , \space} \sum_{c}=\sum_{n'}-\sum_{v''},
\end{align*}
where both $n$ and $n'$ range over all states; noting that the sums over $n$ and $n'$ give no net contribution, we have 
\begin{align*}
& \check{\alpha}_{\text{CS}}^{il}=\frac{e^{2}}{\hbar c}\text{Re}\left[i\epsilon^{lab}\sum_{vv'v''}x_{vv''}^{i}x_{v''v'}^{a}x_{v'v}^{b}\right],
\end{align*}
where the quantity in brackets is real. In three dimensions at least two of $i,l,a,b$ must be identical; if $i\neq l$ the expression is found to vanish, and in general we can write 
\begin{align}
& \check{\alpha}_{\text{CS}}^{il}=\delta^{il}\frac{ie^{2}}{3\hbar c}\epsilon^{cab}\sum_{vv'v''}x_{vv''}^{c}x_{v''v'}^{a}x_{v'v}^{b}.\label{eq:alphabar_CS}
\end{align}

In the presence of uniform fields $\boldsymbol{E}$ and $\boldsymbol{B}$, and in the frozen-ion approximation, each magnetoelectric tensor component $\check{\alpha}^{il}$ of a molecule can then be written (\ref{eq:alphabar_decomp}) as the sum of a Chern-Simons-like term (\ref{eq:alphabar_CS}) and a cross-gap-like term (\ref{eq:alphabar_CG}). As in a solid, the Chern-Simons-like contribution depends only on the occupied orbitals, and describes an isotropic modification, regardless of how complicated might be the structure of the molecule. Note also that in a situation where all relevant initially occupied (unoccupied) orbitals are degenerate, $E_{c}=E_{c'}$ ($E_{v}=E_{v'}),$ the cross-gap-like term vanishes, in line with the vanishing of the cross-gap term in a solid when all relevant initially occupied (unoccupied) bands are degenerate \cite{Essin2010}.

\subsubsection{The molecular crystal limit}

We can now construct the ``molecular crystal limit,'' in which we consider a periodic array of molecules where the orbitals associated with a molecule at a given lattice site share no common support with those of molecules associated with other lattice sites; again, we take the electric and magnetic fields that modify the electronic properties of the molecules to be the macroscopic Maxwell fields. Since the first-order modifications to the electric and magnetic moments associated with each molecule are given by (\ref{eq:atom_moments}), using the expressions (\ref{PM}) for $\boldsymbol{P}$ and $\boldsymbol{M}$, together with the defining equation (\ref{OMP}) for the orbital magnetoelectric polarizability tensor, from (\ref{eq:alphabar_decomp},\ref{eq:alphabar_CG},\ref{eq:alphabar_CS}) we have 
\begin{align}
\mathring{\alpha}^{il}=\mathring{\alpha}_{\text{G}}^{il}+\mathring{\alpha}_{\text{CS}}^{il},\label{eq:alphadecomp_MC}
\end{align}
where simply
\begin{align}
\mathring{\alpha}_{\text{G}}^{il}=\frac{\check{\alpha}^{il}_{\text{G}}}{\Omega_{uc}}\text{\space , \space} \mathring{\alpha}_{\text{CS}}^{il}=\frac{\check{\alpha}^{il}_{\text{CS}}}{\Omega_{uc}},\label{eq:alphaCS_MC}
\end{align}
with $\check{\alpha}^{il}_{\text{G}}$ and $\check{\alpha}^{il}_{\text{CS}}$ given by (\ref{eq:alphabar_CG}); here the circle accent identifies that the molecular crystal limit has been taken. 

Alternately, rather than building up the molecular crystal limit by assembling a collection of molecules, we can imagine starting with a full bandstructure calculation and taking the limit where the Wannier functions associated with each lattice site have no common support with the Wannier functions associated with a different lattice site. We also take the ELWFs to be eigenfunctions of the unperturbed Hamiltonian, (\ref{H}), which requires taking $E_{n\boldsymbol{k}}\rightarrow E_{n}$. These conditions lead to simplifications in the general expressions (\ref{alphaCS},\ref{alphaG}), and when they are employed the result should reproduce the molecular crystal limit (\ref{eq:alphadecomp_MC},\ref{eq:alphaCS_MC}). A first simplification is that, since the bands are flat, $\partial_a(E_{m\boldsymbol{k}}+E_{n\boldsymbol{k}})\rightarrow0$. Further, taking the orbitals introduced in (\ref{eq:orbital_expand}) to be the ELWFs $W_{v\boldsymbol{0}}(\boldsymbol{x})$ and $W_{c\boldsymbol{0}}(\boldsymbol{x})$, the flat bands can be identified by taking $U_{n\alpha}(\boldsymbol{k})=\delta_{n\alpha}$ in (\ref{WF}) relating $\ket{\alpha\boldsymbol{R}}$ with $\ket{\psi_{n\boldsymbol{k}}}$. Hence the Hermitian matrices $\mathcal{W}^{a}$ (\ref{W}) vanish, and we need not distinguish between the connections (\ref{connectionWannier},\ref{eq:Bloch_connection}); the lack of common support for orbitals associated with different lattice sites, together with (\ref{firstMoment}), also implies that $\xi_{cv}^{a}$, etc., are independent of $\boldsymbol{k}$, and from (\ref{firstMoment},\ref{connectionWannier},\ref{eq:Bloch_connection}) we take
\begin{align*}
\xi_{cv}^{a}(\boldsymbol{k})=i\left(c\boldsymbol{k}|\partial_av\boldsymbol{k}\right)\rightarrow x_{cv}^{a},
\end{align*}
where we have used (\ref{eq:xcv}) and 
\begin{align*}
\Omega_{uc}\int_{\text{BZ}}\frac{d\boldsymbol{k}}{(2\pi)^{3}}=1.
\end{align*}
Applying these molecular crystal conditions to the general expressions (\ref{alphaCS},\ref{alphaG}), we indeed recover (\ref{eq:alphaCS_MC}), as expected. Note that in this limit there is no gauge dependence in $\mathring{\alpha}_{\text{CS}}^{il}$, since the bands are isolated, and as well both $\mathring{\alpha}_{\text{CS}}^{il}$ and $\mathring{\alpha}_{\text{G}}^{il}$ arise solely from dynamical contributions. For while a calculation of the magnetic susceptibility \textit{would} involve a compositional contribution due to the modification of the diamagnetic term, even in the molecular crystal limit (see the discussion around (\ref{eq:diamagnetic})), the contributions to the magnetoelectric effect in that limit are purely dynamical; were the calculation for a single molecule here done in terms of the Green function strategy used for a crystal, the modification would result from changes in $\eta_{\alpha\boldsymbol{0};\beta\boldsymbol{0}}$. Of course, in the molecular crystal limit all contributions are what we have called atomic-like rather than itinerant.

\subsection{Microscopic origin of $\alpha_{\text{G}}^{il}$ and $\alpha_{\text{CS}}^{il}$}

While the qualitative features of the two contributions to the OMP tensor have been discussed earlier \cite{Essin2010}, the microscopic nature of the approach implemented here can provide further insight into the character of the cross-gap and Chern-Simons contributions.

Since both of these contributions are nonvanishing in the molecular crystal limit, neither can simply be understood as entirely a consequence of the delocalized nature of Bloch electrons. For the Chern-Simons contribution, this is in agreement with earlier work \cite{Essin2010} where a particular model for a molecule at a lattice site was constructed that exhibits a Chern-Simons-like modification; our expression (\ref{eq:alphabar_CS}) for the Chern-Simons-like modification of an arbitrary molecule generalizes that result. However, neither contribution can be understood as a purely ``localized molecule-like contribution'' either, because the full expressions for $\alpha_{\text{CS}}^{il}$ (\ref{alphaCS}) and $\alpha_{\text{G}}^{il}$ (\ref{alphaG}) contain terms that vanish in the molecular crystal limit; this is again in agreement with earlier arguments \cite{Essin2010}.

When moving from the molecular crystal limit of the Chern-Simons and cross-gap tensors, where the only contributions are atomic-like, to the full crystal expressions, both acquire itinerant contributions. In addition, while the cross-gap tensor is purely dynamical in nature, both in the molecular crystal limit and more generally, the Chern-Simons tensor acquires a compositional modification when moving from the molecular crystal limit to the general expression for a crystal. This suggests that perhaps it is through this compositional modification that a bulk interpretation of the discrete ambiguity associated with the Chern-Simons tensor can be constructed. We plan to explore this conjecture in a future publication.

\section{Conclusion}
\label{Sect:Conclusion}

We have implemented a previously developed \cite{Mahon2019} microscopic theory of polarization and magnetization to study modifications of the orbital electronic properties of a class of insulators due to uniform dc electric and magnetic fields, at zero temperature. To first-order in the Maxwell fields, the free charge and current densities vanish \cite{Mahon2019}; in future work we plan to extend this type of investigation to metallic systems for which these quantities would be relevant. Thus the perturbative modifications of both the electronic charge and current density expectation values due to the Maxwell fields can be found directly from the corresponding modifications of the microscopic polarization and magnetization fields. Associated with the dipole moment of the site microscopic polarization (magnetization) field is a macroscopic polarization (magnetization), for which we extract various tensors relating it to the Maxwell fields.

A quantity central in any calculation implementing this microscopic formalism is the single-particle density matrix. We began by re-expressing the equation governing its dynamics to include an arbitrary lattice site, $\boldsymbol{R}_{\text{a}}$, which is to be used as a reference site for the electric and magnetic fields when calculating modifications to site quantities due to those fields. This strategy will be useful in future work, where we plan to take into account the spatial variations of the electric and magnetic fields. However, in the calculation reported here we have restricted ourselves to the limit of uniform dc electric and magnetic fields. We began by reproducing the usual electric susceptibility \cite{Aversa}, and then found the orbital magnetoelectric polarizability tensor. Generally, the OMP tensor is written as a combination of the isotropic, Chern-Simons contribution and the cross-gap contribution; this microscopic theory reproduces the usual result \cite{Qi2008,Vanderbilt2009,Essin2010,Malashevich2010}. 

In the course of the perturbative analysis it became evident that there are generally two distinct types of modifications that contribute to the tensors that relate site quantities to the Maxwell fields. The first type arose from modifications to the single-particle density matrix due to the electromagnetic field, and were termed ``dynamical.'' The other type arose from modifications to the diagonal elements of site quantity matrices, and were termed ``compositional.'' The electric susceptibility was found to arise from only a dynamical modification. In our analysis of the magnetoelectric effect, we found three terms with distinct microscopic origins combine to form the OMP tensor: a dynamical modification to the atomic-like contribution, and dynamical and compositional modifications to the itinerant contribution. We found the Chern-Simons contribution to arise from a combination of parts of the atomic-like dynamical modification, and the itinerant compositional modification. The cross-gap contribution was found to arise from the remainder of the atomic-like dynamical modification, and the itinerant dynamical modification. We have also compared our expressions with those that arise in the molecular crystal limit, a model in which a periodic array of isolated molecules is considered.

As is well known, the linearly induced macroscopic bulk charge and current densities that result from uniform dc electric and magnetic fields are gauge invariant, but at a surface where the bulk arguments underlying this result are not valid, gauge dependent currents emerge. The ambiguity associated with such surface currents can be studied in detail via the microscopic theory that underpins the approach taken here; we plan to consider this in future work. We also intend to extend the calculations presented in this paper to take into account the frequency dependence of response tensors in general. This will lead to a description of frequency-dependent magnetoelectric effects, such as optical activity, for which the breaking of both spatial-inversion symmetry and time-reversal symmetry in the unperturbed material system is not required. A microscopic understanding of the mechanisms giving rise to such effects is now accessible via this formalism.

\section{Acknowledgments}

We thank Ivo Souza, Rodrigo A. Muniz, Sylvia Swiecicki, and Julen Ibanez-Azpiroz for insightful discussions. This work was supported by the Natural Sciences and Engineering Research Council of Canada (NSERC). P.T.M. acknowledges a PGS-D scholarship from NSERC.

\section{Appendices}
\appendix

\begin{widetext}

\section{Introducing arbitrary lattice sites}
\label{AppendixA}

Here we work out an expression for 
\begin{align} e^{i\Delta(\boldsymbol{R},\boldsymbol{R}'',\boldsymbol{R}';t)}\bar{H}_{\alpha\boldsymbol{R};\eta\boldsymbol{R}''}(t)=\mathscr{A}+\mathscr{B}+\mathscr{C}+\mathscr{D} \label{eq:workout_Hbar}
\end{align}
(see Eq.~(35) of \cite{Mahon2019} for the definition of $\bar{H}_{\alpha\boldsymbol{R};\eta\boldsymbol{R}''}(t)$), where
\begin{align*}
& \mathscr{A}=\frac{1}{2}e^{i\Delta(\boldsymbol{R},\boldsymbol{R}'',\boldsymbol{R}';t)}\int\chi_{\alpha\boldsymbol{R}}^{*}(\boldsymbol{x},t)e^{i\Delta(\boldsymbol{R},\boldsymbol{x},\boldsymbol{R}'';t)}H_{0}(\boldsymbol{x},\boldsymbol{\mathfrak{p}}(\boldsymbol{x},\boldsymbol{R}'';t))\chi_{\eta\boldsymbol{R}''}(\boldsymbol{x},t)d\boldsymbol{x},\\
& \mathscr{B}=\frac{1}{2}e^{i\Delta(\boldsymbol{R},\boldsymbol{R}'',\boldsymbol{R}';t)}\int\left(H_{0}(\boldsymbol{x},\boldsymbol{\mathfrak{p}}(\boldsymbol{x},\boldsymbol{R};t))\chi_{\alpha\boldsymbol{R}}(\boldsymbol{x},t)\right)^{*}\chi_{\eta\boldsymbol{R}''}(\boldsymbol{x},t)e^{i\Delta(\boldsymbol{R},\boldsymbol{x},\boldsymbol{R}'';t)}d\boldsymbol{x},\\
& \mathscr{C}=-\frac{e}{2}e^{i\Delta(\boldsymbol{R},\boldsymbol{R}'',\boldsymbol{R}';t)}\int e^{i\Delta(\boldsymbol{R},\boldsymbol{x},\boldsymbol{R}'';t)}\chi_{\alpha\boldsymbol{R}}^{*}(\boldsymbol{x},t)\left(\Omega_{\boldsymbol{R}''}^{0}(\boldsymbol{x},t)+\Omega_{\boldsymbol{R}}^{0}(\boldsymbol{x},t)\right)\chi_{\eta\boldsymbol{R}''}(\boldsymbol{x},t)d\boldsymbol{x},\\
& \mathscr{D}=-\frac{i\hbar}{2}e^{i\Delta(\boldsymbol{R},\boldsymbol{R}'',\boldsymbol{R}';t)}\int e^{i\Delta(\boldsymbol{R},\boldsymbol{x},\boldsymbol{R}'';t)}\left(\chi_{\alpha\boldsymbol{R}}^{*}(\boldsymbol{x},t)\frac{\partial\chi_{\eta\boldsymbol{R}''}(\boldsymbol{x},t)}{\partial t}-\frac{\partial\chi_{\alpha\boldsymbol{R}}^{*}(\boldsymbol{x},t)}{\partial t}\chi_{\eta\boldsymbol{R}''}(\boldsymbol{x},t)\right)d\boldsymbol{x}.
\end{align*}
Looking at the first of these terms, we note that 
\begin{align*} \boldsymbol{\mathfrak{p}}(\boldsymbol{x},\boldsymbol{R}'';t)=e^{-i\Delta(\boldsymbol{R}_{\text{a}},\boldsymbol{x},\boldsymbol{R}'';t)}\boldsymbol{\mathfrak{p}}(\boldsymbol{x},\boldsymbol{R}_{\text{a}};t)e^{i\Delta(\boldsymbol{R}_{\text{a}},\boldsymbol{x},\boldsymbol{R}'';t)},
\end{align*}
so 
\begin{align*} \mathscr{A}=\frac{1}{2}e^{i\Delta(\boldsymbol{R},\boldsymbol{R}'',\boldsymbol{R}';t)}\int\chi_{\alpha\boldsymbol{R}}^{*}(\boldsymbol{x},t)e^{i\Delta(\boldsymbol{R},\boldsymbol{x},\boldsymbol{R}'';t)}e^{-i\Delta(\boldsymbol{R}_{\text{a}},\boldsymbol{x},\boldsymbol{R}'';t)}H_{0}(\boldsymbol{x},\boldsymbol{\mathfrak{p}}(\boldsymbol{x},\boldsymbol{R}_{\text{a}};t))e^{i\Delta(\boldsymbol{R}_{\text{a}},\boldsymbol{x},\boldsymbol{R}'';t)}\chi_{\eta\boldsymbol{R}''}(\boldsymbol{x},t)d\boldsymbol{x}.
\end{align*}
Now 
\begin{align*}
\Delta(\boldsymbol{R},\boldsymbol{R}'',\boldsymbol{R}';t)+\Delta(\boldsymbol{R},\boldsymbol{x},\boldsymbol{R}'';t)-\Delta(\boldsymbol{R}_{\text{a}},\boldsymbol{x},\boldsymbol{R}'';t) &=\Delta(\boldsymbol{R},\boldsymbol{x},\boldsymbol{R}_{\text{a}},\boldsymbol{R}'',\boldsymbol{R}';t)\\
&=\Delta(\boldsymbol{R},\boldsymbol{x},\boldsymbol{R}_{\text{a}};t)+\Delta(\boldsymbol{R},\boldsymbol{R}_{\text{a}},\boldsymbol{R}'',\boldsymbol{R}';t),
\end{align*}
and so 
\begin{align*} \mathscr{A}=\frac{1}{2}e^{i\Delta(\boldsymbol{R},\boldsymbol{R}_{\text{a}},\boldsymbol{R}'',\boldsymbol{R}';t)}\int\chi_{\alpha\boldsymbol{R}}^{*}(\boldsymbol{x},t)e^{i\Delta(\boldsymbol{R},\boldsymbol{x},\boldsymbol{R}_{\text{a}};t)}H_{0}(\boldsymbol{x},\boldsymbol{\mathfrak{p}}(\boldsymbol{x},\boldsymbol{R}_{\text{a}};t))e^{i\Delta(\boldsymbol{R}_{\text{a}},\boldsymbol{x},\boldsymbol{R}'';t)}\chi_{\eta\boldsymbol{R}''}(\boldsymbol{x},t)d\boldsymbol{x}.
\end{align*}
Similarly, since 
\begin{align*} \boldsymbol{\mathfrak{p}}(\boldsymbol{x},\boldsymbol{R};t)=e^{-i\Delta(\boldsymbol{R}_{\text{a}},\boldsymbol{x},\boldsymbol{R};t)}\boldsymbol{\mathfrak{p}}(\boldsymbol{x},\boldsymbol{R}_{\text{a}};t)e^{i\Delta(\boldsymbol{R}_{\text{a}},\boldsymbol{x},\boldsymbol{R};t)},
\end{align*}
we have 
\begin{align*}
\mathscr{B}&=\frac{1}{2}e^{i\Delta(\boldsymbol{R},\boldsymbol{R}'',\boldsymbol{R}';t)}\int\left(e^{-i\Delta(\boldsymbol{R}_{\text{a}},\boldsymbol{x},\boldsymbol{R};t)}H_{0}(\boldsymbol{x},\boldsymbol{\mathfrak{p}}(\boldsymbol{x},\boldsymbol{R}_{\text{a}};t))e^{i\Delta(\boldsymbol{R}_{\text{a}},\boldsymbol{x},\boldsymbol{R};t)}\chi_{\alpha\boldsymbol{R}}(\boldsymbol{x},t)\right)^{*}\chi_{\eta\boldsymbol{R}''}(\boldsymbol{x},t)e^{i\Delta(\boldsymbol{R},\boldsymbol{x},\boldsymbol{R}'';t)}d\boldsymbol{x}\\
& =\frac{1}{2}e^{i\Delta(\boldsymbol{R},\boldsymbol{R}'',\boldsymbol{R}';t)}\int\left(H_{0}(\boldsymbol{x},\boldsymbol{\mathfrak{p}}(\boldsymbol{x},\boldsymbol{R}_{\text{a}};t))e^{i\Delta(\boldsymbol{R}_{\text{a}},\boldsymbol{x},\boldsymbol{R};t)}\chi_{\alpha\boldsymbol{R}}(\boldsymbol{x},t)\right)^{*}\chi_{\eta\boldsymbol{R}''}(\boldsymbol{x},t)e^{i\Delta(\boldsymbol{R},\boldsymbol{x},\boldsymbol{R}'';t)}e^{i\Delta(\boldsymbol{R}_{\text{a}},\boldsymbol{x},\boldsymbol{R};t)}d\boldsymbol{x}.
\end{align*}
Now 
\begin{align*}
\Delta(\boldsymbol{R},\boldsymbol{R}'',\boldsymbol{R}';t)+\Delta(\boldsymbol{R},\boldsymbol{x},\boldsymbol{R}'';t)+\Delta(\boldsymbol{R}_{\text{a}},\boldsymbol{x},\boldsymbol{R};t)
&=\Delta(\boldsymbol{R},\boldsymbol{R}_{\text{a}},\boldsymbol{x},\boldsymbol{R}'',\boldsymbol{R}';t)\\
&=\Delta(\boldsymbol{R}_{\text{a}},\boldsymbol{x},\boldsymbol{R}'';t)+\Delta(\boldsymbol{R},\boldsymbol{R}_{\text{a}},\boldsymbol{R}'',\boldsymbol{R}';t),
\end{align*}
so
\begin{align*}
\mathscr{B}&=\frac{1}{2}e^{i\Delta(\boldsymbol{R},\boldsymbol{R}_{\text{a}},\boldsymbol{R}'',\boldsymbol{R}';t)}\int\left(H_{0}(\boldsymbol{x},\boldsymbol{\mathfrak{p}}(\boldsymbol{x},\boldsymbol{R}_{\text{a}};t))e^{i\Delta(\boldsymbol{R}_{\text{a}},\boldsymbol{x},\boldsymbol{R};t)}\chi_{\alpha\boldsymbol{R}}(\boldsymbol{x},t)\right)^{*}\chi_{\eta\boldsymbol{R}''}(\boldsymbol{x},t)e^{i\Delta(\boldsymbol{R}_{\text{a}},\boldsymbol{x},\boldsymbol{R}'';t)}d\boldsymbol{x}\\
&=\frac{1}{2}e^{i\Delta(\boldsymbol{R},\boldsymbol{R}_{\text{a}},\boldsymbol{R}'',\boldsymbol{R}';t)}\int\left(H_{0}^{*}(\boldsymbol{x},\boldsymbol{\mathfrak{p}}(\boldsymbol{x},\boldsymbol{R}_{\text{a}};t))\left(e^{i\Delta(\boldsymbol{R},\boldsymbol{x},\boldsymbol{R}_{\text{a}};t)}\chi_{\alpha\boldsymbol{R}}^{*}(\boldsymbol{x},t)\right)\right)e^{i\Delta(\boldsymbol{R}_{\text{a}},\boldsymbol{x},\boldsymbol{R}'';t)}\chi_{\eta\boldsymbol{R}''}(\boldsymbol{x},t)d\boldsymbol{x}.
\end{align*}
And then 
\begin{align*}
\mathscr{A}+\mathscr{B}&=\frac{1}{2}e^{i\Delta(\boldsymbol{R},\boldsymbol{R}_{\text{a}},\boldsymbol{R}'',\boldsymbol{R}';t)}\int\chi_{\alpha\boldsymbol{R}}^{*}(\boldsymbol{x},t)e^{i\Delta(\boldsymbol{R},\boldsymbol{x},\boldsymbol{R}_{\text{a}};t)}H_{0}(\boldsymbol{x},\boldsymbol{\mathfrak{p}}(\boldsymbol{x},\boldsymbol{R}_{\text{a}};t))e^{i\Delta(\boldsymbol{R}_{\text{a}},\boldsymbol{x},\boldsymbol{R}'';t)}\chi_{\eta\boldsymbol{R}''}(\boldsymbol{x},t)d\boldsymbol{x}\\
& +\frac{1}{2}e^{i\Delta(\boldsymbol{R},\boldsymbol{R}_{\text{a}},\boldsymbol{R}'',\boldsymbol{R}';t)}\int\left(H_{0}^{*}(\boldsymbol{x},\boldsymbol{\mathfrak{p}}(\boldsymbol{x},\boldsymbol{R}_{\text{a}};t))\left(e^{i\Delta(\boldsymbol{R},\boldsymbol{x},\boldsymbol{R}_{\text{a}};t)}\chi_{\alpha\boldsymbol{R}}^{*}(\boldsymbol{x},t)\right)\right)e^{i\Delta(\boldsymbol{R}_{\text{a}},\boldsymbol{x},\boldsymbol{R}'';t)}\chi_{\eta\boldsymbol{R}''}(\boldsymbol{x},t)d\boldsymbol{x}\\
& =e^{i\Delta(\boldsymbol{R},\boldsymbol{R}_{\text{a}},\boldsymbol{R}'',\boldsymbol{R}';t)}\int\chi_{\alpha\boldsymbol{R}}^{*}(\boldsymbol{x},t)e^{i\Delta(\boldsymbol{R},\boldsymbol{x},\boldsymbol{R}_{\text{a}};t)}H_{0}(\boldsymbol{x},\boldsymbol{\mathfrak{p}}(\boldsymbol{x},\boldsymbol{R}_{\text{a}};t))e^{i\Delta(\boldsymbol{R}_{\text{a}},\boldsymbol{x},\boldsymbol{R}'';t)}\chi_{\eta\boldsymbol{R}''}(\boldsymbol{x},t)d\boldsymbol{x}.
\end{align*}
The last form is not ``explicitly Hermitian,'' but it will be convenient.
Next, since 
\begin{align} \Delta(\boldsymbol{R},\boldsymbol{R}_{\text{a}},\boldsymbol{R}'',\boldsymbol{R}';t)+\Delta(\boldsymbol{R},\boldsymbol{x},\boldsymbol{R}_{\text{a}};t)+\Delta(\boldsymbol{R}_{\text{a}},\boldsymbol{x},\boldsymbol{R}'';t)=\Delta(\boldsymbol{R},\boldsymbol{R}'',\boldsymbol{R}';t)+\Delta(\boldsymbol{R},\boldsymbol{x},\boldsymbol{R}'';t),\label{eq:delta_identity} 
\end{align}
we can write 
\begin{align*}
\mathscr{A}+\mathscr{B}+\mathscr{C}&=
e^{i\Delta(\boldsymbol{R},\boldsymbol{R}_{\text{a}},\boldsymbol{R}'',\boldsymbol{R}';t)}\int\chi_{\alpha\boldsymbol{R}}^{*}(\boldsymbol{x},t)e^{i\Delta(\boldsymbol{R},\boldsymbol{x},\boldsymbol{R}_{\text{a}};t)}\mathcal{H}_{\boldsymbol{R}_{\text{a}}}(\boldsymbol{x},t)e^{i\Delta(\boldsymbol{R}_{\text{a}},\boldsymbol{x},\boldsymbol{R}'';t)}\chi_{\eta\boldsymbol{R}''}(\boldsymbol{x},t)d\boldsymbol{x}\\
&+e^{i\Delta(\boldsymbol{R},\boldsymbol{R}'',\boldsymbol{R}';t)}\int e^{i\Delta(\boldsymbol{R},\boldsymbol{x},\boldsymbol{R}'';t)}\chi_{\alpha\boldsymbol{R}}^{*}(\boldsymbol{x},t)\left(e\Omega_{\boldsymbol{R}_{\text{a}}}^{0}(\boldsymbol{x},t)-\frac{e}{2}\Omega_{\boldsymbol{R}''}^{0}(\boldsymbol{x},t)-\frac{e}{2}\Omega_{\boldsymbol{R}}^{0}(\boldsymbol{x},t)\right)\chi_{\eta\boldsymbol{R}''}(\boldsymbol{x},t)d\boldsymbol{x},
\end{align*}
where we have used (\ref{Hcal}). Now 
\begin{align*}
 e\Omega_{\boldsymbol{R}_{\text{a}}}^{0}(\boldsymbol{x},t)-\frac{e}{2}\Omega_{\boldsymbol{R}''}^{0}(\boldsymbol{x},t)-\frac{e}{2}\Omega_{\boldsymbol{R}}^{0}(\boldsymbol{x},t)
& =\frac{e}{2}\left(\Omega_{\boldsymbol{R}_{\text{a}}}^{0}(\boldsymbol{x},t)-\Omega_{\boldsymbol{R}''}^{0}(\boldsymbol{x},t)\right)+\frac{e}{2}\left(\Omega_{\boldsymbol{R}_{\text{a}}}^{0}(\boldsymbol{x},t)-\Omega_{\boldsymbol{R}}^{0}(\boldsymbol{x},t)\right)\\
& =\frac{e}{2}\left(\Omega_{\boldsymbol{R}_{\text{a}}}^{0}(\boldsymbol{x},t)+\Omega_{\boldsymbol{x}}^{0}(\boldsymbol{R}'',t)\right)+\frac{e}{2}\left(\Omega_{\boldsymbol{R}_{\text{a}}}^{0}(\boldsymbol{x},t)+\Omega_{\boldsymbol{x}}^{0}(\boldsymbol{R},t)\right)\\
& =\frac{e}{2}\left(\Omega_{\boldsymbol{R}_{\text{a}}}^{0}(\boldsymbol{x},t)+\Omega_{\boldsymbol{x}}^{0}(\boldsymbol{R}'',t)+\Omega_{\boldsymbol{R}''}^{0}(\boldsymbol{R}_{\text{a}},t)\right)\\
& +\frac{e}{2}\left(\Omega_{\boldsymbol{R}_{\text{a}}}^{0}(\boldsymbol{x},t)+\Omega_{\boldsymbol{x}}^{0}(\boldsymbol{R},t)+\Omega_{\boldsymbol{R}}^{0}(\boldsymbol{R}_{\text{a}},t)\right)-\frac{e}{2}\left(\Omega_{\boldsymbol{R}''}^{0}(\boldsymbol{R}_{\text{a}},t)+\Omega_{\boldsymbol{R}}^{0}(\boldsymbol{R}_{\text{a}},t)\right)\\
& =\frac{\hbar}{2}\frac{\partial\Delta(\boldsymbol{R}_{\text{a}},\boldsymbol{x},\boldsymbol{R}'';t)}{\partial t}+\frac{\hbar}{2}\frac{\partial\Delta(\boldsymbol{R}_{\text{a}},\boldsymbol{x},\boldsymbol{R};t)}{\partial t}-\frac{e}{2}\left(\Omega_{\boldsymbol{R}''}^{0}(\boldsymbol{R}_{\text{a}},t)+\Omega_{\boldsymbol{R}}^{0}(\boldsymbol{R}_{\text{a}},t)\right),
\end{align*}
and recalling (\ref{eq:delta_identity}) we have 
\begin{align*}
\mathscr{A}+\mathscr{B}+\mathscr{C}&=
e^{i\Delta(\boldsymbol{R},\boldsymbol{R}_{\text{a}},\boldsymbol{R}'',\boldsymbol{R}';t)}\\
& \times\int\chi_{\alpha\boldsymbol{R}}^{*}(\boldsymbol{x},t)e^{i\Delta(\boldsymbol{R},\boldsymbol{x},\boldsymbol{R}_{\text{a}};t)}\left(\mathcal{H}_{\boldsymbol{R}_{\text{a}}}(\boldsymbol{x},t)+\frac{\hbar}{2}\frac{\partial\Delta(\boldsymbol{R}_{\text{a}},\boldsymbol{x},\boldsymbol{R}'';t)}{\partial t}+\frac{\hbar}{2}\frac{\partial\Delta(\boldsymbol{R}_{\text{a}},\boldsymbol{x},\boldsymbol{R};t)}{\partial t}\right)e^{i\Delta(\boldsymbol{R}_{\text{a}},\boldsymbol{x},\boldsymbol{R}'';t)}\chi_{\eta\boldsymbol{R}''}(\boldsymbol{x},t)d\boldsymbol{x}\\
& -\frac{e}{2}e^{i\Delta(\boldsymbol{R},\boldsymbol{R}'',\boldsymbol{R}';t)}\left(\Omega_{\boldsymbol{R}''}^{0}(\boldsymbol{R}_{\text{a}},t)+\Omega_{\boldsymbol{R}}^{0}(\boldsymbol{R}_{\text{a}},t)\right)\int e^{i\Delta(\boldsymbol{R},\boldsymbol{x},\boldsymbol{R}'';t)}\chi_{\alpha\boldsymbol{R}}^{*}(\boldsymbol{x},t)\chi_{\eta\boldsymbol{R}''}(\boldsymbol{x},t)d\boldsymbol{x}.
\end{align*}
As the function $\chi_{\alpha\boldsymbol{R}}(\boldsymbol{x},t)$ satisfies the modified orthogonality relation \cite{Mahon2019}
\begin{align*}
\int e^{i\Delta(\boldsymbol{R},\boldsymbol{x},\boldsymbol{R}'';t)}\chi_{\alpha\boldsymbol{R}}^{*}(\boldsymbol{x},t)\chi_{\eta\boldsymbol{R}''}(\boldsymbol{x},t)d\boldsymbol{x}=\delta_{\alpha\eta}\delta_{\boldsymbol{R}\boldsymbol{R}''},
\end{align*}
we find
\begin{align*}
\mathscr{A}+\mathscr{B}+\mathscr{C}&=e^{i\Delta(\boldsymbol{R},\boldsymbol{R}_{\text{a}},\boldsymbol{R}'',\boldsymbol{R}';t)}\\
& \times\int\chi_{\alpha\boldsymbol{R}}^{*}(\boldsymbol{x},t)e^{i\Delta(\boldsymbol{R},\boldsymbol{x},\boldsymbol{R}_{\text{a}};t)}\left(\mathcal{H}_{\boldsymbol{R}_{\text{a}}}(\boldsymbol{x},t)+\frac{\hbar}{2}\frac{\partial\Delta(\boldsymbol{R}_{\text{a}},\boldsymbol{x},\boldsymbol{R}'';t)}{\partial t}+\frac{\hbar}{2}\frac{\partial\Delta(\boldsymbol{R}_{\text{a}},\boldsymbol{x},\boldsymbol{R};t)}{\partial t}\right)e^{i\Delta(\boldsymbol{R}_{\text{a}},\boldsymbol{x},\boldsymbol{R}'';t)}\chi_{\eta\boldsymbol{R}''}(\boldsymbol{x},t)d\boldsymbol{x}\\
& -e\Omega_{\boldsymbol{R}}^{0}(\boldsymbol{R}_{\text{a}},t)\delta_{\alpha\eta}\delta_{\boldsymbol{R}\boldsymbol{R}''},
\end{align*}
and so from (\ref{eq:workout_Hbar}) we have (\ref{RaRefSite}), where we have defined $\bar{H}_{\alpha\boldsymbol{R};\eta\boldsymbol{R}''}(\boldsymbol{R}_{\text{a}},t)$ as in (\ref{barHelements}). We also require
\begin{align*}
e^{i\Delta(\boldsymbol{R},\boldsymbol{R}'',\boldsymbol{R}';t)}\bar{H}_{\eta\boldsymbol{R}'';\beta\boldsymbol{R}'}(t)&=e^{i\Delta(\boldsymbol{R}'',\boldsymbol{R}',\boldsymbol{R};t)}\bar{H}_{\eta\boldsymbol{R}'';\beta\boldsymbol{R}'}(t)\\
& =e^{i\Delta(\boldsymbol{R}'',\boldsymbol{R}_{\text{a}},\boldsymbol{R}',\boldsymbol{R};t)}\bar{H}_{\eta\boldsymbol{R}'';\beta\boldsymbol{R}'}(\boldsymbol{R}_{\text{a}},t)-e\Omega_{\boldsymbol{R}'}^{0}(\boldsymbol{R}_{\text{a}},t)\delta_{\eta\beta}\delta_{\boldsymbol{R}'\boldsymbol{R}''}\\
& =e^{i\Delta(\boldsymbol{R},\boldsymbol{R}'',\boldsymbol{R}_{\text{a}},\boldsymbol{R}';t)}\bar{H}_{\eta\boldsymbol{R}'';\beta\boldsymbol{R}'}(\boldsymbol{R}_{\text{a}},t)-e\Omega_{\boldsymbol{R}'}^{0}(\boldsymbol{R}_{\text{a}},t)\delta_{\eta\beta}\delta_{\boldsymbol{R}'\boldsymbol{R}''},
\end{align*}
where we have used (\ref{barHelements}). Since the lattice site $\boldsymbol{R}_{\text{a}}$
is arbitrary, we can as well write (\ref{RbRefSite}) for any lattice site $\boldsymbol{R}_{b}$. 
\end{widetext}

\section{Perturbative modifications of the single-particle density matrix}
\label{AppendixD}

Beginning with the equation of motion for the single-particle density matrix
\begin{align}
i\hbar\frac{\partial\eta_{\alpha\boldsymbol{R}'';\beta\boldsymbol{R}'}(t)}{\partial t}=\sum_{\mu\nu\boldsymbol{R}_{1}\boldsymbol{R}_{2}}\mathfrak{F}_{\alpha\boldsymbol{R}'';\beta\boldsymbol{R}'}^{\mu\boldsymbol{R}_{1};\nu\boldsymbol{R}_{2}}(t)\eta_{\mu\boldsymbol{R}_{1};\nu\boldsymbol{R}_{2}}(t),
\label{EDMeom}
\end{align}
then expanding all quantities in powers of the electromagnetic field, and then matching powers, at zeroth-order we find
\begin{align*}
&i\hbar\frac{\partial\eta^{(0)}_{\alpha\boldsymbol{R}'';\beta\boldsymbol{R}'}(t)}{\partial t}=\\
&\quad\sum_{\lambda\boldsymbol{R}_3}\Big({H}^{(0)}_{\alpha\boldsymbol{R}'';\lambda\boldsymbol{R}_3}\eta^{(0)}_{\lambda\boldsymbol{R}_3;\beta\boldsymbol{R}'}(t)-\eta^{(0)}_{\alpha\boldsymbol{R}'';\lambda\boldsymbol{R}_3}{H}^{(0)}_{\lambda\boldsymbol{R}_3;\beta\boldsymbol{R}'}\Big),\nonumber 
\end{align*}
where ${H}^{(0)}_{\alpha\boldsymbol{R}'';\lambda\boldsymbol{R}_3}$ is given by (\ref{unpertHelements}). Thus $\eta^{(0)}_{\alpha\boldsymbol{R}'';\beta\boldsymbol{R}'}(t)$ evolves as the unperturbed single-particle density matrix, and consequently
\begin{align*}
\eta^{(0)}_{\alpha\boldsymbol{R}'';\beta\boldsymbol{R}'}(t)=f_\alpha\delta_{\alpha\beta}\delta_{\boldsymbol{R}''\boldsymbol{R}'},
\end{align*}
as expected. 

From (\ref{EDMeom}) it is found that the first-order modification to the single-particle density matrix due to the electromagnetic field evolves according to
\begin{align}
&i\hbar\frac{\partial\eta^{(1)}_{\alpha\boldsymbol{R}'';\beta\boldsymbol{R}'}(t)}{\partial t}=\nonumber\\
&\quad\sum_{\lambda\boldsymbol{R}_3}\Big({H}^{(0)}_{\alpha\boldsymbol{R}'';\lambda\boldsymbol{R}_3}\eta^{(1)}_{\lambda\boldsymbol{R}_3;\beta\boldsymbol{R}'}(t)-\eta^{(1)}_{\alpha\boldsymbol{R}'';\lambda\boldsymbol{R}_3}{H}^{(0)}_{\lambda\boldsymbol{R}_3;\beta\boldsymbol{R}'}\Big)\nonumber\\
&\quad+\sum_{\mu\boldsymbol{R}_{1}}f_{\mu}\mathfrak{F}_{\alpha\boldsymbol{R}'';\beta\boldsymbol{R}'}^{\mu\boldsymbol{R}_{1};\mu\boldsymbol{R}_{1}(1)}(t), \label{EDMeom1}
\end{align}
the final term of which is found to be
\begin{align*}
\sum_{\mu\boldsymbol{R}_{1}}f_{\mu}\mathfrak{F}_{\alpha\boldsymbol{R}'';\beta\boldsymbol{R}'}^{\mu\boldsymbol{R}_{1};\mu\boldsymbol{R}_{1}(1)}(t)=f_{\beta\alpha}\bar{H}_{\alpha\boldsymbol{R}'';\beta\boldsymbol{R}'}^{(1)}(\boldsymbol{R}_{\text{a}},t).
\end{align*}
It is useful to define the intermediate quantity
\begin{align} \eta_{m\boldsymbol{k};n\boldsymbol{k}'}(t)=\sum_{\mu\nu\boldsymbol{R}_{1}\boldsymbol{R}_{2}}\braket{\psi_{m\boldsymbol{k}}}{\mu\boldsymbol{R}_{1}} \eta_{\mu\boldsymbol{R}_{1};\nu\boldsymbol{R}_{2}}(t)\braket{ \nu\boldsymbol{R}_{2}}{\psi_{n\boldsymbol{k}'}},\label{transf}
\end{align}
for which, from (\ref{EDMeom1}), we find
\begin{align*}
& i\hbar\frac{\partial\eta_{m\boldsymbol{k};n\boldsymbol{k}'}^{(1)}(t)}{\partial t}=\big(E_{m\boldsymbol{k}}-E_{n\boldsymbol{k}'}\big)\eta_{m\boldsymbol{k};n\boldsymbol{k}'}^{(1)}(t)\\
&\quad+f_{nm}\sum_{\mu\nu\boldsymbol{R}_{1}\boldsymbol{R}_{2}}\braket{\psi_{m\boldsymbol{k}}}{\mu\boldsymbol{R}_{1}}\bar{H}_{\mu\boldsymbol{R}_{1};\nu\boldsymbol{R}_{2}}^{(1)}(\boldsymbol{R}_{\text{a}},t)\braket{\nu\boldsymbol{R}_{2}}{\psi_{n\boldsymbol{k}'}}.
\end{align*}
Then, implementing the usual Fourier analysis via (\ref{Fouier}), we find
\begin{align*}
& \eta_{m\boldsymbol{k};n\boldsymbol{k}'}^{(1)}(\omega)=\\
&\quad-f_{nm}\sum_{\mu\nu\boldsymbol{R}_{1}\boldsymbol{R}_{2}}\frac{\braket{\psi_{m\boldsymbol{k}}}{\mu\boldsymbol{R}_{1}}\bar{H}_{\mu\boldsymbol{R}_{1};\nu\boldsymbol{R}_{2}}^{(1)}(\boldsymbol{R}_{\text{a}},\omega)\braket{\nu\boldsymbol{R}_{2}}{\psi_{n\boldsymbol{k}'}}}{E_{m\boldsymbol{k}}-E_{n\boldsymbol{k}'}-\hbar(\omega+i0^{+})},
\end{align*}
where $0^{+}$ entering in the denominator describes the ``turning on'' of the electromagnetic field at $t>-\infty$. Finally, using (\ref{barHelements}) and the inverse of (\ref{transf}), we find
\begin{widetext}
\begin{align}
\eta_{\alpha\boldsymbol{R}'';\beta\boldsymbol{R}'}^{(1)}(\omega)&= -\sum_{\mu\nu\boldsymbol{R}_{1}\boldsymbol{R}_{2}}\sum_{mn}f_{nm}\int_{\text{BZ}}d\boldsymbol{k}d\boldsymbol{k}'\frac{\braket{\alpha\boldsymbol{R}''}{\psi_{m\boldsymbol{k}}}\braket{\psi_{m\boldsymbol{k}}}{\mu\boldsymbol{R}_{1}}H_{\mu\boldsymbol{R}_{1};\nu\boldsymbol{R}_{2}}^{(1)}(\boldsymbol{R}_{\text{a}},\omega)\braket{\nu\boldsymbol{R}_{2}}{\psi_{n\boldsymbol{k}'}}\braket{\psi_{n\boldsymbol{k}'}}{\beta\boldsymbol{R}'}}{E_{m\boldsymbol{k}}-E_{n\boldsymbol{k}'}-\hbar(\omega+i0^{+})}\nonumber\\
&+\frac{i}{2}f_{\beta\alpha}\int W_{\alpha\boldsymbol{R}''}^{*}(\boldsymbol{x})\Big(\Delta(\boldsymbol{R}'',\boldsymbol{x},\boldsymbol{R}_{\text{a}};\omega)+\Delta(\boldsymbol{R}',\boldsymbol{x},\boldsymbol{R}_{\text{a}};\omega)\Big)W_{\beta\boldsymbol{R}'}(\boldsymbol{x})d\boldsymbol{x}, \label{EDMpt}
\end{align}
where $H_{\mu\boldsymbol{R}_{1};\nu\boldsymbol{R}_{2}}^{(1)}(\boldsymbol{R}_{\text{a}},\omega)$ is given by (\ref{Hcorr}). We now apply this result to the case of uniform dc fields and find the first-order modification to the single-particle density matrix due to a electric field, Eq.~(\ref{EDMe}). The second term of (\ref{EDMpt}) vanishes trivially, and for the first we make use of (\ref{Hcorr},\ref{Hcal1}) to find
\begin{align}
H_{\mu\boldsymbol{R}_{1};\nu\boldsymbol{R}_{2}}^{(E)}(\boldsymbol{R}_{\text{a}},\omega)=-eE^l\int W^*_{\mu\boldsymbol{R}_{1}}(\boldsymbol{x})\big(x^l-R^l_{\text{a}}\big)W_{\nu\boldsymbol{R}_{2}}(\boldsymbol{x})d\boldsymbol{x},
\end{align}
where $\boldsymbol{E}\equiv\boldsymbol{E}(\boldsymbol{R}_{\text{a}},\omega=0)$ for any $\boldsymbol{R}_{\text{a}}$, in this limit. Then, recalling (\ref{WF}), we find
\begin{align*}
\eta_{\alpha\boldsymbol{R}'';\beta\boldsymbol{R}'}^{(E)}&=eE^l\frac{\Omega^2_{uc}}{(2\pi)^{6}}\sum_{mn}f_{nm}\int_{\text{BZ}}d\boldsymbol{k}d\boldsymbol{k}'\frac{e^{i(\boldsymbol{k}\boldsymbol{\cdot}\boldsymbol{R}''-\boldsymbol{k}'\boldsymbol{\cdot}\boldsymbol{R}')}U^{\dagger}_{\alpha m}(\boldsymbol{k})U_{n\beta}(\boldsymbol{k}')}{E_{m\boldsymbol{k}}-E_{n\boldsymbol{k}'}}\\
&\quad\qquad\qquad\qquad\qquad\times\sum_{\mu\nu\boldsymbol{R}_{1}\boldsymbol{R}_{2}}e^{-i\boldsymbol{k}\boldsymbol{\cdot}\boldsymbol{R}_1}e^{i\boldsymbol{k}'\boldsymbol{\cdot}\boldsymbol{R}_2}U_{m\mu}(\boldsymbol{k})\left(\int W^*_{\mu\boldsymbol{R}_{1}-\boldsymbol{R}_2}(\boldsymbol{x})x^lW_{\nu\boldsymbol{0}}(\boldsymbol{x})d\boldsymbol{x}\right)U^\dagger_{\nu n}(\boldsymbol{k}')\nonumber\\
&=eE^l\frac{\Omega^3_{uc}}{(2\pi)^{9}}\sum_{mn}f_{nm}\int_{\text{BZ}}d\boldsymbol{k}d\boldsymbol{k}'d\boldsymbol{k}_1\frac{e^{i(\boldsymbol{k}\boldsymbol{\cdot}\boldsymbol{R}''-\boldsymbol{k}'\boldsymbol{\cdot}\boldsymbol{R}')}U^{\dagger}_{\alpha m}(\boldsymbol{k})U_{n\beta}(\boldsymbol{k}')}{E_{m\boldsymbol{k}}-E_{n\boldsymbol{k}'}}\\
&\quad\qquad\qquad\qquad\qquad\times\sum_{\mu\nu}U_{m\mu}(\boldsymbol{k})\tilde{\xi}^l_{\mu\nu}(\boldsymbol{k}_1)U^\dagger_{\nu n}(\boldsymbol{k}')\sum_{\boldsymbol{R}_{1}}e^{-i(\boldsymbol{k}-\boldsymbol{k_1})\boldsymbol{\cdot}\boldsymbol{R}_1}\sum_{\boldsymbol{R}_{2}}e^{i(\boldsymbol{k}'-\boldsymbol{k}_1)\boldsymbol{\cdot}\boldsymbol{R}_2}\nonumber\\
&=eE^l\frac{\Omega_{uc}}{(2\pi)^3}\sum_{mn}f_{nm}\int_{\text{BZ}}d\boldsymbol{k}\frac{e^{i\boldsymbol{k}\boldsymbol{\cdot}(\boldsymbol{R}''-\boldsymbol{R}')}U^\dagger_{\alpha m}(\boldsymbol{k})\xi^l_{mn}(\boldsymbol{k})U_{n\beta}(\boldsymbol{k})}{E_{m\boldsymbol{k}}-E_{n\boldsymbol{k}}},
\end{align*}
\end{widetext}
where we have used the identity
\begin{align}
\frac{\Omega_{uc}}{(2\pi)^{3}}\sum_{\boldsymbol{R}}e^{i(\boldsymbol{k}-\boldsymbol{k}')\boldsymbol{\cdot}\boldsymbol{R}}=\delta(\boldsymbol{k}-\boldsymbol{k}'),
\end{align}
as well as (\ref{firstMoment}) and (\ref{connection}), in going to the final expression. Notice that the approximation of an electric field that varies little on the scale of the lattice constant gives rise to a simplified form of (\ref{Hcorr}), which in turn allows $\eta_{\alpha\boldsymbol{R}'';\beta\boldsymbol{R}'}^{(E)}$ to be written as a single Brillouin zone integral. To derive the expression for $\eta_{\alpha\boldsymbol{R}'';\beta\boldsymbol{R}'}^{(B)}$ a similar procedure is followed; however, one must also use (\ref{pMatrixElements}). In this case, the approximation that the magnetic field varies little on the scale of the lattice constant allows $\eta_{\alpha\boldsymbol{R}'';\beta\boldsymbol{R}'}^{(B)}$ to be written as a single Brillouin zone integral.

\section{Nearly uniform electromagnetic fields}
\label{AppendixB} 

We now work out some of the general expressions for our relators and quantities dependent on them in the limit of nearly uniform electromagnetic fields. By this we mean that we keep the electric field and its first derivatives at the ``expansion point,'' but only the magnetic field at that point. We use a straight-line path; see \cite{Mahon2019}, where we find 
\begin{align}
s^{i}(\boldsymbol{w};\boldsymbol{x},\boldsymbol{y}) & =\int_{0}^{1}(x^{i}-y^{i})\delta(\boldsymbol{w}-\boldsymbol{y}-u(\boldsymbol{x}-\boldsymbol{y}))du,\nonumber \\
\alpha^{lj}(\boldsymbol{w};\boldsymbol{x},\boldsymbol{y}) & =\epsilon^{lmj}\int_{0}^{1}(x^{m}-y^{m})\delta(\boldsymbol{w}-\boldsymbol{y}-u(\boldsymbol{x}-\boldsymbol{y}))udu,\label{eq:straight-line_use}
\end{align}
for our relators, and we will consider the quantities 
\begin{align}
& \Omega_{\boldsymbol{y}}^{j}(\boldsymbol{x},t)=\int\alpha^{lj}(\boldsymbol{w};\boldsymbol{x},\boldsymbol{y})B^{l}(\boldsymbol{w},t)d\boldsymbol{w},\\
& \Omega_{\boldsymbol{y}}^{0}(\boldsymbol{x},t)=\int s^{i}(\boldsymbol{w};\boldsymbol{x},\boldsymbol{y})E^{i}(\boldsymbol{w},t)d\boldsymbol{w}, \label{eq:Omega_0} 
\end{align}
and 
\begin{align}
& \frac{\hbar c}{e}\Delta(\boldsymbol{x},\boldsymbol{z},\boldsymbol{y};t)= \nonumber\\
& \int s^{i}(\boldsymbol{w};\boldsymbol{x},\boldsymbol{z})A^{i}(\boldsymbol{w},t)d\boldsymbol{w}+\int s^{i}(\boldsymbol{w};\boldsymbol{y},\boldsymbol{x})A^{i}(\boldsymbol{w},t)d\boldsymbol{w}\nonumber\\
& +\int s^{i}(\boldsymbol{w};\boldsymbol{z},\boldsymbol{y})A^{i}(\boldsymbol{w},t)d\boldsymbol{w}.
\end{align}
The ``expansion point'' here is $\boldsymbol{y}$. We first consider
\begin{align*}
\Omega_{\boldsymbol{y}}^{j}(\boldsymbol{x},t)\simeq B^{l}(\boldsymbol{y},t)\int\alpha^{lj}(\boldsymbol{w};\boldsymbol{x},\boldsymbol{y})d\boldsymbol{w}. 
\end{align*}
Now 
\begin{align*}
\int\alpha^{lj}(\boldsymbol{w};\boldsymbol{x},\boldsymbol{y})d\boldsymbol{w} & =\epsilon^{lmj}\int_{0}^{1}(x^{m}-y^{m})udu\\
& =\frac{1}{2}\epsilon^{lmj}(x^{m}-y^{m}),
\end{align*}
so 
\begin{align*}
\boldsymbol{\Omega}_{\boldsymbol{y}}(\boldsymbol{x},t)\simeq\frac{1}{2}\boldsymbol{B}(\boldsymbol{y},t)\cross(\boldsymbol{x}-\boldsymbol{y}).
\end{align*}
Next, we consider 
\begin{align*}
\Omega_{\boldsymbol{y}}^{0}(\boldsymbol{x},t)& \simeq E^{i}(\boldsymbol{y},t)\int s^{i}(\boldsymbol{w};\boldsymbol{x},\boldsymbol{y})d\boldsymbol{w}\nonumber \\
& +\frac{\partial E^{i}(\boldsymbol{y},t)}{\partial y^{k}}\int s^{i}(\boldsymbol{w};\boldsymbol{x},\boldsymbol{y})(w^{k}-y^{k})d\boldsymbol{w}.
\end{align*}
The terms we need are 
\begin{align}
\int s^{i}(\boldsymbol{w};\boldsymbol{x},\boldsymbol{y})d\boldsymbol{w} & =\int_{0}^{1}(x^{i}-y^{i})du=x^{i}-y^{i},\label{eq:s_zero_moment}
\end{align}
and 
\begin{align}
\int s^{i}(\boldsymbol{w};\boldsymbol{x},\boldsymbol{y})(w^{k}-y^{k})d\boldsymbol{w}&=\int_{0}^{1}(x^{i}-y^{i})(x^{k}-y^{k})udu\nonumber \\
& =\frac{1}{2}(x^{i}-y^{i})(x^{k}-y^{k}) \label{eq:s_one_moment}
\end{align}
so we have 
\begin{align*}
& \Omega_{\boldsymbol{y}}^{0}(\boldsymbol{x},t)\\
& \simeq(x^{i}-y^{i})E^{i}(\boldsymbol{y},t)+\frac{1}{2}(x^{i}-y^{i})(x^{k}-y^{k})\frac{\partial E^{i}(\boldsymbol{y},t)}{\partial y^{k}}\\
& =(x^{i}-y^{i})E^{i}(\boldsymbol{y},t)\\
& +\frac{1}{2}(x^{i}-y^{i})(x^{k}-y^{k})\left(\frac{1}{2}\frac{\partial E^{i}(\boldsymbol{y},t)}{\partial y^{k}}+\frac{1}{2}\frac{\partial E^{k}(\boldsymbol{y},t)}{\partial y^{i}}\right)\\
& =\left(x^i-y^i\right)E^i(\boldsymbol{y},t)+\frac{1}{2}(x^{i}-y^{i})(x^{k}-y^{k})F^{ik}(\boldsymbol{y},t),
\end{align*}
where we have used (\ref{eq:Fdef}). Finally, we look at $\Delta(\boldsymbol{x},\boldsymbol{z},\boldsymbol{y};t)$.
This can be done ``by hand'' when the magnetic field is uniform,
but in what follows we work it out formally. 
\begin{align*}
&\frac{\hbar c}{e}\Delta(\boldsymbol{x},\boldsymbol{z},\boldsymbol{y};t)\\
& \simeq A^{i}(\boldsymbol{y},t)\int\left(s^{i}(\boldsymbol{w};\boldsymbol{x},\boldsymbol{z})+s^{i}(\boldsymbol{w};\boldsymbol{y},\boldsymbol{x})+s^{i}(\boldsymbol{w};\boldsymbol{z},\boldsymbol{y})\right)d\boldsymbol{w}\\
& +\frac{\partial A^{i}(\boldsymbol{y},t)}{\partial y^{j}}\int(w^{j}-y^{j})\Big(s^{i}(\boldsymbol{w};\boldsymbol{x},\boldsymbol{z})\\
& \qquad\qquad\qquad\qquad+s^{i}(\boldsymbol{w};\boldsymbol{y},\boldsymbol{x})+s^{i}(\boldsymbol{w};\boldsymbol{z},\boldsymbol{y})\Big)d\boldsymbol{w}+\ldots
\end{align*}
From (\ref{eq:s_zero_moment}) we have 
\begin{align*}
& \int\left(s^{i}(\boldsymbol{w};\boldsymbol{x},\boldsymbol{z})+s^{i}(\boldsymbol{w};\boldsymbol{y},\boldsymbol{x})+s^{i}(\boldsymbol{w};\boldsymbol{z},\boldsymbol{y})\right)d\boldsymbol{w}=0,
\end{align*}
while 
\begin{align*}
& \int s^{i}(\boldsymbol{w};\boldsymbol{x},\boldsymbol{z})(w^{j}-y^{j})d\boldsymbol{w}\\
& =\int s^{i}(\boldsymbol{w};\boldsymbol{x},\boldsymbol{z})(z^{j}-y^{j})d\boldsymbol{w}+\int s^{i}(\boldsymbol{w};\boldsymbol{x},\boldsymbol{z})(w^{j}-z^{j})d\boldsymbol{w}\\
& =(x^{i}-z^{i})(z^{j}-y^{j})+\frac{1}{2}(x^{i}-z^{i})(x^{j}-z^{j})\\
& =((x^{i}-y^{i})-(z^{i}-y^{i}))(z^{j}-y^{j})\\
& +\frac{1}{2}((x^{i}-y^{i})-(z^{i}-y^{i}))((x^{j}-y^{j})-(z^{j}-y^{j}))\\
& =\left(\frac{1}{2}(x^{i}-y^{i})(z^{j}-y^{j})-\frac{1}{2}(z^{i}-y^{i})(x^{j}-y^{j})\right)\\
& +\frac{1}{2}(x^{i}-y^{i})(x^{j}-y^{j})-\frac{1}{2}(z^{i}-y^{i})(z^{j}-y^{j}),
\end{align*}
where in the third line we have used (\ref{eq:s_zero_moment},\ref{eq:s_one_moment});
similarly 
\begin{align*}
\int s^{i}(\boldsymbol{w};\boldsymbol{y},\boldsymbol{x})(w^{j}-y^{j})d\boldsymbol{w}=-\frac{1}{2}(x^{i}-y^{i})(x^{j}-y^{j}),
\end{align*}
and finally 
\begin{align*}
\int s^{i}(\boldsymbol{w};\boldsymbol{z},\boldsymbol{y})(w^{j}-y^{j})d\boldsymbol{w}=\frac{1}{2}(z^{i}-y^{i})(z^{j}-y^{j}).
\end{align*}
So, in all 
\begin{align*}
& \int(w^{j}-R^{j})\left(s^{i}(\boldsymbol{w};\boldsymbol{x},\boldsymbol{z})+s^{i}(\boldsymbol{w};\boldsymbol{y},\boldsymbol{x})+s^{i}(\boldsymbol{w};\boldsymbol{z},\boldsymbol{y})\right)d\boldsymbol{w}\\
& =\frac{1}{2}(x^{i}-y^{i})(z^{j}-y^{j})-\frac{1}{2}(z^{i}-y^{i})(x^{j}-y^{j}),
\end{align*}
and 
\begin{align*}
& \frac{\hbar c}{e}\Delta(\boldsymbol{x},\boldsymbol{z},\boldsymbol{y};t)\\
& \simeq\frac{1}{2}\frac{\partial A^{i}(\boldsymbol{y},t)}{\partial y^{j}}\left((x^{i}-y^{i})(z^{j}-y^{j})-(z^{i}-y^{i})(x^{j}-y^{j})\right)\\
& =\frac{1}{2}\left(\frac{\partial A^{i}(\boldsymbol{y},t)}{\partial y^{j}}-\frac{\partial A^{j}(\boldsymbol{y},t)}{\partial y^{i}}\right)(x^{i}-y^{i})(z^{j}-y^{j}),
\end{align*}
and since 
\begin{align*}
& \frac{\partial A^{j}(\boldsymbol{y},t)}{\partial y^{i}}-\frac{\partial A^{i}(\boldsymbol{y},t)}{\partial y^{j}}=\epsilon^{kij}B^{k}(\boldsymbol{y},t),
\end{align*}
we have 
\begin{align*}
\frac{\hbar c}{e}\Delta(\boldsymbol{x},\boldsymbol{z},\boldsymbol{y};t) & \simeq-\frac{1}{2}\boldsymbol{B}(\boldsymbol{y},t)\boldsymbol{\cdot}\big[\left(\boldsymbol{x}-\boldsymbol{y}\right)\cross(\boldsymbol{z}-\boldsymbol{y})\big].
\end{align*}
We collect all of these approximate expressions in (\ref{omega0},\ref{omegaVec},\ref{delta}). The expansions of $\Omega_{\boldsymbol{y}}^{j}(\boldsymbol{x},t)$
and $\Omega_{\boldsymbol{y}}^{0}(\boldsymbol{x},t)$ derived here
can also be derived using a formal expansion of the relators (\ref{eq:straight-line_use})
about $u=0$.

\begin{widetext}
\section{Constructing the Chern-Simons contribution to the OMP tensor}
\label{AppendixCS}

Here we outline the steps in going from the first to the second equality of Eq.~(\ref{alphaCS}). The final term of the first line of (\ref{alphaCS}) can be re-expressed, using (\ref{connection}), as proportional to the Brillouin zone integral of
\begin{align*}
&\epsilon^{lab}\sum_{\alpha\gamma}f_\alpha\tilde{\xi}^i_{\alpha\gamma}\partial_b \tilde{\xi}^a_{\gamma\alpha}\\
&=\epsilon^{lab}\sum_{\alpha\gamma}\sum_{nmps}f_n(\xi^i_{ps}+\mathcal{W}^i_{ps})U_{s\gamma}
\partial_b\big(U^\dagger_{\gamma m}(\xi^a_{mn}+\mathcal{W}^a_{mn})U_{n\alpha}\big)U^\dagger_{\alpha p} \\
&=\epsilon^{lab}\sum_{nm}f_n(\xi^i_{nm}+\mathcal{W}^i_{nm})\partial_b(\xi^a_{mn}+\mathcal{W}^a_{mn})+\epsilon^{lab}\sum_{nmps}f_n(\xi^i_{ps}+\mathcal{W}^i_{ps})(\xi^a_{mn}+\mathcal{W}^a_{mn})\big(i\mathcal{W}^b_{sm}\delta_{np}-i\delta_{sm}\mathcal{W}^b_{np}\big) \\
&=\epsilon^{lab}\sum_{nms}f_n(\xi^i_{nm}+\mathcal{W}^i_{nm})(i\xi^b_{ms}\xi^a_{sn}-i\mathcal{W}^b_{ms}\mathcal{W}^a_{sn})\\
&+i\epsilon^{lab}\sum_{nms}f_n\Big((\xi^i_{ns}+\mathcal{W}^i_{ns})\mathcal{W}^b_{sm}(\xi^a_{mn}+\mathcal{W}^a_{mn})-(\xi^i_{sm}+\mathcal{W}^i_{sm})(\xi^a_{mn}+\mathcal{W}^a_{mn})\mathcal{W}^b_{ns}\Big),
\end{align*}
where we have used the identities $\epsilon^{iab}\partial_b\xi^a_{mn}=i\epsilon^{iab}\sum_{s}\xi^b_{ms}\xi^a_{sn}$ and $\epsilon^{iab}\partial_b\mathcal{W}^a_{mn}=-i\epsilon^{iab}\sum_{s}\mathcal{W}^b_{ms}\mathcal{W}^a_{sn}$; in the following we will also often use $\mathcal{W}^a_{mn}\neq0$ only if $f_m=f_n$. We now consider the contributions to (\ref{alphaCS}) at each order in $\mathcal{W}$. We first consider the terms quadratic in $\mathcal{W}$; their contribution to (\ref{alphaCS}) is proportional to the Brillouin zone integral of
\begin{align*}
&\epsilon^{lab}\text{Re}\sum_{nms}if_n\Big(-\xi^i_{nm}\mathcal{W}^b_{ms}\mathcal{W}^a_{sn}+\mathcal{W}^i_{ns}\mathcal{W}^b_{sm}\xi^a_{mn}+\xi^i_{ns}\mathcal{W}^b_{sm}\mathcal{W}^a_{mn}-\mathcal{W}^i_{sm}\xi^a_{mn}\mathcal{W}^b_{ns}-\xi^i_{sm}\mathcal{W}^a_{mn}\mathcal{W}^b_{ns}\Big) \\
&=\epsilon^{lab}\text{Re}\sum_{nms}if_n\Big(\mathcal{\xi}^a_{mn}\mathcal{W}^i_{ns}\mathcal{W}^b_{sm}+\xi^b_{mn}\mathcal{W}^a_{ns}\mathcal{W}^i_{sm}+\xi^i_{sm}\mathcal{W}^b_{mn}\mathcal{W}^a_{ns}\Big) \\
&=\delta^{il}\epsilon^{lab}\text{Re}\sum_{nms}if_n\Big(\mathcal{\xi}^a_{mn}\mathcal{W}^l_{ns}\mathcal{W}^b_{sm}+\xi^b_{mn}\mathcal{W}^a_{ns}\mathcal{W}^l_{sm}+\xi^l_{sm}\mathcal{W}^b_{mn}\mathcal{W}^a_{ns}\Big) \\
&=\delta^{il}\epsilon^{cab}\text{Re}\sum_{nms}if_n\mathcal{\xi}^a_{nm}\mathcal{W}^c_{ms}\mathcal{W}^b_{sn},
\end{align*}
where in going from the second to third line we have used the fact that, in three-dimensions, at least two of $i,l,a,b$ must be identical; if $i\neq l$ the expression is found to vanish. This sort of argument is often used in what follows. The contribution to (\ref{alphaCS}) that is linear in $\mathcal{W}$ (notice the penultimate term of (\ref{alphaCS}) also contributes here) is proportional to the Brillouin zone integral of
\begin{align*}
&\epsilon^{lab}\text{Re}\sum_{nms}if_n\Big(\mathcal{W}^i_{nm}\xi^b_{ms}\xi^a_{sn}+\big(\xi^i_{ns}\mathcal{W}^b_{sm}\xi^a_{mn}-\xi^i_{sm}\xi^a_{mn}\mathcal{W}^b_{ns}\big)\Big)-\epsilon^{lab}\sum_{mns}f_{nm}\text{Re}\big[i\xi^i_{nm}\mathcal{W}^b_{ms}\xi^a_{sn}\big] \\
&=\epsilon^{lab}\text{Re}\sum_{nms}if_n\Big(\mathcal{W}^i_{nm}\xi^b_{ms}\xi^a_{sn}-\xi^i_{sm}\xi^a_{mn}\mathcal{W}^b_{ns}\Big)+\epsilon^{lab}\sum_{mns}f_{m}\text{Re}\big[i\xi^i_{nm}\mathcal{W}^b_{ms}\xi^a_{sn}\big] \\
&=\delta^{il}\epsilon^{lab}\text{Re}\sum_{nms}if_n\Big(\xi^b_{ms}\xi^a_{sn}\mathcal{W}^l_{nm}+\xi^l_{sm}\xi^b_{mn}\mathcal{W}^a_{ns}+\xi^a_{sm}\xi^l_{mn}\mathcal{W}^b_{ns}\Big) \\
&=\delta^{il}\epsilon^{cab}\text{Re}\sum_{nms}if_n\xi^b_{nm}\xi^a_{ms}\mathcal{W}^c_{sn}.
\end{align*}
Now the combined contribution of the linear and quadratic in $\mathcal{W}$ terms is proportional to
\begin{align*}
&\delta^{il}\epsilon^{abc}\sum_{nms}f_n\int_{\text{BZ}}d\boldsymbol{k}\text{Re}\Big[i\xi^b_{nm}\xi^a_{ms}\mathcal{W}^c_{sn}+i\mathcal{\xi}^a_{nm}\mathcal{W}^c_{ms}\mathcal{W}^b_{sn}\Big] \\
&=\delta^{il}\epsilon^{abc}\sum_{ns}f_n\int_{\text{BZ}}d\boldsymbol{k}\text{Re}\Big[\partial_b\xi^a_{ns}\mathcal{W}^c_{sn}+i\sum_{m}\mathcal{\xi}^a_{nm}\mathcal{W}^c_{ms}\mathcal{W}^b_{sn}\Big] \\
&=\delta^{il}\epsilon^{abc}\sum_{ns}f_n\int_{\text{BZ}}d\boldsymbol{k}\text{Re}\Big[-\xi^a_{ns}\partial_b\mathcal{W}^c_{sn}+i\sum_{m}\mathcal{\xi}^a_{nm}\mathcal{W}^c_{ms}\mathcal{W}^b_{sn}\Big] \\
&=0,
\end{align*}
where we have used an integration by parts on the initially linear in $\mathcal{W}$ term. This ``miraculous cancellation'' is also presented in Appendix C of Vanderbilt \cite{VanderbiltBook}. Thus, the contribution cubic in $\mathcal{W}$ is the only gauge dependent term that has not vanished, or been canceled. Its contribution to (\ref{alphaCS}) is proportional to the Brillouin zone integral of 
\begin{align}
&\epsilon^{lab}\text{Re}\sum_{nms}if_n\Big(-\mathcal{W}^i_{nm}\mathcal{W}^b_{ms}\mathcal{W}^a_{sn}+\mathcal{W}^i_{ns}\mathcal{W}^b_{mn}\mathcal{W}^a_{sm}-\mathcal{W}^i_{sm}\mathcal{W}^a_{mn}\mathcal{W}^b_{ns}\Big) \nonumber\\
&=-\epsilon^{lab}\text{Re}\sum_{vv'v_1}i\mathcal{W}^i_{vv'}\mathcal{W}^a_{v'v_1}\mathcal{W}^b_{v_1v} \nonumber\\
&=-\delta^{il}\frac{\epsilon^{cab}}{3}\text{Re}\sum_{vv'v_1}i\mathcal{W}^c_{vv'}\mathcal{W}^a_{v'v_1}\mathcal{W}^b_{v_1v} \nonumber\\
&=-\delta^{il}\epsilon^{abc}\text{Re}\left[\sum_{vv_1}\big(\partial_b\mathcal{W}^a_{vv_1}\big)\mathcal{W}^c_{v_1v}-\frac{2i}{3}\sum_{vv_1v'}\mathcal{W}^a_{vv'}\mathcal{W}^b_{v'v_1}\mathcal{W}^c_{v_1v}\right].
\label{CSgauge}
\end{align}
The contribution of this term to (\ref{alphaCS}) is proportional to the Brillouin zone integral of the well-known term arising from the gauge-transformation of the Chern-Simons 3-form (see, e.g., Eq.~C.19 of Vanderbilt \cite{VanderbiltBook}). The contribution independent of $\mathcal{W}$ is proportional (notice the first term of (\ref{alphaCS}) contributes here) to the Brillouin zone integral of
\begin{align}
&\epsilon^{lab}\text{Re}\sum_{nms}if_n\xi^i_{nm}\xi^b_{ms}\xi^a_{sn}+\epsilon^{lab}\Bigg\llbracket 2\sum_{cvv'}\text{Re}\big[\left(\partial_iv|c\right)\left(c|\partial_av'\right)\left(v'|\partial_bv\right)\big]+\sum_{cv}\text{Re}\big[\left(\partial_iv|c\right)\left(\partial_ac|\partial_bv\right)\big]\Bigg\rrbracket \nonumber\\
&=\epsilon^{lab}\text{Re}\sum_{vm}\big[\left(\partial_iv|m\right)\left(\partial_bm|\partial_av\right)-\left(\partial_iv|c\right)\left(\partial_bc|\partial_av\right)\big]+2\epsilon^{lab}\sum_{cvv'}\text{Re}\big[\left(\partial_iv|c\right)\left(c|\partial_av'\right)\left(v'|\partial_bv\right)\big] \nonumber\\
&=\epsilon^{lab}\text{Re}\sum_{vv'}\big[\left(\partial_iv|v'\right)\left(\partial_bv'|\partial_av\right)\big]+2\epsilon^{lab}\sum_{cvv'}\text{Re}\big[\left(\partial_iv|c\right)\left(c|\partial_av'\right)\left(v'|\partial_bv\right)\big],
\label{CS}
\end{align}
which is equivalent to Eqs.~(A11a)+(A11b) of Essin \textit{et al.}~\cite{Essin2010}. Then, in all, (\ref{alphaCS}) results from the combination of (\ref{CSgauge})+(\ref{CS}).
\end{widetext}

\bibliographystyle{apsrev4-1}
\bibliography{dcResponse_Updated}

\begin{thebibliography}{46}%
\makeatletter
\providecommand \@ifxundefined [1]{%
 \@ifx{#1\undefined}
}%
\providecommand \@ifnum [1]{%
 \ifnum #1\expandafter \@firstoftwo
 \else \expandafter \@secondoftwo
 \fi
}%
\providecommand \@ifx [1]{%
 \ifx #1\expandafter \@firstoftwo
 \else \expandafter \@secondoftwo
 \fi
}%
\providecommand \natexlab [1]{#1}%
\providecommand \enquote  [1]{``#1''}%
\providecommand \bibnamefont  [1]{#1}%
\providecommand \bibfnamefont [1]{#1}%
\providecommand \citenamefont [1]{#1}%
\providecommand \href@noop [0]{\@secondoftwo}%
\providecommand \href [0]{\begingroup \@sanitize@url \@href}%
\providecommand \@href[1]{\@@startlink{#1}\@@href}%
\providecommand \@@href[1]{\endgroup#1\@@endlink}%
\providecommand \@sanitize@url [0]{\catcode `\\12\catcode `\$12\catcode
  `\&12\catcode `\#12\catcode `\^12\catcode `\_12\catcode `\%12\relax}%
\providecommand \@@startlink[1]{}%
\providecommand \@@endlink[0]{}%
\providecommand \url  [0]{\begingroup\@sanitize@url \@url }%
\providecommand \@url [1]{\endgroup\@href {#1}{\urlprefix }}%
\providecommand \urlprefix  [0]{URL }%
\providecommand \Eprint [0]{\href }%
\providecommand \doibase [0]{http://dx.doi.org/}%
\providecommand \selectlanguage [0]{\@gobble}%
\providecommand \bibinfo  [0]{\@secondoftwo}%
\providecommand \bibfield  [0]{\@secondoftwo}%
\providecommand \translation [1]{[#1]}%
\providecommand \BibitemOpen [0]{}%
\providecommand \bibitemStop [0]{}%
\providecommand \bibitemNoStop [0]{.\EOS\space}%
\providecommand \EOS [0]{\spacefactor3000\relax}%
\providecommand \BibitemShut  [1]{\csname bibitem#1\endcsname}%
\let\auto@bib@innerbib\@empty
\bibitem [{\citenamefont {Lorentz}(1909)}]{Lorentz}%
  \BibitemOpen
  \bibfield  {author} {\bibinfo {author} {\bibfnamefont {H.}~\bibnamefont
  {Lorentz}},\ }\href@noop {} {\emph {\bibinfo {title} {The Theory of
  Electrons}}}\ (\bibinfo  {publisher} {Columbia University Press},\ \bibinfo
  {year} {1909})\BibitemShut {NoStop}%
\bibitem [{\citenamefont {Jackson}(1999)}]{Jackson}%
  \BibitemOpen
  \bibfield  {author} {\bibinfo {author} {\bibfnamefont {J.~D.}\ \bibnamefont
  {Jackson}},\ }\href {http://cdsweb.cern.ch/record/490457} {\emph {\bibinfo
  {title} {Classical electrodynamics}}},\ \bibinfo {edition} {3rd}\ ed.\
  (\bibinfo  {publisher} {Wiley},\ \bibinfo {address} {New York, {NY}},\
  \bibinfo {year} {1999})\BibitemShut {NoStop}%
\bibitem [{\citenamefont {Resta}(1994)}]{Resta1994}%
  \BibitemOpen
  \bibfield  {author} {\bibinfo {author} {\bibfnamefont {R.}~\bibnamefont
  {Resta}},\ }\href {\doibase 10.1103/RevModPhys.66.899} {\bibfield  {journal}
  {\bibinfo  {journal} {Rev. Mod. Phys.}\ }\textbf {\bibinfo {volume} {66}},\
  \bibinfo {pages} {899} (\bibinfo {year} {1994})}\BibitemShut {NoStop}%
\bibitem [{\citenamefont {Thonhauser}\ \emph {et~al.}(2005)\citenamefont
  {Thonhauser}, \citenamefont {Ceresoli}, \citenamefont {Vanderbilt},\ and\
  \citenamefont {Resta}}]{Resta2005}%
  \BibitemOpen
  \bibfield  {author} {\bibinfo {author} {\bibfnamefont {T.}~\bibnamefont
  {Thonhauser}}, \bibinfo {author} {\bibfnamefont {D.}~\bibnamefont
  {Ceresoli}}, \bibinfo {author} {\bibfnamefont {D.}~\bibnamefont
  {Vanderbilt}}, \ and\ \bibinfo {author} {\bibfnamefont {R.}~\bibnamefont
  {Resta}},\ }\href {\doibase 10.1103/PhysRevLett.95.137205} {\bibfield
  {journal} {\bibinfo  {journal} {Phys. Rev. Lett.}\ }\textbf {\bibinfo
  {volume} {95}},\ \bibinfo {pages} {137205} (\bibinfo {year}
  {2005})}\BibitemShut {NoStop}%
\bibitem [{\citenamefont {Ceresoli}\ \emph {et~al.}(2006)\citenamefont
  {Ceresoli}, \citenamefont {Thonhauser}, \citenamefont {Vanderbilt},\ and\
  \citenamefont {Resta}}]{Resta2006}%
  \BibitemOpen
  \bibfield  {author} {\bibinfo {author} {\bibfnamefont {D.}~\bibnamefont
  {Ceresoli}}, \bibinfo {author} {\bibfnamefont {T.}~\bibnamefont
  {Thonhauser}}, \bibinfo {author} {\bibfnamefont {D.}~\bibnamefont
  {Vanderbilt}}, \ and\ \bibinfo {author} {\bibfnamefont {R.}~\bibnamefont
  {Resta}},\ }\href {\doibase 10.1103/PhysRevB.74.024408} {\bibfield  {journal}
  {\bibinfo  {journal} {Phys. Rev. B}\ }\textbf {\bibinfo {volume} {74}},\
  \bibinfo {pages} {024408} (\bibinfo {year} {2006})}\BibitemShut {NoStop}%
\bibitem [{\citenamefont {Shi}\ \emph {et~al.}(2007)\citenamefont {Shi},
  \citenamefont {Vignale}, \citenamefont {Xiao},\ and\ \citenamefont
  {Niu}}]{Niu2007}%
  \BibitemOpen
  \bibfield  {author} {\bibinfo {author} {\bibfnamefont {J.}~\bibnamefont
  {Shi}}, \bibinfo {author} {\bibfnamefont {G.}~\bibnamefont {Vignale}},
  \bibinfo {author} {\bibfnamefont {D.}~\bibnamefont {Xiao}}, \ and\ \bibinfo
  {author} {\bibfnamefont {Q.}~\bibnamefont {Niu}},\ }\href {\doibase
  10.1103/PhysRevLett.99.197202} {\bibfield  {journal} {\bibinfo  {journal}
  {Phys. Rev. Lett.}\ }\textbf {\bibinfo {volume} {99}},\ \bibinfo {pages}
  {197202} (\bibinfo {year} {2007})}\BibitemShut {NoStop}%
\bibitem [{\citenamefont {Resta}(1998)}]{RestaX}%
  \BibitemOpen
  \bibfield  {author} {\bibinfo {author} {\bibfnamefont {R.}~\bibnamefont
  {Resta}},\ }\href {\doibase 10.1103/PhysRevLett.80.1800} {\bibfield
  {journal} {\bibinfo  {journal} {Phys. Rev. Lett.}\ }\textbf {\bibinfo
  {volume} {80}},\ \bibinfo {pages} {1800} (\bibinfo {year}
  {1998})}\BibitemShut {NoStop}%
\bibitem [{\citenamefont {King-Smith}\ and\ \citenamefont
  {Vanderbilt}(1993)}]{KingSmith1993}%
  \BibitemOpen
  \bibfield  {author} {\bibinfo {author} {\bibfnamefont {R.~D.}\ \bibnamefont
  {King-Smith}}\ and\ \bibinfo {author} {\bibfnamefont {D.}~\bibnamefont
  {Vanderbilt}},\ }\href {\doibase 10.1103/PhysRevB.47.1651} {\bibfield
  {journal} {\bibinfo  {journal} {Phys. Rev. B}\ }\textbf {\bibinfo {volume}
  {47}},\ \bibinfo {pages} {1651} (\bibinfo {year} {1993})}\BibitemShut
  {NoStop}%
\bibitem [{\citenamefont {Vanderbilt}(2018)}]{VanderbiltBook}%
  \BibitemOpen
  \bibfield  {author} {\bibinfo {author} {\bibfnamefont {D.}~\bibnamefont
  {Vanderbilt}},\ }\href {\doibase 10.1017/9781316662205} {\emph {\bibinfo
  {title} {Berry Phases in Electronic Structure Theory: Electric Polarization,
  Orbital Magnetization and Topological Insulators}}}\ (\bibinfo  {publisher}
  {Cambridge University Press},\ \bibinfo {year} {2018})\BibitemShut {NoStop}%
\bibitem [{\citenamefont {Mahon}\ \emph {et~al.}(2019)\citenamefont {Mahon},
  \citenamefont {Muniz},\ and\ \citenamefont {Sipe}}]{Mahon2019}%
  \BibitemOpen
  \bibfield  {author} {\bibinfo {author} {\bibfnamefont {P.~T.}\ \bibnamefont
  {Mahon}}, \bibinfo {author} {\bibfnamefont {R.~A.}\ \bibnamefont {Muniz}}, \
  and\ \bibinfo {author} {\bibfnamefont {J.~E.}\ \bibnamefont {Sipe}},\ }\href
  {\doibase 10.1103/PhysRevB.99.235140} {\bibfield  {journal} {\bibinfo
  {journal} {Phys. Rev. B}\ }\textbf {\bibinfo {volume} {99}},\ \bibinfo
  {pages} {235140} (\bibinfo {year} {2019})}\BibitemShut {NoStop}%
\bibitem [{\citenamefont {Healy}(1982)}]{Healybook}%
  \BibitemOpen
  \bibfield  {author} {\bibinfo {author} {\bibfnamefont {W.}~\bibnamefont
  {Healy}},\ }\href@noop {} {\emph {\bibinfo {title} {Non-relativistic quantum
  electrodynamics}}}\ (\bibinfo  {publisher} {Academic Press},\ \bibinfo {year}
  {1982})\BibitemShut {NoStop}%
\bibitem [{\citenamefont {Cohen-Tannoudji}\ \emph {et~al.}(1989)\citenamefont
  {Cohen-Tannoudji}, \citenamefont {Dupont-Roc},\ and\ \citenamefont
  {Grynberg}}]{PZW}%
  \BibitemOpen
  \bibfield  {author} {\bibinfo {author} {\bibfnamefont {C.}~\bibnamefont
  {Cohen-Tannoudji}}, \bibinfo {author} {\bibfnamefont {J.}~\bibnamefont
  {Dupont-Roc}}, \ and\ \bibinfo {author} {\bibfnamefont {G.}~\bibnamefont
  {Grynberg}},\ }\href@noop {} {\emph {\bibinfo {title} {Photons and Atoms:
  Introduction to Quantum Electrodynamics}}}\ (\bibinfo  {publisher} {John
  Wiley and Sons, Inc.},\ \bibinfo {year} {1989})\BibitemShut {NoStop}%
\bibitem [{\citenamefont {Marzari}\ and\ \citenamefont
  {Vanderbilt}(1997)}]{WF1}%
  \BibitemOpen
  \bibfield  {author} {\bibinfo {author} {\bibfnamefont {N.}~\bibnamefont
  {Marzari}}\ and\ \bibinfo {author} {\bibfnamefont {D.}~\bibnamefont
  {Vanderbilt}},\ }\href@noop {} {\bibfield  {journal} {\bibinfo  {journal}
  {Phys. Rev. B}\ }\textbf {\bibinfo {volume} {56}},\ \bibinfo {pages} {12847}
  (\bibinfo {year} {1997})}\BibitemShut {NoStop}%
\bibitem [{\citenamefont {Souza}\ \emph {et~al.}(2001)\citenamefont {Souza},
  \citenamefont {Marzari},\ and\ \citenamefont {Vanderbilt}}]{WF2}%
  \BibitemOpen
  \bibfield  {author} {\bibinfo {author} {\bibfnamefont {I.}~\bibnamefont
  {Souza}}, \bibinfo {author} {\bibfnamefont {N.}~\bibnamefont {Marzari}}, \
  and\ \bibinfo {author} {\bibfnamefont {D.}~\bibnamefont {Vanderbilt}},\
  }\href@noop {} {\bibfield  {journal} {\bibinfo  {journal} {Phys. Rev. B}\
  }\textbf {\bibinfo {volume} {65}},\ \bibinfo {pages} {035109} (\bibinfo
  {year} {2001})}\BibitemShut {NoStop}%
\bibitem [{Note1()}]{Note1}%
  \BibitemOpen
  \bibinfo {note} {While in past work \cite {Mahon2019} and in this paper we
  treat the electromagnetic field classically, quantum mechanical effects can,
  in principle, be taken into account.}\BibitemShut {Stop}%
\bibitem [{Note2()}]{Note2}%
  \BibitemOpen
  \bibinfo {note} {This includes both ordinary and $\protect \mathbb {Z}_2$
  topological insulators. We discuss this further below.}\BibitemShut {Stop}%
\bibitem [{\citenamefont {Qi}\ \emph {et~al.}(2008)\citenamefont {Qi},
  \citenamefont {Hughes},\ and\ \citenamefont {Zhang}}]{Qi2008}%
  \BibitemOpen
  \bibfield  {author} {\bibinfo {author} {\bibfnamefont {X.-L.}\ \bibnamefont
  {Qi}}, \bibinfo {author} {\bibfnamefont {T.~L.}\ \bibnamefont {Hughes}}, \
  and\ \bibinfo {author} {\bibfnamefont {S.-C.}\ \bibnamefont {Zhang}},\ }\href
  {\doibase 10.1103/PhysRevB.78.195424} {\bibfield  {journal} {\bibinfo
  {journal} {Phys. Rev. B}\ }\textbf {\bibinfo {volume} {78}},\ \bibinfo
  {pages} {195424} (\bibinfo {year} {2008})}\BibitemShut {NoStop}%
\bibitem [{\citenamefont {Essin}\ \emph {et~al.}(2009)\citenamefont {Essin},
  \citenamefont {Moore},\ and\ \citenamefont {Vanderbilt}}]{Vanderbilt2009}%
  \BibitemOpen
  \bibfield  {author} {\bibinfo {author} {\bibfnamefont {A.~M.}\ \bibnamefont
  {Essin}}, \bibinfo {author} {\bibfnamefont {J.~E.}\ \bibnamefont {Moore}}, \
  and\ \bibinfo {author} {\bibfnamefont {D.}~\bibnamefont {Vanderbilt}},\
  }\href {\doibase 10.1103/PhysRevLett.102.146805} {\bibfield  {journal}
  {\bibinfo  {journal} {Phys. Rev. Lett.}\ }\textbf {\bibinfo {volume} {102}},\
  \bibinfo {pages} {146805} (\bibinfo {year} {2009})}\BibitemShut {NoStop}%
\bibitem [{\citenamefont {Malashevich}\ \emph {et~al.}(2010)\citenamefont
  {Malashevich}, \citenamefont {Souza}, \citenamefont {Coh},\ and\
  \citenamefont {Vanderbilt}}]{Malashevich2010}%
  \BibitemOpen
  \bibfield  {author} {\bibinfo {author} {\bibfnamefont {A.}~\bibnamefont
  {Malashevich}}, \bibinfo {author} {\bibfnamefont {I.}~\bibnamefont {Souza}},
  \bibinfo {author} {\bibfnamefont {S.}~\bibnamefont {Coh}}, \ and\ \bibinfo
  {author} {\bibfnamefont {D.}~\bibnamefont {Vanderbilt}},\ }\href {\doibase
  10.1088/1367-2630/12/5/053032} {\bibfield  {journal} {\bibinfo  {journal}
  {New Journal of Physics}\ }\textbf {\bibinfo {volume} {12}},\ \bibinfo
  {pages} {053032} (\bibinfo {year} {2010})}\BibitemShut {NoStop}%
\bibitem [{\citenamefont {Essin}\ \emph {et~al.}(2010)\citenamefont {Essin},
  \citenamefont {Turner}, \citenamefont {Moore},\ and\ \citenamefont
  {Vanderbilt}}]{Essin2010}%
  \BibitemOpen
  \bibfield  {author} {\bibinfo {author} {\bibfnamefont {A.~M.}\ \bibnamefont
  {Essin}}, \bibinfo {author} {\bibfnamefont {A.~M.}\ \bibnamefont {Turner}},
  \bibinfo {author} {\bibfnamefont {J.~E.}\ \bibnamefont {Moore}}, \ and\
  \bibinfo {author} {\bibfnamefont {D.}~\bibnamefont {Vanderbilt}},\ }\href
  {\doibase 10.1103/PhysRevB.81.205104} {\bibfield  {journal} {\bibinfo
  {journal} {Phys. Rev. B}\ }\textbf {\bibinfo {volume} {81}},\ \bibinfo
  {pages} {205104} (\bibinfo {year} {2010})}\BibitemShut {NoStop}%
\bibitem [{\citenamefont {Swiecicki}\ and\ \citenamefont
  {Sipe}(2014)}]{Swiecicki2014}%
  \BibitemOpen
  \bibfield  {author} {\bibinfo {author} {\bibfnamefont {S.~D.}\ \bibnamefont
  {Swiecicki}}\ and\ \bibinfo {author} {\bibfnamefont {J.~E.}\ \bibnamefont
  {Sipe}},\ }\href {\doibase 10.1103/PhysRevB.90.125115} {\bibfield  {journal}
  {\bibinfo  {journal} {Phys. Rev. B}\ }\textbf {\bibinfo {volume} {90}},\
  \bibinfo {pages} {125115} (\bibinfo {year} {2014})}\BibitemShut {NoStop}%
\bibitem [{Note3()}]{Note3}%
  \BibitemOpen
  \bibinfo {note} {More precisely, the OMP tensor vanishes modulo a discrete
  ambiguity when time-reversal or inversion symmetry are present in the
  unperturbed system.}\BibitemShut {Stop}%
\bibitem [{\citenamefont {Hasan}\ and\ \citenamefont {Kane}(2010)}]{Kane2010}%
  \BibitemOpen
  \bibfield  {author} {\bibinfo {author} {\bibfnamefont {M.~Z.}\ \bibnamefont
  {Hasan}}\ and\ \bibinfo {author} {\bibfnamefont {C.~L.}\ \bibnamefont
  {Kane}},\ }\href {\doibase 10.1103/RevModPhys.82.3045} {\bibfield  {journal}
  {\bibinfo  {journal} {Rev. Mod. Phys.}\ }\textbf {\bibinfo {volume} {82}},\
  \bibinfo {pages} {3045} (\bibinfo {year} {2010})}\BibitemShut {NoStop}%
\bibitem [{\citenamefont {Tokura}\ \emph {et~al.}(2019)\citenamefont {Tokura},
  \citenamefont {Yasuda},\ and\ \citenamefont {Tsukazaki}}]{Tokura2019}%
  \BibitemOpen
  \bibfield  {author} {\bibinfo {author} {\bibfnamefont {Y.}~\bibnamefont
  {Tokura}}, \bibinfo {author} {\bibfnamefont {K.}~\bibnamefont {Yasuda}}, \
  and\ \bibinfo {author} {\bibfnamefont {A.}~\bibnamefont {Tsukazaki}},\ }\href
  {\doibase 10.1038/s42254-018-0011-5} {\bibfield  {journal} {\bibinfo
  {journal} {Nature Reviews Physics}\ }\textbf {\bibinfo {volume} {1}},\
  \bibinfo {pages} {126} (\bibinfo {year} {2019})}\BibitemShut {NoStop}%
\bibitem [{\citenamefont {Brouder}\ \emph {et~al.}(2007)\citenamefont
  {Brouder}, \citenamefont {Panati}, \citenamefont {Calandra}, \citenamefont
  {Mourougane},\ and\ \citenamefont {Marzari}}]{Brouder}%
  \BibitemOpen
  \bibfield  {author} {\bibinfo {author} {\bibfnamefont {C.}~\bibnamefont
  {Brouder}}, \bibinfo {author} {\bibfnamefont {G.}~\bibnamefont {Panati}},
  \bibinfo {author} {\bibfnamefont {M.}~\bibnamefont {Calandra}}, \bibinfo
  {author} {\bibfnamefont {C.}~\bibnamefont {Mourougane}}, \ and\ \bibinfo
  {author} {\bibfnamefont {N.}~\bibnamefont {Marzari}},\ }\href@noop {}
  {\bibfield  {journal} {\bibinfo  {journal} {Phys. Rev. Lett.}\ }\textbf
  {\bibinfo {volume} {98}},\ \bibinfo {pages} {046402} (\bibinfo {year}
  {2007})}\BibitemShut {NoStop}%
\bibitem [{\citenamefont {Marzari}\ \emph {et~al.}(2012)\citenamefont
  {Marzari}, \citenamefont {Mostofi}, \citenamefont {Yates}, \citenamefont
  {Souza},\ and\ \citenamefont {Vanderbilt}}]{Marzari2012}%
  \BibitemOpen
  \bibfield  {author} {\bibinfo {author} {\bibfnamefont {N.}~\bibnamefont
  {Marzari}}, \bibinfo {author} {\bibfnamefont {A.~A.}\ \bibnamefont
  {Mostofi}}, \bibinfo {author} {\bibfnamefont {J.~R.}\ \bibnamefont {Yates}},
  \bibinfo {author} {\bibfnamefont {I.}~\bibnamefont {Souza}}, \ and\ \bibinfo
  {author} {\bibfnamefont {D.}~\bibnamefont {Vanderbilt}},\ }\href {\doibase
  10.1103/RevModPhys.84.1419} {\bibfield  {journal} {\bibinfo  {journal} {Rev.
  Mod. Phys.}\ }\textbf {\bibinfo {volume} {84}},\ \bibinfo {pages} {1419}
  (\bibinfo {year} {2012})}\BibitemShut {NoStop}%
\bibitem [{\citenamefont {Soluyanov}\ and\ \citenamefont
  {Vanderbilt}(2012)}]{Vanderbilt2012}%
  \BibitemOpen
  \bibfield  {author} {\bibinfo {author} {\bibfnamefont {A.~A.}\ \bibnamefont
  {Soluyanov}}\ and\ \bibinfo {author} {\bibfnamefont {D.}~\bibnamefont
  {Vanderbilt}},\ }\href {\doibase 10.1103/PhysRevB.85.115415} {\bibfield
  {journal} {\bibinfo  {journal} {Phys. Rev. B}\ }\textbf {\bibinfo {volume}
  {85}},\ \bibinfo {pages} {115415} (\bibinfo {year} {2012})}\BibitemShut
  {NoStop}%
\bibitem [{\citenamefont {Panati}\ and\ \citenamefont
  {Pisante}(2013)}]{Panati}%
  \BibitemOpen
  \bibfield  {author} {\bibinfo {author} {\bibfnamefont {G.}~\bibnamefont
  {Panati}}\ and\ \bibinfo {author} {\bibfnamefont {A.}~\bibnamefont
  {Pisante}},\ }\href@noop {} {\bibfield  {journal} {\bibinfo  {journal}
  {Commun. Math. Phys.}\ }\textbf {\bibinfo {volume} {322}},\ \bibinfo {pages}
  {835} (\bibinfo {year} {2013})}\BibitemShut {NoStop}%
\bibitem [{\citenamefont {Winkler}\ \emph {et~al.}(2016)\citenamefont
  {Winkler}, \citenamefont {Soluyanov},\ and\ \citenamefont
  {Troyer}}]{Troyer2016}%
  \BibitemOpen
  \bibfield  {author} {\bibinfo {author} {\bibfnamefont {G.~W.}\ \bibnamefont
  {Winkler}}, \bibinfo {author} {\bibfnamefont {A.~A.}\ \bibnamefont
  {Soluyanov}}, \ and\ \bibinfo {author} {\bibfnamefont {M.}~\bibnamefont
  {Troyer}},\ }\href {\doibase 10.1103/PhysRevB.93.035453} {\bibfield
  {journal} {\bibinfo  {journal} {Phys. Rev. B}\ }\textbf {\bibinfo {volume}
  {93}},\ \bibinfo {pages} {035453} (\bibinfo {year} {2016})}\BibitemShut
  {NoStop}%
\bibitem [{Note4()}]{Note4}%
  \BibitemOpen
  \bibinfo {note} {More precisely, $\ket {\psi _{n\protect \boldsymbol
  {k}+\protect \boldsymbol {G}}}=\ket {\psi _{n\protect \boldsymbol {k}}}$ and
  $U(\protect \boldsymbol {k}+\protect \boldsymbol {G})=U(\protect \boldsymbol
  {k})$ for any reciprocal lattice vector $\protect \boldsymbol {G}$. For more
  details, see, e.g., Vanderbilt \cite {VanderbiltBook}.}\BibitemShut {Stop}%
\bibitem [{Note5()}]{Note5}%
  \BibitemOpen
  \bibinfo {note} {The derived expressions can later be applied to lower
  dimensional systems by confining the Bloch and Wannier functions to the
  appropriate subspace of $\protect \mathbb {R}^3$. However, in systems with
  spatial dimension less than three, the Chern-Simons contribution vanishes.
  Thus, three-dimensional systems are of primary interest here.}\BibitemShut
  {Stop}%
\bibitem [{Note6()}]{Note6}%
  \BibitemOpen
  \bibinfo {note} {By ``ordinary insulator'' we mean crystalline insulators
  supporting Bloch energy eigenvectors for which there exists no topological
  obstruction to choosing a smooth gauge that can respect some underlying
  symmetry of the system. For instance, there exists no obstruction to choosing
  a time-reversal or inversion symmetric gauge for a system with the same
  discrete symmetry.}\BibitemShut {Stop}%
\bibitem [{Note7()}]{Note7}%
  \BibitemOpen
  \bibinfo {note} {See, e.g., Peskin and Schroeder \cite {Peskin}.}\BibitemShut
  {Stop}%
\bibitem [{Note8()}]{Note8}%
  \BibitemOpen
  \bibinfo {note} {Rodrigo A. Muniz, J. L. Cheng, and J. E. Sipe, in
  preparation}\BibitemShut {NoStop}%
\bibitem [{Note9()}]{Note9}%
  \BibitemOpen
  \bibinfo {note} {For a review and references to original work see Ref.~\cite
  {PZW}.}\BibitemShut {Stop}%
\bibitem [{\citenamefont {van Driel}\ and\ \citenamefont
  {Sipe}(2001)}]{SipeBook}%
  \BibitemOpen
  \bibfield  {author} {\bibinfo {author} {\bibfnamefont {H.~M.}\ \bibnamefont
  {van Driel}}\ and\ \bibinfo {author} {\bibfnamefont {J.~E.}\ \bibnamefont
  {Sipe}},\ }\href@noop {} {\emph {\bibinfo {title} {Coherence Control of
  Photocurrents in Semiconductors}}}\ (\bibinfo  {publisher} {Springer, New
  York},\ \bibinfo {year} {2001})\ Chap.~\bibinfo {chapter} {5}\BibitemShut
  {NoStop}%
\bibitem [{Note10()}]{Note10}%
  \BibitemOpen
  \bibinfo {note} {As previously discussed, in this paper, we move the gauge
  freedom of the energy eigenvectors, and thus the gauge dependence of the
  connections $\xi ^a_{mn}$, into the $U_{n\alpha }$ matrices.}\BibitemShut
  {Stop}%
\bibitem [{Note11()}]{Note11}%
  \BibitemOpen
  \bibinfo {note} {The relator $\alpha ^{ib}(\protect \boldsymbol {x};\protect
  \boldsymbol {y},\protect \boldsymbol {R})$ is not to be confused with the OMP
  tensor $\alpha ^{il}$ introduced in Eq.~(\ref {OMP}).}\BibitemShut {Stop}%
\bibitem [{\citenamefont {Resta}(2010)}]{Resta2010}%
  \BibitemOpen
  \bibfield  {author} {\bibinfo {author} {\bibfnamefont {R.}~\bibnamefont
  {Resta}},\ }\href {\doibase 10.1088/0953-8984/22/12/123201} {\bibfield
  {journal} {\bibinfo  {journal} {Journal of Physics: Condensed Matter}\
  }\textbf {\bibinfo {volume} {22}},\ \bibinfo {pages} {123201} (\bibinfo
  {year} {2010})}\BibitemShut {NoStop}%
\bibitem [{\citenamefont {Aversa}\ and\ \citenamefont {Sipe}(1995)}]{Aversa}%
  \BibitemOpen
  \bibfield  {author} {\bibinfo {author} {\bibfnamefont {C.}~\bibnamefont
  {Aversa}}\ and\ \bibinfo {author} {\bibfnamefont {J.~E.}\ \bibnamefont
  {Sipe}},\ }\href {\doibase 10.1103/PhysRevB.52.14636} {\bibfield  {journal}
  {\bibinfo  {journal} {Phys. Rev. B}\ }\textbf {\bibinfo {volume} {52}},\
  \bibinfo {pages} {14636} (\bibinfo {year} {1995})}\BibitemShut {NoStop}%
\bibitem [{Note12()}]{Note12}%
  \BibitemOpen
  \bibinfo {note} {As previously discussed, in this paper, we move the gauge
  freedom of the energy eigenvectors, and thus the gauge dependence of the
  connections $\xi ^a_{mn}$, into the $U_{n\alpha }$ matrices.}\BibitemShut
  {Stop}%
\bibitem [{Note13()}]{Note13}%
  \BibitemOpen
  \bibinfo {note} {See, e.g., Zhong \protect \textit {et al.}~\cite
  {Souza2016}.}\BibitemShut {Stop}%
\bibitem [{\citenamefont {Chen}\ and\ \citenamefont {Lee}(2012)}]{Lee2012}%
  \BibitemOpen
  \bibfield  {author} {\bibinfo {author} {\bibfnamefont {K.-T.}\ \bibnamefont
  {Chen}}\ and\ \bibinfo {author} {\bibfnamefont {P.~A.}\ \bibnamefont {Lee}},\
  }\href {\doibase 10.1103/PhysRevB.86.195111} {\bibfield  {journal} {\bibinfo
  {journal} {Phys. Rev. B}\ }\textbf {\bibinfo {volume} {86}},\ \bibinfo
  {pages} {195111} (\bibinfo {year} {2012})}\BibitemShut {NoStop}%
\bibitem [{\citenamefont {Olsen}\ \emph {et~al.}(2017)\citenamefont {Olsen},
  \citenamefont {Taherinejad}, \citenamefont {Vanderbilt},\ and\ \citenamefont
  {Souza}}]{Souza2017}%
  \BibitemOpen
  \bibfield  {author} {\bibinfo {author} {\bibfnamefont {T.}~\bibnamefont
  {Olsen}}, \bibinfo {author} {\bibfnamefont {M.}~\bibnamefont {Taherinejad}},
  \bibinfo {author} {\bibfnamefont {D.}~\bibnamefont {Vanderbilt}}, \ and\
  \bibinfo {author} {\bibfnamefont {I.}~\bibnamefont {Souza}},\ }\href
  {\doibase 10.1103/PhysRevB.95.075137} {\bibfield  {journal} {\bibinfo
  {journal} {Phys. Rev. B}\ }\textbf {\bibinfo {volume} {95}},\ \bibinfo
  {pages} {075137} (\bibinfo {year} {2017})}\BibitemShut {NoStop}%
\bibitem [{\citenamefont {Peskin}\ and\ \citenamefont
  {Schroeder}(1995)}]{Peskin}%
  \BibitemOpen
  \bibfield  {author} {\bibinfo {author} {\bibfnamefont {M.~E.}\ \bibnamefont
  {Peskin}}\ and\ \bibinfo {author} {\bibfnamefont {D.~V.}\ \bibnamefont
  {Schroeder}},\ }\href@noop {} {\emph {\bibinfo {title} {{An Introduction to
  quantum field theory}}}}\ (\bibinfo  {publisher} {Addison-Wesley},\ \bibinfo
  {address} {Reading, USA},\ \bibinfo {year} {1995})\BibitemShut {NoStop}%
\bibitem [{\citenamefont {Zhong}\ \emph {et~al.}(2016)\citenamefont {Zhong},
  \citenamefont {Moore},\ and\ \citenamefont {Souza}}]{Souza2016}%
  \BibitemOpen
  \bibfield  {author} {\bibinfo {author} {\bibfnamefont {S.}~\bibnamefont
  {Zhong}}, \bibinfo {author} {\bibfnamefont {J.~E.}\ \bibnamefont {Moore}}, \
  and\ \bibinfo {author} {\bibfnamefont {I.}~\bibnamefont {Souza}},\ }\href
  {\doibase 10.1103/PhysRevLett.116.077201} {\bibfield  {journal} {\bibinfo
  {journal} {Phys. Rev. Lett.}\ }\textbf {\bibinfo {volume} {116}},\ \bibinfo
  {pages} {077201} (\bibinfo {year} {2016})}\BibitemShut {NoStop}%
\end{thebibliography}%

\end{document}